\documentclass[12pt]{article}

\usepackage[letterpaper,hmargin=27mm,vmargin=23mm]{geometry}
\usepackage{amsfonts,amsmath,amsthm,amssymb,bm,bbm,dsfont}
\usepackage[utf8]{inputenc}
\usepackage{lmodern}
\usepackage{upgreek}
\usepackage{enumitem}
\usepackage{natbib,color}
\usepackage[pdftex]{graphicx}
\usepackage{subfigure,float,booktabs}
\usepackage{collcell}
\definecolor{winered}{rgb}{0.5,0,0}
\usepackage[pdfstartview=FitH,  bookmarks=true,  bookmarksopen=true,  breaklinks=true,  colorlinks=true,
linkcolor=winered,anchorcolor=my,  citecolor=blue, filecolor=blue,  menucolor=blue,  urlcolor=black]{hyperref} 
\usepackage{array,makecell}
\usepackage{booktabs,makecell}

\numberwithin{equation}{section}
\newtheorem{theorem}{Theorem}[section]
\newtheorem{lemma}[theorem]{Lemma}

\newtheorem{corollary}[theorem]{Corollary}

\newtheorem{remark}{Remark}

\theoremstyle{definition}

{\theoremstyle{plain}
	\newtheorem{assumption}{Assumption}}
\definecolor{my}{rgb}{0.05,0.05,0.5}
\definecolor{myBlue}{rgb}{.1,.1,.5}
\definecolor{myGreen}{rgb}{0,.4,0}
\definecolor{myRed}{rgb}{.25,0.15,.5}

\definecolor{my}{rgb}{0.05,0.05,0.5}
\newcommand{\eps}{\varepsilon}
\newcommand{\cond}{\displaystyle \stackrel{d}{\longrightarrow}}

\newcommand{\conas}{\stackrel{a.s.}{\longrightarrow}}
\newcommand{\conp}{\stackrel{p}{\longrightarrow}}

\makeatletter
\renewcommand\paragraph{%
	\@startsection{paragraph}{4}{0mm}%
	{-\baselineskip}%
	{.2\baselineskip}%
	{\normalfont\normalsize\bfseries}}
\makeatother

\renewcommand{\limsup}{\displaystyle \operatornamewithlimits{\lim\sup \ }}
\renewcommand{\liminf}{\displaystyle \operatornamewithlimits{\lim\inf \ }}
\newcommand{\ipart}[1]{\left \lfloor{#1}\right \rfloor}
\newcommand{\Id}{\text{Id}}

\newcommand{\Norm}[1]{\mathcal{N}\left(#1\right)}
\newcommand{\eqd}{\overset{d}{=}}

\newcommand{\indep}{\perp\!\!\!\perp}
\newcommand{\N}{\mathbb{N}}
\newcommand{\R}{\mathbb{R}}
\newcommand{\C}{\operatorname{Cov}}
\newcommand{\G}{\mathbb{G}}
\newcommand{\E}{\operatorname{E}}
\newcommand{\V}{\operatorname{Var}}

\newcommand{\Rmnum}[1]{\expandafter\@slowromancap\romannumeral #1@}
\begin{document}
	
	{\title{{A Robust Permutation Test for Subvector Inference in Linear Regressions}\thanks{We thank three anonymous referees and conference and seminar participants at CIREQ, TSE and York for their useful comments.}}
	\date{}
	\author{
		Xavier D'Haultf\oe{}uille\thanks{CREST-ENSAE, xavier.dhaultfoeuille@ensae.fr. Xavier D'Haultfoeuille thanks the hospitality of PSE where part of this research was  conducted.}
		\and
		Purevdorj Tuvaandorj\thanks{York University, {tpujee@yorku.ca.}}
	}
}
	\maketitle
	\begin{abstract}
We develop a new permutation test for inference on a subvector of coefficients in linear models. The test is exact when the regressors and the error terms are independent. Then, we show that the test is asymptotically of correct level, consistent and has power against local alternatives when the independence condition is relaxed, under two main conditions. The first is a slight reinforcement of the usual absence of correlation between the regressors and the error term. The second is that the number of strata, defined by values of the regressors not involved in the subvector test, is small compared to the sample size. The latter implies that the vector of nuisance regressors is discrete. Simulations and  empirical illustrations suggest that the test has good power in practice if, indeed, the number of strata is small compared to the sample size. 

		\begin{description}
			\item[Keywords:] Linear regressions, permutation tests, exact tests, asymptotic validity, heteroskedasticity.
		\end{description}
	\end{abstract}
	\section{Introduction}
	
Inference in linear regressions is one of the oldest problems in statistics. The first tests, developed by Student and Fisher, are still in use today, with their nice features of being  exact with independent, normally distributed unobserved terms but also asymptotically valid under homoskedasticity only. Heteroskedasticity is a common phenomenon, however. As a result, many applied researchers nowadays rely instead on robust t- and Wald tests based on robust variance estimators \citep{White(1980)}. These tests are not the panacea, yet: they are only asymptotically valid, and may suffer from important distortions in finite samples \citep{MacKinnon(2013)}, even if the unobserved terms are independent of the regressors.

\par
Recently, \cite{DR2017} show that it is actually possible to construct  a test sharing the desirable properties of both approaches. Specifically, they develop a permutation test that is exact in finite samples under independence, but also asymptotically valid under weak exogeneity conditions, allowing in particular for heteroskedasticity. However, under general conditions, the exact validity of the test only holds for inference on the whole vector of parameters. Most often, researchers are interested in subvector (e.g., scalar) inference. When applied to such subvectors, their test is exact only if the regressors corresponding to the subvector that is tested and other regressors are independent. This condition is seldom satisfied in practice.\footnote{An important exception is randomized experiments, where the treatment is often drawn independently of all observed variables.} \cite{DR2017} also develop a partial correlation test for specific components, analogous to the residual permutation test of \cite{Freedman-Lane(1983)} (see \cite{Toulis(2022)} for an extension). Their test is asymptotically valid but is not exact in finite samples, even if the unobserved terms are independent of the regressors.
	 	
\par
The objective of this paper is to extend the results of \cite{DR2017}  by developing a test on subvectors that is exact under a conditional independence assumption but also consistent under a weaker exogeneity condition. To this end, we consider a new, ``stratified randomization'' test (SR test hereafter) for a parameter vector $\beta$ based on a heteroskedasticity-robust Wald statistic in the partitioned regression
\begin{equation}
y=X\beta+Z\gamma+u,	
	\label{E1}
\end{equation}
where  $X$ is the regressors matrix of interest, $Z$ is the matrix of the nuisance regressors and $u$ is the vector of error terms. The key idea is to stratify the data according to the different values of $Z$ and permute the set of observations within each stratum. It is exact if $X$ is independent of the error term $u$ conditional on $Z$, without any restriction on the dependence between $X$ and $Z$.
	
\par
The test is also asymptotically valid and has power against local alternatives under a weak exogeneity condition. Specifically, we assume that $X$ and $u$ are uncorrelated conditonal on $Z$.  This condition is stronger than the usual condition of no correlation between $u$ and $(X,Z)$, but weaker than the mean independence condition $\E[u|X,Z]=0$. To obtain this result, we assume in particular  that the number of strata $S$, equivalently the cardinality of the empirical support of $Z$, is small compared to $n$. This condition fails if $Z$ has a continuous distribution. But it holds if the distribution of $Z$ is discrete, with a finite support or even an infinite one (as with, e.g., a Poisson distribution or a multivariate extension of it) provided that some moment of $Z$ is finite.

\par
The main technical challenge for proving our asymptotic results is that as the sample size tends to infinity, there may be a growing number of strata, some being large and others small. We then consider separately ``large'' and ``small'' strata. For large strata, whose number may still tend to infinity, we establish a permutation central limit theorem using Stein's method \citep[see, e.g.][for an exposition of this method]{CGS(2011)}. To this end, we derive a permutation version of the Marcinkiewicz-Zygmund inequality which, to our knowledge, is also new. For small strata, the combinatorial central limit theorem does not apply because the strata sizes may not tend to infinity. Instead, we use independence of units belonging to different strata, and the central limit theorem for triangular arrays.

\par
 We also study the performance of the SR test and other tests, through simulations. We show that the exactness of the SR test may endow it with a power advantage in some DGPs where other tests are underpowered, at least for small to moderate $n$. Otherwise, the SR test seems to have comparable power to other tests when $S/n$ is small. In the heteroskedastic case we explore, the SR test has a level closer to the nominal level than most of the other tests we consider. We also study in simulations an approximate version of the SR test, which may be useful when $Z$ is not discrete or $S/n$ is large. The approximate SR test is the SR test based on the strata obtained by discretizing the index $Z\widehat{\gamma}$, where $\widehat{\gamma}$ is the OLS estimator of $\gamma$. The approximate SR test displays some level of distortions, but still outperforms the partial correlation permutation test or the standard heteroskedasticity-robust test in some cases.

\par
Finally, we consider two applications. The first studies the effect of some policies on traffic fatalities in the US, whereas the second revisits the effect of class size on students' achievement, using data from the project STAR (Student-Teacher Achievement Ratio). In both cases, confidence intervals on the effect of the evaluated policy obtained by inverting the SR test are informative. In the second application, the SR confidence intervals are always smaller than the usual, so-called HC3, confidence intervals. We also present evidence that inference based on the SR test may be preferable than that based on robust standard errors with these data.

\medskip\noindent
{\bf Related literature.} As mentioned above, the paper is most closely related to \cite{DR2017}. We extend their work by developing subvector inference that is exact under conditional independence between the covariates of interest and the unobserved term. Another related work is \cite{Lei-Bickel(2019)}, who develop a cyclic permutation test on subvectors. Their test is exact under a stronger independence condition than ours, but without any restriction on the regressors. On the other hand, they do not study the asymptotic behavior of their test under weaker conditions, and our simulations strongly suggest that it is not asymptotically valid under heteroskedasticity. 

\par
The idea of a stratified (or a restricted) randomization appears in the context of experimental designs \citep{Edgington(1983), Good(2013)} and evaluation of treatment with randomly assigned instruments \citep{Imbens-Rosenbaum(2005)} and treatments \citep{Bugni-Canay-Shaikh(2018)}.
While the theoretical results developed in this paper are for observed units randomly sampled from some population, they are also directly applicable in experimental settings where units are randomly assigned to treatments as considered by the aforementioned studies.\footnote{\cite{Lehmann(1975)} calls the former the \emph{population model} and the latter the \emph{randomization model}. Regarding the terms ``permutation test" and ``randomization test" which are often used interchangeably, \cite{Edgington-Onghena(2007)} indicate that the former typically refers to permutation methods in population model while the latter is used for permutation methods in randomization model. Related to this distinction, see \cite{AAIW(2020)} for obtaining standard errors in regressions in the presence of design-based and/or sampling-based uncertainties.} In particular, when applied to \cite{Imbens-Rosenbaum(2005)}'s setup, our results allow one to fully characterize the asymptotic distribution of their test statistic without restricting the strata sizes and, at the same time, render it heteroskedasticity-robust.

\par
Yet another approach to constructing exact permutation tests in the presence of nuisance parameters is to condition, whenever available, on a sufficient statistic for the nuisance parameters. This approach has been pursued by \cite{Rosenbaum(1984)} for testing sharp null of no treatment effect under the logit assumption for the propensity score. However, in the absence of parametric assumptions as in our context, a low-dimensional sufficient statistic cannot be obtained in general. The approximate SR test we consider in the simulations is nonetheless related to \cite{Rosenbaum(1984)}, in the sense that it depends only on the index $Z\widehat{\gamma}$, rather than on the full set of regressors $Z$.

\par
\medskip\noindent
{\bf Organization of the paper.} The paper is organized as follows. Section \ref{MS} introduces the set-up and develops the test. Section \ref{sec:prop} studies the finite-sample and asymptotic properties of the test. In Section \ref{sec:MC}, we compare the performance of our test with alternative procedures through simulations. The two applications are considered in Section \ref{sec:applications}, while Section \ref{sec:concl} concludes. The appendix gathers all proofs and additional results on the project STAR.


	
\section{The set-up and definition of the test}\label{MS}

\subsection{Construction of the test} 
\label{sub:construction_of_the_test}
	
	Consider \eqref{E1}, where $y=[y_{1},\dots, y_{n}]^{\prime}$ is a $n\times 1$ vector of dependent variables, $X=[X_1,\dots, X_n]^{\prime}$ and $Z=[Z_1,\dots, Z_n]^{\prime}$ are $n\times k$  and $n\times p$  regressors, where $Z$ is assumed to include the intercept, and ${u}=[u_1,\dots, u_n]^{\prime}$ is a $n\times 1$ vector of exogenous error terms (see  Assumptions \ref{A1} and \ref{A2}\ref{2cu} below). $\beta$ and $\gamma$ are $k\times 1$ and $p\times 1$ vector of unknown regression coefficients, respectively.  We consider tests of the restriction\footnote{We thus do not consider tests on the intercept. These would require a different approach from that considered below.}
	\begin{equation}\label{H}
		H_0:\beta=\beta_0.
	\end{equation}
	
\begin{remark}
The tests of \eqref{H} are useful not only for usual linear models with exogenous regressors, but also in the case of endogenous regressors. In the latter case, we have a model $y=Y\delta+Z\gamma+u$, where the endogenous set of regressors $Y$ is instrumented by $X$. Then,  following the approach by \cite{Anderson-Rubin(1949)}, we can test for $\delta=\delta_0$ by testing that the $X$ coefficients in the regression of $y-Y\delta_0$ on $X$ and $Z$ are equal to 0 \citep[see][for such an approach with permutation tests]{Pujee2021}. 
\end{remark}
	
\par
To formally define our test, we introduce the following notation. Let $W_i=(X_i^{\prime}, Z_i^{\prime})^{\prime}$, $W= [X, Z]$ and for any $n\times m$ matrix $A$, $M_{A}=I_{n}-A(A^{\prime}A)^{-1}A^{\prime}$. The set of all permutations of $\{1,\dots, n\}$ is  denoted by $\G_n$,  with $\Id\in\mathbb{G}_n$ corresponding to the identity permutation. For any $\pi\in\G_n$ and vector $c=[c_1,\dots,c_n]'$, we let $c_\pi=[c_{\pi(1)},\dots,c_{\pi(n)}]'$. Similarly, for a matrix $A$ with $n$ rows, $A_\pi$ is the matrix obtained by permuting the rows according to $\pi$.\par
	
	Our test differs from that of \cite{DR2017} in that instead of considering $\mathcal{W}^\pi$ for all $\pi\in\G_n$, we focus on a subset of $\G_n$. Let $\{z_1,\dots,z_S\}$ denote the set of  (distinct) observed values in the sample. Let $n_s=\vert \{i:Z_i=z_s\}\vert$ for $s\in\{1,\dots,S\}$. Let also $y^s$ denote the subvector of $y$ including the rows $i$ satisfying $Z_i=z_s$, and define $X^s$ and $Z^s$ similarly. Without loss of generality, assume that the vector $y$ and matrices $X$ and $Z$ are arranged such that
	\begin{equation}\label{eq: partitioning}
		[y, X, Z]=
		\begin{bmatrix}
			y^1& X^1& Z^1\\
			\vdots&\vdots&\vdots\\
			y^S& X^S& Z^S
		\end{bmatrix}.
	\end{equation}
	Let the corresponding partitioning of the error terms be $u=[u^1,\dots, u^S]'$, and
	$\tilde{X}=[\tilde{X}^{1\prime},\dots, \tilde{X}^{S\prime}]'$, where $\tilde{X}^s=M_{\bm{1}_s}X^s$ and $\bm{1}_s$ denotes the $n_s\times 1$ vector of ones.\footnote{On the other hand, we do not modify the labels $1,\dots,n$ of the units. This means that $y\ne [y_1,\dots,y_n]'$, for instance ($y$ is a permuted version of $[y_1,\dots,y_n]'$).} Also, for any vector $v=[v_1,\dots,v_n]'$, let $\Sigma(v)$ denote the diagonal matrix with $(i,i)$ element equal to $[Dv]_i^2$, with $D\equiv \mathrm{diag}(M_{{\bm{1}_1}},\dots, M_{{\bm{1}_S}})$ and define
	\begin{equation*}
		g(W,v)=v'\tilde{X}\left(\tilde{X}^{\prime}\Sigma(v)\tilde{X}\right)^{-1}\tilde{X}^{\prime}v.
	\end{equation*}
	We consider the following heteroskedasticity-robust Wald statistic for $H_0$:
	\begin{equation}\label{Wald}
		\mathcal{W}=g(W,y-X\beta_0).
	\end{equation}
	We now construct a permutation test based on $\mathcal{W}$. The idea behind permutations tests is that the distribution of some test statistic ($\mathcal{W}$ here)  remains invariant under $H_0$ if we permute the data in an appropriate way. Then, we reject $H_0$ at the level $\alpha\in(0,1)$ if the test statistic on the initial data is larger than $100\times(1-\alpha)\%$ of the test statistics obtained on all possible permuted data.
	
	\par
	First, we define the permuted version of $\mathcal{W}$ as
	\begin{equation}\label{RWald}
		\mathcal{W}^\pi=g(W, (y-X\beta_0)_\pi).
	\end{equation}
	Second, we focus on permutations $\pi$ such that for any $k\in \{1,\dots,n\}$, we have, for all $t\in \{1,\dots,S\}$,
	$$\sum_{s=1}^{t-1} n_s < k\leq \sum_{s=1}^{t} n_s \Rightarrow \sum_{s=1}^{t-1} n_s < \pi(k)\leq \sum_{s=1}^{t} n_s \quad \left(\text{with } \sum_{s=1}^0 n_s=0\right).$$
	That is, $\pi$ shuffles rows only within each of the $S$ strata, ensuring that $Z_\pi=Z$. We denote by $\mathbb{S}_n$ the set of such ``stratified'' permutations. Clearly, $\Id\in\mathbb{S}_n$.\par
	
	With the test statistic $\mathcal{W}$, its permuted version $\mathcal{W}^\pi$ and the set of admissible permutations $\mathbb{S}_n$ in hand, we define a level-$\alpha$  stratified randomization test  by following the general construction of permutation tests. Let $N\in\{1,\dots,|\mathbb{S}_n|\}$, possibly random but independent of $y$ conditional on $W$ and let $\mathbb{S}'_n \subset \mathbb{S}_n$ be such that (i) $\Id\in\mathbb{S}'_n$; (ii) $\mathbb{S}'_n\backslash \{\Id\}$ is  obtained by simple random sampling  without replacement of size $N-1$ from $\mathbb{S}_n\backslash \{\Id\}$. Note that if $N=|\mathbb{S}_n|$, we simply have $\mathbb{S}'_n = \mathbb{S}_n$.\footnote{\label{foot:Sprime} A practical way to obtain $\mathbb{S}'_n$ is (i) to pick $N'-1$ permutations at random from $\mathbb{S}_n$, with replacement and with equal probability, (ii) to add $\Id$ to this initial set, (iii) to delete the duplicates (if any) from this set.} Then, let
	\begin{equation*}
		\mathcal{W}^{(1)}\leq \mathcal{W}^{(2)}\leq \dots \leq \mathcal{W}^{(N)},
	\end{equation*}
	be the order statistics of $(\mathcal{W}^\pi)_{\pi\in \mathbb{S}'_n}$. Let $q=N-\ipart{N\alpha}$ (with $\ipart{x}$ the integer part of $x$), $N^{+}=\vert \{i\in \{1, \dots, N\}: \mathcal{W}^{(i)}> \mathcal{W}^{(q)}\}\vert $ and $N^{0}
	=\vert \{i\in \{1, \dots, N\}: \mathcal{W}^{(i)}= \mathcal{W}^{(q)}\}\vert$.  We define the level-$\alpha$ test function $\phi_\alpha$  by
	\begin{equation}
		\phi_\alpha=
		\begin{cases}
			1& \text{if}\ \mathcal{W}>\mathcal{W}^{(q)},\\
			\frac{N\alpha-N^{+}}{N^0} & \text{if}\ \mathcal{W}=\mathcal{W}^{(q)},\\
			0& \text{if}\ \mathcal{W}<\mathcal{W}^{(q)}.
		\end{cases}
		\label{eq:def_test}
	\end{equation}
Increasing $N$ reduces the role of randomness but increases the computational cost of the test. The exact result (Theorem \ref{RPS} below) holds for any $N$ but for the asymptotic results, we will assume that $N\conp\infty$ as $n\to\infty$.
	
	\par
	The power of the test is directly related to $|\mathbb{S}_n|=\prod_{s=1}^S n_s!$. If $|\mathbb{S}_n|=1$, which occurs if $Z$ includes a continuous component,\footnote{This is at least the case if the data are i.i.d.. If not, we may have $S<n$ with positive probability even if $Z$ includes a continuous component.} the test becomes trivial: $\phi_\alpha=\alpha$. On the other hand, if $|\mathbb{S}_n|>1$, the test is non-trivial, and we may have $\phi_\alpha=1$ as soon as $|\mathbb{S}_n|\ge 1/\alpha$: this occurs if $\mathcal{W}>\max_{\pi \in\mathbb{S}_n:\pi\ne\Id} \mathcal{W}^\pi$. For instance, $S=n-4$, $n_{s_1}=3$, $n_{s_2}=n_{s_3}=2$ for some $(s_1,s_2,s_3)$ and $n_s=1$ otherwise is sufficient to induce $|\mathbb{S}_n|\ge 1/\alpha$ for $\alpha\ge 0.05$. More generally, even if $Z$ has an infinite support, as with count data (e.g., Poisson distributed variables), $|\mathbb{S}_n|$ may be large, and $S/n$ small. We refer to Lemma \ref{lem:suff_cond_stratasize} below for a formal result along these lines.


\subsection{Approximate test} 
\label{sub:approximate_test}

As explained above, our test is trivial when $Z$ is continuous. More generally, it may also have low power with many strata. To circumvent these issues, we consider here an approximate version of our initial proposal. Specifically, instead of constructing strata based on $\{Z_i\}_{i=1}^n$, we rely on a discretization of $I_i\equiv Z'_i\widehat{\gamma}$, with $\widehat{\gamma}$ the OLS estimator of $\gamma$. Letting $u^1=\min_{i=1,\dots,n} I_i$, $u^{S+1} =\max_{i=1,\dots,n} I_i$ and $u^s = u^1 + (u^S-u^1) (s-1)/S$ for $s=2,\dots,S$, one such discretization is simply
$$\sum_{s=1}^S s 1(I_i\in [u^{s}, u^{s+1})),$$
where $1(\cdot)$ denotes the indicator function. Intuitively, the number of strata $S$ that one chooses trades off size distortion and power. With a low $S$, the test is distorted because there are still variations in $\{I_i\}$ within each stratum, but we can expect larger power since the effective sample size $n-S$ is larger.

\par
Because of the discretization, the test is not exact in general, even if $X$ and $u$ are conditionally independent. While we leave the study of its asymptotic validity for future research, we evaluate in Subsection \ref{sub:MC_app} its performances and the impact of the choice of $S$ through Monte Carlo simulations.


\subsection{Confidence regions} 
\label{sub:confidence_regions}

The confidence region for the parameters can be obtained by test inversion. Given the duality between tests and confidence sets, the finite and large sample validity of the confidence regions follow from the corresponding test results.

\par
We recommend using the same set of permutations $\mathbb{S}'_n$ for different tested values $\beta_0$ and the same random variable in case randomization is required for the test (namely, when $\mathcal{W}=\mathcal{W}^{(q)}$ in \eqref{eq:def_test} above), to avoid random fluctuations that could create ``holes'' in the confidence region.\footnote{\label{foot:trivial_test} If the test is trivial ($S=n$), confidence intervals obtained by test inversion may be empty. To avoid this issue, one can define the confidence interval as $(-\infty,\infty)$ in such cases, instead of randomizing.} Even if we do so, inverting the test may not lead to a convex confidence region. In this case, taking the convex hull of the corresponding region is a simple and conservative fix. This issue is not specific to our test but potentially arises with all permutation tests for which the critical value depends on the parameter value that is tested ($\beta_0$ in our context). For discussions about permutation confidence intervals and finding their endpoints, we refer to  \cite{Garthwaite(1996)}, \cite{Good(2013)} and \cite{Wang-Rosenberger(2020)}.


	\section{Statistical properties of the test}
	\label{sec:prop}
	
	\subsection{Finite sample validity under conditional independence} 
	\label{sub:finite_sample_validity_under_independence}

	First, we prove that the SR test is exact under conditional independence between $X$ and $u$. We rely on the following conditions.

	\begin{assumption}[Finite sample validity]\label{A1}
		\leavevmode
		\begin{enumerate}[label={(\alph*)}]
			\item \label{1ex}
			For all $s=1,...,S$, the vector $u^s$ is exchangeable.
			\item \label{1indep}
			Conditional on $Z$, $X$ and $u$ are independently distributed.
			\item \label{1rank}
			$\mathrm{rank}(W)=k+p$ with probability one.
		\end{enumerate}
	\end{assumption}
	
Because all observations $i$ in stratum $s$ are such that $Z_i=z_s$, the first condition allows for any dependence between $Z$ and $u$. The first two conditions are thus weaker than unconditional exchangeability of $u$ and independence between $u$ and $W$, the conditions imposed by \cite{Lei-Bickel(2019)} and \citeauthor{DR2017} (\citeyear{DR2017}, test statistic $U_n(X,Y)$) to establish the exactness of their tests. A particular case where Condition \ref{1indep}	holds is stratified randomized experiment with homogeneous treatment effects. Let $X_i\in\{0,1\}$ denote the treatment variable of individual $i$, $Z_i$ be the vector of strata dummies in and $Y_i(x)$ the potential outcome corresponding to treatment value $x\in\{0,1\}$. With homogeneous treatment effects, we have  $Y_i(1)= \beta + Y_i(0)$ for all $i$, with $Y_i(0)\indep X_i | Z_i$. Then, letting $u_i=Y_i(0)-\E[Y_i(0)|Z_i]$, we obtain \eqref{E1} and Condition \ref{1indep}. Condition \ref{1rank} is maintained for convenience, as the Wald statistic (\ref{Wald}) and its permutation versions would still be exchangeable when defined using a generalized inverse of the covariance matrix estimator.

	\begin{theorem}[Finite sample validity]\label{RPS}
		Let us suppose that \eqref{E1}, Assumption \ref{A1} and $H_0$ hold. Then, for any  $0< \alpha< 1$,
		\begin{equation}
			\E[\phi_\alpha|W]=\alpha,\label{PWaldFSD}
		\end{equation}
	\end{theorem}
	
	Hence, the SR test is exact in finite samples. In particular, in stratified randomized experiments, the test is exact for testing sharp null hypotheses of the kind $Y(1)- Y(0)= \beta$. By focusing on permutations in $\mathbb{S}_n$, we therefore extend the result of \cite{DR2017} to designs where $X$ and $Z$ are not independent. While the details of the proofs are given in the appendix, the intuition of the result is as follows. First, one can show that for the test to be exact, it suffices to prove that the $(\mathcal{W}^\pi)_{\pi \in\mathbb{S}'_n}$ are exchangeable, conditional on $W$. Now, if $H_0$ holds, $D(y-X\beta_0)=D u$. Then, because $v$ in $g(W,v)$ is always premultiplied by $D$, we have $\mathcal{W}= g(W, u)$. Next, for any $\pi\in\mathbb{S}_n$, we have, still under $H_0$,
	$$D(y-X\beta_0)_{\pi}=D(Z_\pi\gamma+u_\pi)=Du_\pi,$$
	since $Z_\pi=Z$. Therefore, $\mathcal{W}^\pi = g(W, u_\pi)$. Conditional independence between $X$ and $u$ and exchangeability of $u^s$ for all $s$ then imply that the $(\mathcal{W}^\pi)_{\pi \in\mathbb{S}'_n}$ are exchangeable, conditional on $W$.
	

\subsection{Asymptotic validity under weaker exogeneity conditions} 
\label{sub:asymptotic_validity_under_weaker_exogeneity_conditions}

Next, we study the asymptotic validity of the SR test under weaker conditions than conditional independence between $X$ and $u$. To this end, we partly build on \cite{DR2017}, who show that a randomization test based on the Wald statistic  ($U_n(X,Y)$ in their paper) is of asymptotically correct level and heteroskedasticity-robust if, in addition to usual moment and nonsingularity conditions, the data are i.i.d. and $\E[W_iu_i]=0$.\footnote{\label{foot:PC}{A close inspection of the proof of Theorem 4.1 in \cite{DR2017} reveals that their partial correlation test, to which we compare our test below, also works if $\E[W_i u_i]=0$.}} 
We present below analogous large sample results for the SR test under similar conditions, displayed in Assumption \ref{A2} below.
	
\par
Before we present these conditions, let us make some preliminary remarks and additional definitions. First, the distribution of $\{(W_i^{\prime}, u_i)^{\prime}\}_{i=1}^n$ and $\beta$ are allowed to change with $n$. For notational convenience, we usually do not make this dependence on $n$ explicit throughout the main text, though we do it in the appendix. 
Second, in order to derive the asymptotic distribution of the stratified randomization statistic, we make a distinction between ``large'' and ``small'' strata as follows. Let $c_n$ be a sequence satisfying	$c_n\geq n^{1/2}$ and $c_n/n\to 0$ as $n\to \infty$ e.g. $c_n=n^{1/2}$. Define
	\begin{align*}
		\mathcal{I}_n
		&\equiv \{s\in\{1,\dots,S\}: n_s\geq c_n\}, \\
		\mathcal{J}_n
		&\equiv \{1,\dots,S\}\setminus \mathcal{I}_n=\{s\in\{1,\dots,S\}: n_s < c_n\}.
	\end{align*}
We explain below on why we separate strata this way. Now, let $X^s=[X_{1}^s,\dots, X_{n_s}^s]'$ with $X_{i}^{s}$ being $k\times 1$, and $u^s=[u_{1}^s,\dots, u_{n_s}^s]$ $(n_s\times 1)$ denote the regressor matrix and the vector of error terms in the $s$th stratum respectively. The covariance matrix for the Wald statistic is defined as
$$	\Omega_n
\equiv n^{-1}\sum_{i=1}^{n}
\E\left[(X_{i}-\E[X_{i}|Z_{i}])(X_{i}-\E[X_{i}|Z_{i}])'u_{i}^{2}|Z_{i}\right].$$
Define also the covariance matrices for the permutation statistic as ${V}_{n\mathcal{I}} \equiv \sum_{s\in \mathcal{I}_n}(n_s/n)\sigma_{u}^{s2}Q^s$ and $V_{n\mathcal{J}}
\equiv \sum_{s\in \mathcal{J}_n}(n_s/n)\sigma_{u}^{s2}Q^s$, where $\sigma_{u}^{s2}\equiv n_s^{-1}\sum_{i=1}^{n_s}\E[u_{i}^{2}|Z_{i}=z_{s}]$ and
$$Q^s	\equiv n_s^{-1}\sum_{i=1}^{n_s}\E[X_{i}X_{i}^{\prime}|Z_{i}=z_{s}]
-n_s^{-1}\sum_{i=1}^{n_s}\E[X_{i}|Z_{i}=z_{s}]\,n_s^{-1}\sum_{i=1}^{n_s}\E[X_{i}^{\prime}|Z_{i}=z_{s}].$$
Let $\lambda_{\min}(\cdot)$ denote the smallest eigenvalue of a symmetric matrix. We impose the following assumption on the strata.

	\begin{assumption}[Large sample validity]\label{A2}
	\leavevmode
	\begin{enumerate}[label={(\alph*)}]
		
		\item \label{2ex}
		$\{(W_{i}^{\prime}, u_{i})^{\prime}\}_{i=1}^{n}$ are independent.
		\item \label{2cu}
		$\E[(X_{i}^{\prime}, 1)' u_{i}|Z_{i}=z_{s}]=0$ for all $n$, $s=1,\dots, S$ and $i=1,\dots, n$.
		\item \label{2mom}
		There exist $m>2$ and $M_0<\infty$ not depending on $n$ such that $\sup_{s,i}\E[\Vert X_{i}\Vert^{2m}|Z_{i}=z_{s}]<M_0$ and $\sup_{s,i}\E[\vert u_{i}\vert^{2m}|Z_{i}=z_{s}]< M_0$.\footnote{Here, $\sup_{s, i}$ is a shortcut for the supremum over $s\in\{1,\dots,S\}$ and $i\in\{1,\dots,n\}$.}
		\item \label{2ns} There exists $\lambda>0$ such that $\liminf_{n\to\infty} \lambda_{\min}(\Omega_n)> \lambda$ a.s.. If $\liminf_{n\to\infty} |\mathcal{I}_n|\ge 1$ a.s. (resp. $|\mathcal{J}_n|\conas \infty$), we also have $ \liminf_{n\to\infty} \lambda_{\min}(V_{n\mathcal{I}})> \lambda$ a.s. (resp. $\liminf_{n\to\infty} \lambda_{\min}(V_{n\mathcal{J}})>\lambda$ a.s.).
		\item\label{stratasize} $n^{-1}S\conas 0$.
	\end{enumerate}
\end{assumption}

Conditions \ref{2ex}-\ref{2mom} are quite standard. Condition \ref{2mom} is a usual moment condition required for the central limit theorem with independent observations \citep{White(2001),Hansen(2021)}. Condition \ref{2cu} is slightly stronger than the usual unconditional absence of correlation between $W_{i}$ and $u_{i}$ but weaker that the usual mean independence condition $\E[u_{i}|W_{i}]=0$, and, as such, allows for any form of heteroskedasticity.\footnote{\label{foot:weaker_cond} {An example where Condition \ref{2cu} holds but $\E[u_{i}|W_{i}]\ne 0$, suppose that $X_{i}$ is continuous, independent of $Z_{i}$ (as in a standard randomized experiment) and has a nonlinear effect on $Y_{i}$, so that $\E[Y_{i}|X_{i}]=g(X_{i})+Z_{i}'\tilde{\gamma}$ for some vector $\tilde{\gamma}$ and nonlinear function $g$. If we  consider the best linear prediction of $Y_{i}$ by $(X_{i}', Z_{i}')'$, it will take the form $X_{i}'\beta+Z_{i}'\gamma$, and $u_{i}\equiv Y_{i}-X_{i}'\beta-Z_{i}'\gamma$ will satisfy Condition \ref{2cu}, though $\E[u_{i}|W_{i}]\ne 0$.}} Consider again the case of stratified randomized experiments, but now assume that treatment effects can be heterogeneous, still with $\E[Y_i(1)-Y_i(0)|Z_i]=\E[Y_i(1)-Y_i(0)]$. Then, \eqref{E1} holds, with $\beta=\E[Y_i(1)-Y_i(0)]$. Besides, $\V[u_{i}|W_{i}]$ depends on $X_{i}$ because of treatment effect heterogeneity. However, $\E[u_{i}|W_{i}]=0$, so Condition \ref{2cu} holds.

\par
Conditions \ref{2ns}-\ref{stratasize} are more specific to our setup. The first imposes that the covariance matrices corresponding to each large stratum are not degenerate. Condition \ref{2ns} also imposes the invertibility of the limit covariance matrix for $\mathcal{J}_n$ when the number of small strata is large. The latter can be replaced by the condition that $\lim_{n\to\infty}V_{n\mathcal{J}}$ exists; then the limit could be degenerate.

\par
To understand Condition \ref{stratasize}, note that because the SR test uses deviations from strata means, the ``effective'' sample size $n-S$ should tend to infinity for the test to be consistent. Condition \ref{stratasize} reinforces this requirement, since under the condition, $n-S=n(1-n^{-1}S)\conas \infty$. Condition \ref{stratasize}  obviously holds if the $\{Z_i\}_{i=1}^n$ are identically distributed; with distribution independent of $n$, and $Z_1$ has finite support. The following lemma shows that it also holds if the support of $Z_{i}$ is in $\N^p$, as with, e.g., Poisson or geometric distributions (or multivariate versions of them), provided that some moments of $Z_{i}$ are finite.
\begin{lemma}
	Suppose that $(Z_{1},\dots,Z_{n})$ are independent and identically distributed with distribution possibly depending on $n$, support included in $\N^p$ and $\E[\|Z_{i}\|^{2p}]<C_0<\infty$, with $C_0$ independent of $n$. Then Assumption \ref{A2}\ref{stratasize} holds.
	\label{lem:suff_cond_stratasize}
\end{lemma}

\begin{remark}
	$S$ does not depend on the exact values in the support of $Z$. Hence, when this support is countable but not in $\N^p$, we can still apply Lemma \ref{lem:suff_cond_stratasize} as follows. Let $\{p_r\}_{r\in\N}$ be the probabilities associated with the support points $\{z_r\}_{r\in\N}$ of $Z$. Let $\sigma$ denote a permutation of $\N$ such that $p_{\sigma(1)}\geq p_{\sigma(2)} \geq \dots$. and define the random variable $\widetilde{Z}$ as $\sigma^{-1}(r)$ when $Z=z_r$. By Lemma \ref{lem:suff_cond_stratasize}, $n^{-1} S \conas 0$ as long as $\E[\widetilde{Z}]<\infty$ or, equivalently, $\sum_{r\geq 0} \sum_{j>r} p_{\sigma(j)} <\infty$.	
	\label{rem:Z_ordered}
\end{remark}
\par
For any finite set $B$, let $\mathcal{U}(B)$ denote the uniform distribution over $B$. To establish the asymptotic properties of the SR test, we first study the asymptotic behavior of $\mathcal{W}^\pi$, with $\pi \sim \mathcal{U}(\mathbb{S}_n)$, conditional on the data.
	\begin{theorem}[Asymptotic behavior of $\mathcal{W}^\pi$]\label{thm:behav_Wpi}
		Let \eqref{E1} and Assumption \ref{A2} hold with $\beta=\beta_n$ such that  $\limsup_{n\to\infty}\Vert\beta_n\Vert<\infty$. Then, conditional on the data,
$\mathcal{W}^\pi \cond \chi^2_k$ with probability tending to one.
	\end{theorem}

The main technical difficulty, and the reason why we cannot apply the same proof as in \cite{DR2017}, is that the number of strata may tend to infinity, and there may be only small strata. To deal with these issues, we consider separately large and small strata. For large strata, we prove the following combinatorial central limit theorem with possibly many strata. Hereafter, we let $P^\pi$ denote the probability measure of $\pi$.

\begin{lemma}[Combinatorial CLT]\label{HoeffdingCLT}
	Let $\mathcal{S}$ be an integer-valued random variable, $\mathcal{S}\ge 1$ and $s=1,\dots,\mathcal{S}$ denote strata of sizes $n_s\ge 2$, with $\sum_{s=1}^{\mathcal{S}} n_s=n$. Let $\pi\sim\mathcal{U}(\mathbb{S}_n)$ and for each $s$, $\{b_{i}^s\}_{i=1}^{n_s}$ and $\{c_{i}^s\}_{i=1}^{n_s}$ be random variables satisfying:\footnote{Again, the distributions of $\mathcal{S}$, $\{b_{i}^s\}_{i=1}^{n_s}$ and $\{c_{i}^s\}_{i=1}^{n_s}$ are allowed to vary with $n$ here.}
	\begin{enumerate}[label=(\alph*)]
		\item\label{cond: centered}$\sum_{i=1}^{n_s}b_{i}^{s}=\sum_{i=1}^{n_s}c_{i}^s=0$ a.s.;
		\item\label{cond: sigcon} For $\sigma_n^2\equiv\sum_{s=1}^{\mathcal{S}} \frac{1}{n_s-1}\left(\sum_{i=1}^{n_s}b_{i}^{s2}\right)\left(\sum_{i=1}^{n_s}c_{i}^{s2}\right)$,
		$\sigma_n^2\conp \sigma^2>0$ as $n\to\infty$;
		\item \label{cond: 3mom} $\sum_{s=1}^{\mathcal{S}}n_s^{-1} \left(\sum_{i=1}^{n_s} |b_{i}^s|^3\right)
		\left(\sum_{i=1}^{n_s} |c_{i}^s|^3\right)\conp 0$ as $n\to\infty$;
		\item\label{cond: Uvarcon}
		$\sum_{s=1}^{\mathcal{S}}(n_s-1)^{-1}\left(\sum_{i=1}^{n_s}b_{i}^{s4}\right)\left(\sum_{i=1}^{n_s}c_{i}^{s4}\right)\conp 0$ as $n\to\infty$.
	\end{enumerate}
	Let $T^\pi\equiv \sum_{s=1}^{\mathcal{S}} \sum_{i=1}^{n_s} b^{s}_{i}c_{\pi(i)}^s/\sigma_n$. Then,
	$P^\pi(T^\pi\leq t)\conp \Phi(t)$ for any $t\in\mathbb{R}$ as $n\to\infty$.
\end{lemma}

The proof of this lemma relies on Stein's method with exchangeable pairs, and on the following permutation version of the Marcinkiewicz-Zygmund inequality, which, to our knowledge, is also new.

\begin{lemma}[Marcinkiewicz-Zygmund inequality for permutation]\label{lem:MZ}
		Let $a_{1},\dots, a_{n}$  and $b_{1},\dots, b_{n}$ be sequences of $d\times 1$ real vectors and scalars, respectively, with $\sum_{i=1}^nb_i=0$. Then, for any $1<r<\infty$, $n>1$, and $\pi\sim\mathcal{U}(\mathbb{G}_n)$ there exists a constant $M_r$ depending only on $r$ such that
		\begin{equation}\label{eq: PMZW ineq}
\E_\pi\left[\left\Vert\sum_{i=1}^na_{i}b_{\pi(i)}\right\Vert^r\right]\leq M_r n^{(r/2)\vee 1 }\left(n^{-1}\sum_{i=1}^n\Vert a_i\Vert^r\right)\left(n^{-1}\sum_{i=1}^n|b_i|^r\right).
		\end{equation}
\end{lemma}		

The usual combinatorial central limit theorem, used in \cite{DR2017}, would apply to a finite number of large strata; but here, the number of large strata, $|\mathcal{I}_n|$, may tend to infinity. We can accomodate that using Lemma \ref{HoeffdingCLT}. Still, to check Condition (c) therein, we need to restrict the growth of $|\mathcal{I}_n|$, by imposing  that $|\mathcal{I}_n|\le C n^{1/2}$ for some $C>0$ (see \eqref{eq: card In bound} in the proof of Theorem \ref{thm:behav_Wpi}). This is why we imposed above $c_n\ge n^{1/2}$.

\par
For small strata, the combinatorial central limit theorem  does not apply because the strata sizes may not tend to infinity. Instead, we use the fact that all strata are independent. Then, we check that the assumptions underlying a conditional version of the Lindeberg CLT for triangular arrays hold. This is technical, however, and we have to rely again on the  Marcinkiewicz-Zygmund inequality, among other tools. Also, we rely therein on the condition that strata are small enough in the sense that $c_n/n\to 0$. A similar condition is used by \cite{Hansen-Lee(2019)} to establish asymptotic normality of a sample mean with potentially many clusters.

\medskip
The asymptotic properties of the SR test are also based on the the asymptotic behavior of $\mathcal{W}$, given in the following theorem. Hereafter, we let $\chi^2_k(c)$ be the noncentral chi-squared distribution with degrees of freedom $k$ and noncentrality parameter $c$, for any $c\ge 0$.

	\begin{theorem}[Asymptotic behavior of $\mathcal{W}$]\label{thm:behav_W}
		Let us suppose that \eqref{E1} and Assumption \ref{A2} hold. Then:
		\begin{enumerate}
			\item If $\beta_n =\beta_0+hn^{-1/2}$ with $h\in\mathbb{R}^{k}$ fixed and $G\equiv \lim_{n\to\infty} \Omega_n^{-1/2}\E[n^{-1}\tilde{X}'\tilde{X}]$ exists, $\mathcal{W}\cond\chi_k^2(\|Gh\|^2)$.\footnote{If $h=0$, we need not assume that $\lim_{n\to\infty} \Omega_n^{-1/2}\E[n^{-1}\tilde{X}'\tilde{X}]$ exists.}
			\item If $n^{1/2}\Vert \E[n^{-1} \tilde{X}'\tilde{X}](\beta_n-\beta_0)\Vert\to \infty$,  $\mathcal{W}\conp \infty$.
		\end{enumerate}
	\end{theorem}

As Theorem \ref{thm:behav_Wpi}, Theorem \ref{thm:behav_W} would be standard with a finite number of strata; but here there may be many small strata. In particular, the result does not immediately follow from the result of \cite{Wooldridge(2001)}, which is derived under the assumption that each stratum frequency has a nondegenerate limit. To prove the result, we show that the assumptions underlying a conditional Lindeberg CLT hold, see Lemma \ref{lem: cond Lindeberg} in Appendix \ref{sub:key_lemmas}. 
\par

\begin{remark}
Theorem \ref{thm:behav_W}  establishes the asymptotic normality of the within OLS estimator for stratified regression models (\cite{Cameron-Trivedi(2009)}, Chapter 24.5), with possibly many small strata. Although we prove it for  linear models only, we expect the result to carry over to general $M$-estimation under stratified sampling.
\label{rem:nonlin}	
\end{remark}

\medskip
The two previous results imply the following asymptotic properties of the SR test.  Below, $q_{1-\alpha}(\chi^2_k)$ denotes the $1-\alpha$ quantile of the $\chi^2_k$ distribution.

\begin{corollary}\label{cor:SR}
	Suppose that \eqref{E1} and Assumption \ref{A2} hold and $N_n=|\mathbb{S}'_n| \conp \infty$. Then:
	\begin{enumerate}
		\item If $\beta_n =\beta_0$, $\lim_{n\to\infty} \E[\phi_\alpha((\mathcal{W}^\pi)_{\pi\in\mathbb{S}'_n})]=\alpha$.
		\item\label{H1loc}
		If $\beta_n=\beta_0+hn^{-1/2}$ and  $G\equiv \lim_{n\to\infty} \Omega_n^{-1/2}\E[n^{-1}\tilde{X}'\tilde{X}]$ exists,
$\lim_{n\to\infty} \E[\phi_\alpha((\mathcal{W}^\pi)_{\pi\in\mathbb{S}'_n})]= P[\mathcal{W}_\infty>q_{1-\alpha}(\chi^2_k)]$, where $\mathcal{W}_\infty\sim \chi_k^2(\|Gh\|^2)$.
		\item\label{H1fix}
		If $ n^{1/2}\Vert \E[n^{-1} \tilde{X}'\tilde{X}](\beta_n-\beta_0)\Vert\to\infty$, $\lim_{n\to\infty} \E[\phi_\alpha((\mathcal{W}^\pi)_{\pi\in\mathbb{S}'_n})]=1$.
	\end{enumerate}
\end{corollary}

The first result shows that the SR test is of asymptotically correct level. Combined with Theorem \ref{PWaldFSD}, this result implies that under i.i.d. sampling and technical restrictions, the SR test is exact under conditional independence between $X_i$ and $u_i$, and asymptotically valid under the weaker exogeneity restriction $\E[(X_{i}^{\prime}, 1)' u_{i}|Z_{i}=z_{s}]=0$. The second result in Corollary \ref{cor:SR} states that the test has nontrivial power to detect local alternatives. Finally, the third result implies that if $\E[n^{-1} \tilde{X}\tilde{X}']$ converges to symmetric positive definite matrix, the test is consistent for fixed alternatives, namely when $\beta_n=\beta\ne \beta_0$.


	
\section{Monte Carlo simulations} 
	\label{sec:MC}	

\subsection{Main test} 
\label{sub:main_test}

We first present some simulation evidence on the performance of the proposed test. We consider the following model:
	\begin{align*}
		y_i&=X_i\beta+\gamma_1 + \sum_{j=2}^p Z_{ij}\gamma_j +u_i, \quad i=1,\ldots, n,
	\end{align*}
	where $\beta=\gamma_1=0$, $\gamma_j=1$ for $j\geq 2$. We consider $p\in\{2,4\}$ and sample sizes $n\in\{50,100,500\}$. Then, we consider three data generating processes (DGPs) with various distributions for $X_i$ and  $u_i$. In the three cases, the $\{Z_{ij}\}_{j=2}^{p}$ are i.i.d. and follow a Poisson distribution with parameter 1. Then:
\begin{itemize}
	\item[-] In DGP1, $u_i|W_i\sim \mathcal{N}(0,1)$ and $X_i=X_i^*$, with
\begin{equation}
X^*_i = \frac{1}{\sqrt{2}}\left[\frac{1}{\sqrt{p-1}}\sum_{j=2}^p Z_{ij} + v_i\right], \quad v_i|Z_i\sim \mathcal{N}(0,1).	
	\label{eq:def_Xstar}
\end{equation}
	\item[-] In DGP2, 
	$$u_i=\left[\frac{1}{\sqrt{p-1}}\sum_{j=2}^p (Z_{ij}-1)\right] v_i, \quad P(v_i=-1|W_i)=P(v_i=1|W_i)=\frac{1}{2}$$ 
	and $X_i=\exp(X_i^*)$, with $X_i^*$ defined by \eqref{eq:def_Xstar}.
	\item[-] In DGP3, $u_i|W_i\sim \mathcal{N}(0,(1+X_i^2)/(1+\exp(2)))$ (the constant $1+\exp(2)$ ensures that $\V(u_i)\simeq 1$) and $X_i=\exp(X_i^*)$.
\end{itemize}

\par
Remark that  in the first two DGPs, $u_i$ is independent of $X_i$ given $Z_i$ and so the SR test has exact size. In DGP3, $u_i$ is heteroskedastic and the SR test may exhibit a finite-sample size distortion. As seen below, DGP3 leads to over-rejection of usual heteroskedasticity-robust test in finite samples. We aim at investigating the properties of the SR test (and others) in these situations. 
	
	\par
	Table \ref{tab:caract_perms} shows some characteristics of $W$ related to the SR test. With one Poisson regressor, the largest stratum roughly corresponds to  40\% of the whole sample, which implies that the number of distinct permutations in $\mathbb{S}_n$ ($\E[|\mathbb{S}_n|]$) is very large, even with $n=50$. With three Poisson regressors, on the other hand, many strata have just size one, and $S$ is quite large compared to $n$. It is then interesting to investigate the power of the test in this more difficult case.\footnote{\label{foot:large_p} When increasing $p$ further, e.g. to 10, strata have only size 1, and the SR test becomes trivial in this case.}

	\par	
	\begin{table}[htbp]
		\caption{Some characteristics of $Z$ related to the SR test.}
		\centering
		\begin{tabular}{l|ccc|ccc}
			 & \multicolumn{3}{c|}{$p=2$} & \multicolumn{3}{c}{$p=4$} \\
			Statistics on strata & $n=50$ & $n=100$ & $n=500$  & $n=50$ & $n=100$ & $n=500$ \\ \midrule
			$\E[|\mathbb{S}_n|]$ &  $1.1\times 10^{43}$ & $4.3\times 10^{112}$ & $>10^{308}$ &   $1.3\times 10^{12}$ &  $3.4\times 10^{36}$ &   $>10^{308}$  \\
					$\E[\max_{s=1,\dots,S} n_s]$  & 20.8 & 40.3 & 191.8 & 5.0 & 8.4 & 32.4 \\
				$\E[S]$  & 4.7 & 5.1 & 6.1 & 29.0 & 41.5 & 76.4 \\
			\bottomrule
		\end{tabular}
{\footnotesize Notes: for each $n$ and $p$, the expectations are estimated using 3,000 simulations from DGP1.}
		\label{tab:caract_perms}
	\end{table}

We consider the power curve on the interval $[-0.5, 0.5]$. We construct $\mathbb{S}'_n$ as explained in Footnote \ref{foot:Sprime} above, with $N'=499$. However, if $|\mathbb{S}_n|<499$, we simply consider all possible permutations, so that $\mathbb{S}'_n=\mathbb{S}_n$.

\par
We then compare our test with an asymptotic version of it (denoted by SRa in the figures below), where the test statistic is the same but we use the $\chi^2_1$ asymptotic critical value instead of the distribution of the permutation statistic. Apart from the SR test, we consider the cyclic permutation (CP) test of \cite{Lei-Bickel(2019)}, which is exact under independence between $W$ and $u$ (but not conditional independence between $X$ and $u$ given $Z$). Following \cite{Lei-Bickel(2019)} (p.406), we use $19$ cyclic permutation samples and a stochastic algorithm to find an ordering of the data that improves power. We also consider the partial correlation permutation test, denoted as PC, proposed by \cite{DR2017}, which is asymptotically valid under simply no correlation between $W$ and $u$. Finally, we consider the sign test of \cite{Toulis(2022)}, which is asymptotically valid under symmetry of $u$, which holds in the three DGPs. For these two permutation tests, we use 499 permutations drawn at random with replacement from $\G_n$. We also compare our test with the usual F-test, which is not heteroskedasticity-robust. We use the $F_{1, n-p-1}$ critical value for this  test. Finally, we compare our test with the heteroskedasticity-robust Wald tests. For the latter, we use the so-called HC1 and HC3 versions of the test. The former is the default in Stata and is widely used, whereas HC3 is often the recommended version of the test \citep[see, e.g.,][]{long2000using}. In both tests, we use the $\chi^2_1$ critical value.

\par
\begin{figure}[H]
\caption{Power curves: DGP1}
{\centering
\includegraphics[clip=true,trim=25mm 60mm 10mm 50mm]{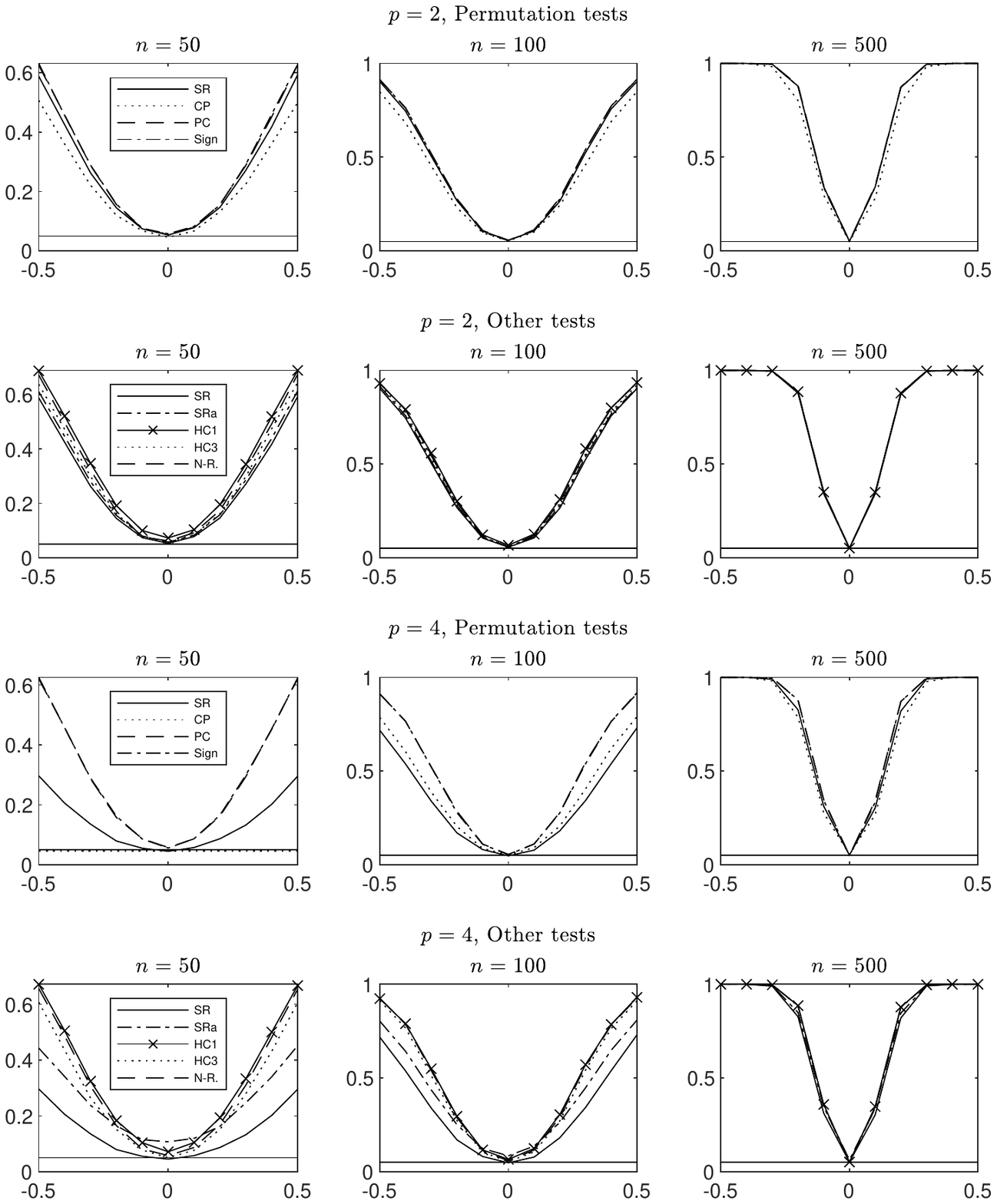}}

{\footnotesize Notes: the horizontal line is at the nominal level of the tests (5\%). ``SR'' stands for the stratified randomization test, ``CP'' is the cyclic permutation test of \cite{Lei-Bickel(2019)}, ``PC'' is the partial correlation test in Section 4 of \cite{DR2017} and ``Sign'' is the sign test in \cite{Toulis(2022)}. ``SRa'' uses the same test statistic as SR but with the asymptotic critical value, ``HC1'' and ``HC3'' are two heteroskedasticity-robust Wald tests and  ``N-R'' is the usual (non-robust) $F$-test. Results based on 3,000 simulations.}
	\label{fig:DGP1}
\end{figure}

In the first DGP and with $p=2$, the performances of the permutation tests are overall similar (see Figure \ref{fig:DGP1}). We just note that the CP test is slightly less powerful than the others. Compared to standard tests, the SR test is slightly less powerful but no difference can be detected when $n=500$. With $p=4$ and $n<500$, the SR test is less powerful than the PC and sign tests. This could be expected because basically, the test relies only on $n-S$ observations, and for $p=4$ and $n\in\{50,100\}$, $n-S$ is much smaller than $n$ (see Table \ref{tab:caract_perms} above).\footnote{\label{foot:tails} The tails of $X$ also seem to affect the power of the SR test. Considering $X_i=\exp(X_i^*)$ instead of $X_i=X_i^*$, the power of the SR deteriorates compared to that of the CP and sign test (but improves compared to the PC test, at least for $p=2$).} However, the difference in power becomes very small when $n=500$. Interestingly, the CP test does not seem to have power for $n=50$. We also notice that in this yet homoskedastic model, the HC1 and HC3 Wald tests overreject when $n=50$.

\par
\begin{figure}[H]
\caption{Power curves: DGP2}
{\centering
\includegraphics[clip=true,trim=25mm 60mm 10mm 50mm]{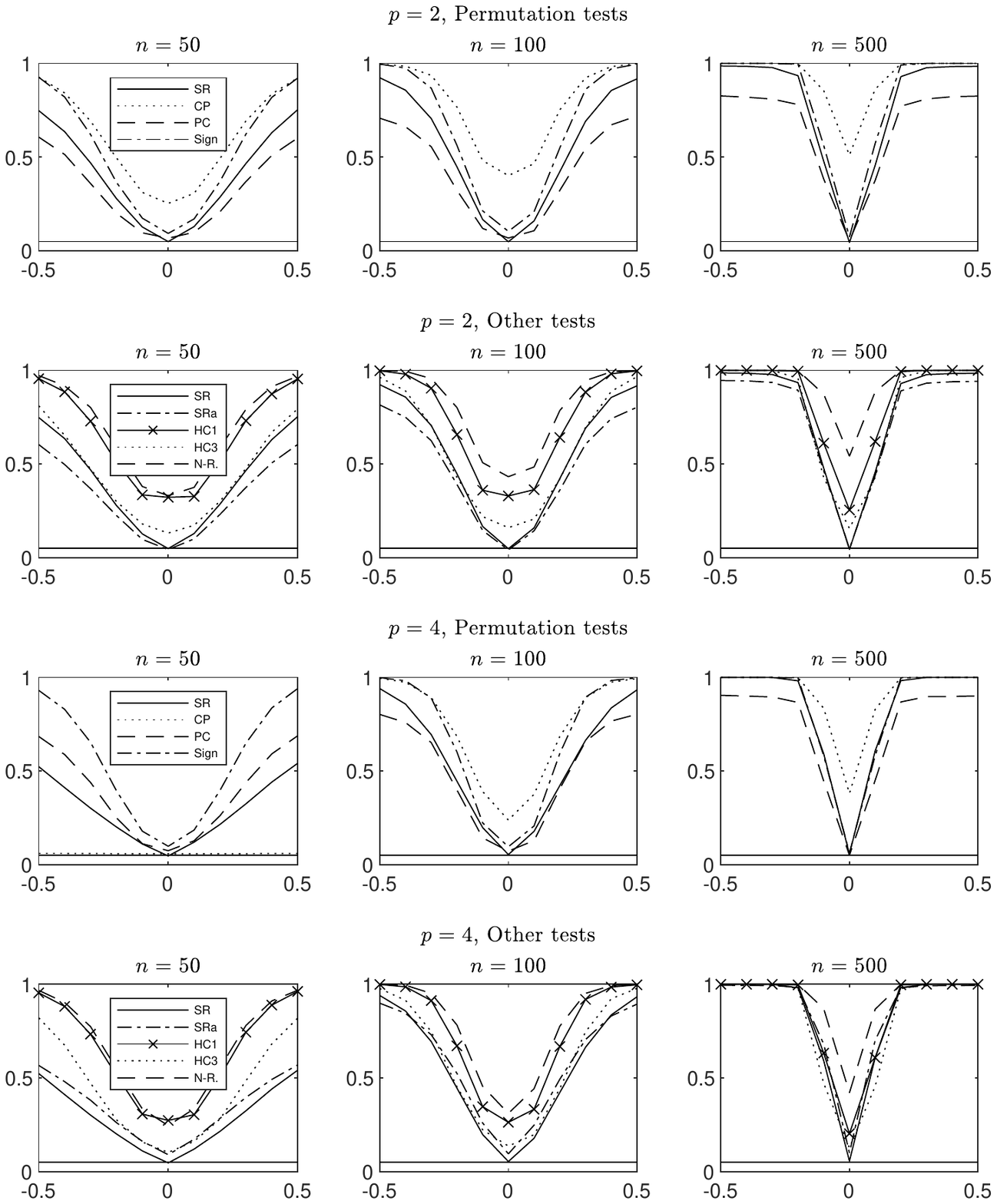}}

\textcolor{white}{.}\hspace{0.5cm} {\footnotesize Notes: same as in Figure \ref{fig:DGP1}.}
	\label{fig:DGP2}
\end{figure}

The second DGP is an instance where the SR test is the only exact one, among all the tests we consider. Even without size correction, it also has generally larger power than the PC test, except for $n=50$ and $p=4$ (see Figure \ref{fig:DGP2}). The sign test has better power (especially when $n\le 100$) but is also distorted, with levels around 10\% for $n\le 100$. The HC1 and HC3 tests exhibit large distortions. With $p=2$, they respectively reject the null hypothesis in 33.0\% and 16.1\% of the samples with $n=100$, and in 25.5\% and 15.7\% of the samples with $n=500$.

\newpage
\thispagestyle{empty}

\begin{figure}[H]
\caption{Power curves: DGP3}
{\centering
\includegraphics[clip=true,trim=25mm 60mm 10mm 50mm]{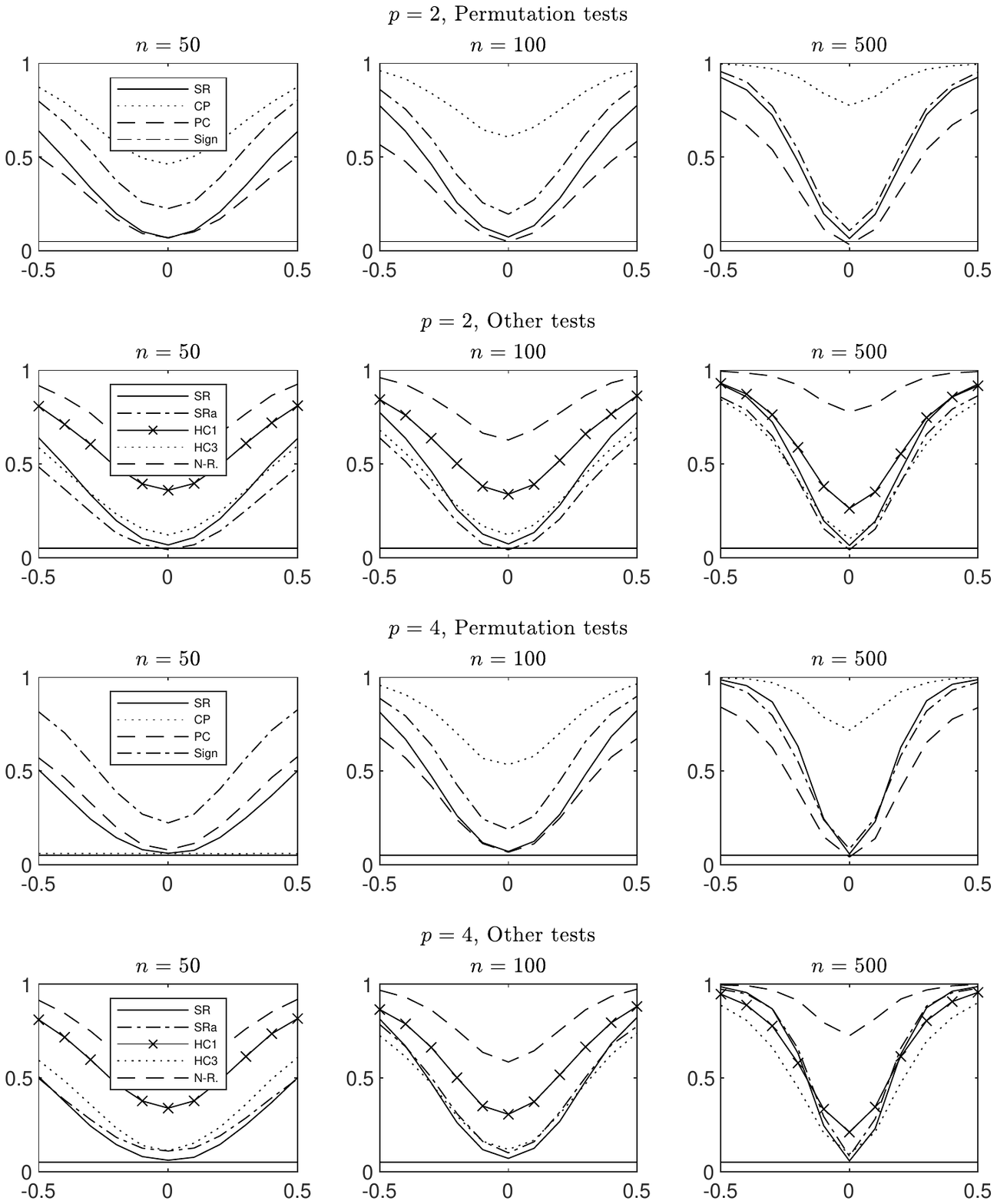}}

\textcolor{white}{.}\hspace{0.5cm} {\footnotesize Notes: same as in Figure \ref{fig:DGP1}.}
	\label{fig:DGP3}
\end{figure}

\par
The last DGP corresponds to a case where no test is exact because of heteroskedasticity, and all the tests we consider overreject in finite samples (see Figure \ref{fig:DGP3}). Overall, the SR test exhibits reasonable level of distortion, with a level that never exceeds 7.4\% over the six combinations of $n$ and $p$, and equal to 6.6\% on average. Other tests have average rejection rates of 52.4\% (CP), 5.7\% (PC), 17.1\% (sign), 30.3\% (HC1), 11.2\% (HC3) and 62.8\% (non-robust). The CP test exhibits high level of distortions, which also increase with $n$, confirming our conjecture that it is not heteroskedasticity-robust. Compared to the PC test, which has a similar level, the SR test has a larger power both when $p=2$ and $p=4$.  

\par
Of course, the ranking between the different tests in terms of level and power may vary for other DGPs. Nevertheless, the simulations above suggest that the SR test can be a good competitor to existing tests, especially in models with few additional covariates.


\subsection{Approximate test} 
\label{sub:MC_app}

We now investigate the approximate version of our SR test which handles continuous $Z$s, see Subsection \ref{sub:approximate_test} above. To this end, we consider the same DGPs as above but now assume that the $\{Z_{ij}\}_{j=2}^{p}$ are i.i.d. and follow a standard normal distribution, instead of a Poisson distribution. We refer to the corresponding DGPs as DGP1', DGP2' and DGP3'. We also have to choose the number of strata $S$. As mentioned above, this involves a trade-off between size distortion and power. Also, one should consider a larger number of strata when the correlation between $X_i$ and $Z_i'\gamma$ is high, since then the correlation between $y_i-X_i\beta_0$ and $X_i$ within strata becomes larger. Guided by this, we use the following data-driven $S$:
\begin{equation}\label{eq:def_S_app}
  S= \left\lceil \frac{n}{\min\left({n}^{1/2},1+2/|\widehat{\text{corr}}(X,Z\widehat{\gamma})|\right)}\right\rceil,
\end{equation}
where $\lceil x\rceil$ is the smallest integer greater than or equal to $x$ and $\widehat{\text{corr}}$ denotes the empirical correlation coefficient. The rule in \eqref{eq:def_S_app} generally leads to small strata sizes, around four in DGP1' and five in DGP2' and DGP3', as Table \ref{tab:S_app} shows. 

\begin{table}[H]
  \caption{Average number of strata in DGP1'-DGP3' when using \eqref{eq:def_S_app}}
  \label{tab:S_app}

  \centering
  \begin{tabular}{l|ccc|ccc}
	 & \multicolumn{3}{c|}{$p=2$} & \multicolumn{3}{c}{$p=4$} \\
DGP & $n=50$ & $n=100$ & $n=500$  & $n=50$ & $n=100$ & $n=500$ \\ \midrule
    1' & 12.5 & 25.6 & 130.1 & 12.4 & 25.4 & 130  \\
    2' & 10.8 & 21.8 & 107.7 & 10.6 & 21.5 & 107.6  \\
    3' & 10.8 & 21.7 & 107.7 & 10.9 & 21.7 & 107.8 \\
    \bottomrule
  \end{tabular}
\end{table}

\begin{figure}[H]
\caption{Power curves: DGP1'}
{\centering
\includegraphics[scale=0.9, clip=true,trim=20mm 100mm 0mm 92mm]{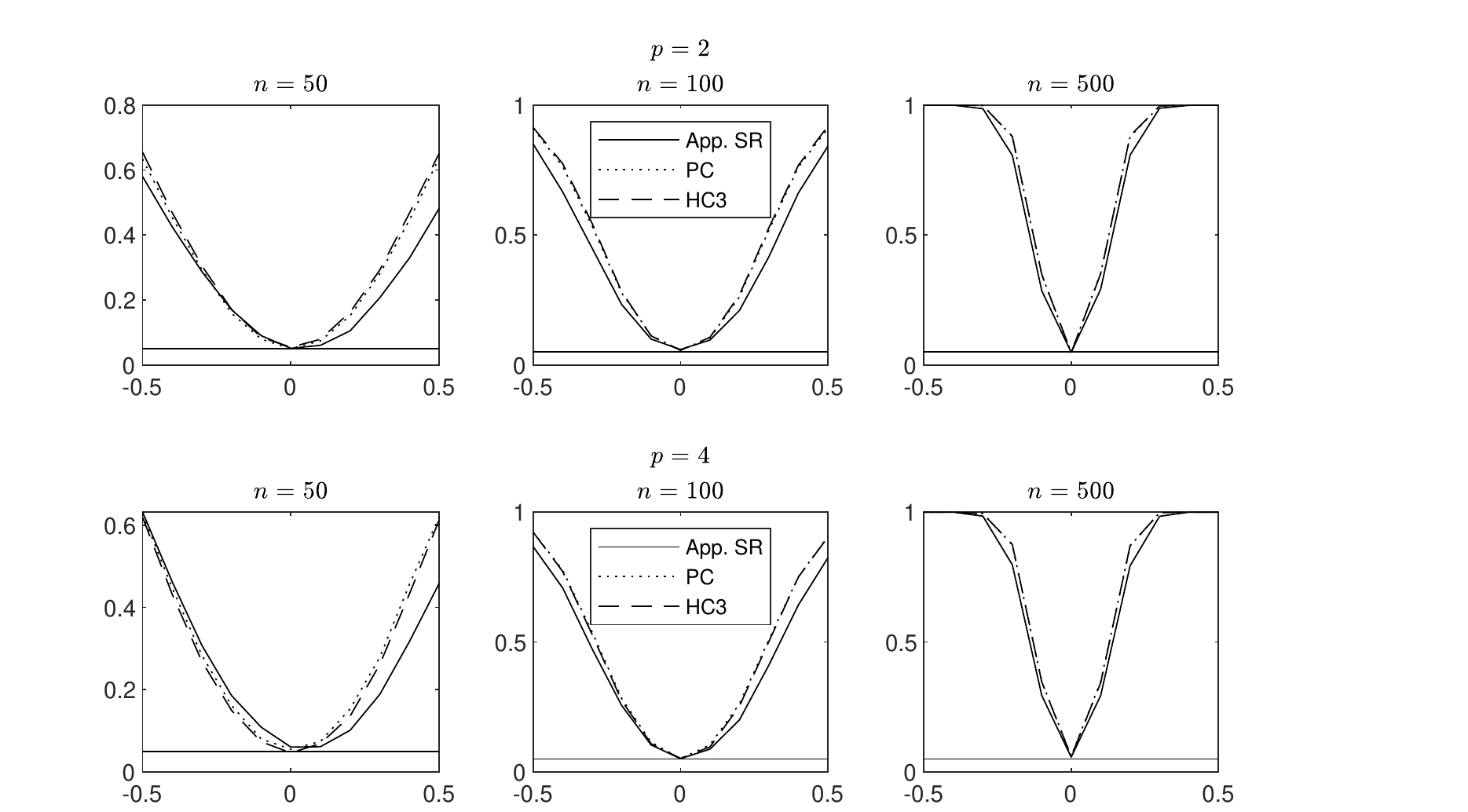}}

{\footnotesize Notes: the horizontal line is at the nominal level of the tests (5\%). ``App. SR'' stands for the approximate stratified randomization test, ``PC'' is the partial correlation test in Section 4 of \cite{DR2017} and ``HC3'' is the heteroskedasticity-robust Wald test. Results based on 3,000 simulations.}
	\label{fig:DGP1p}
\end{figure}

\begin{figure}[H]
\caption{Power curves: DGP2'}
{\centering
\includegraphics[scale=0.9, clip=true,trim=20mm 100mm 0mm 92mm]{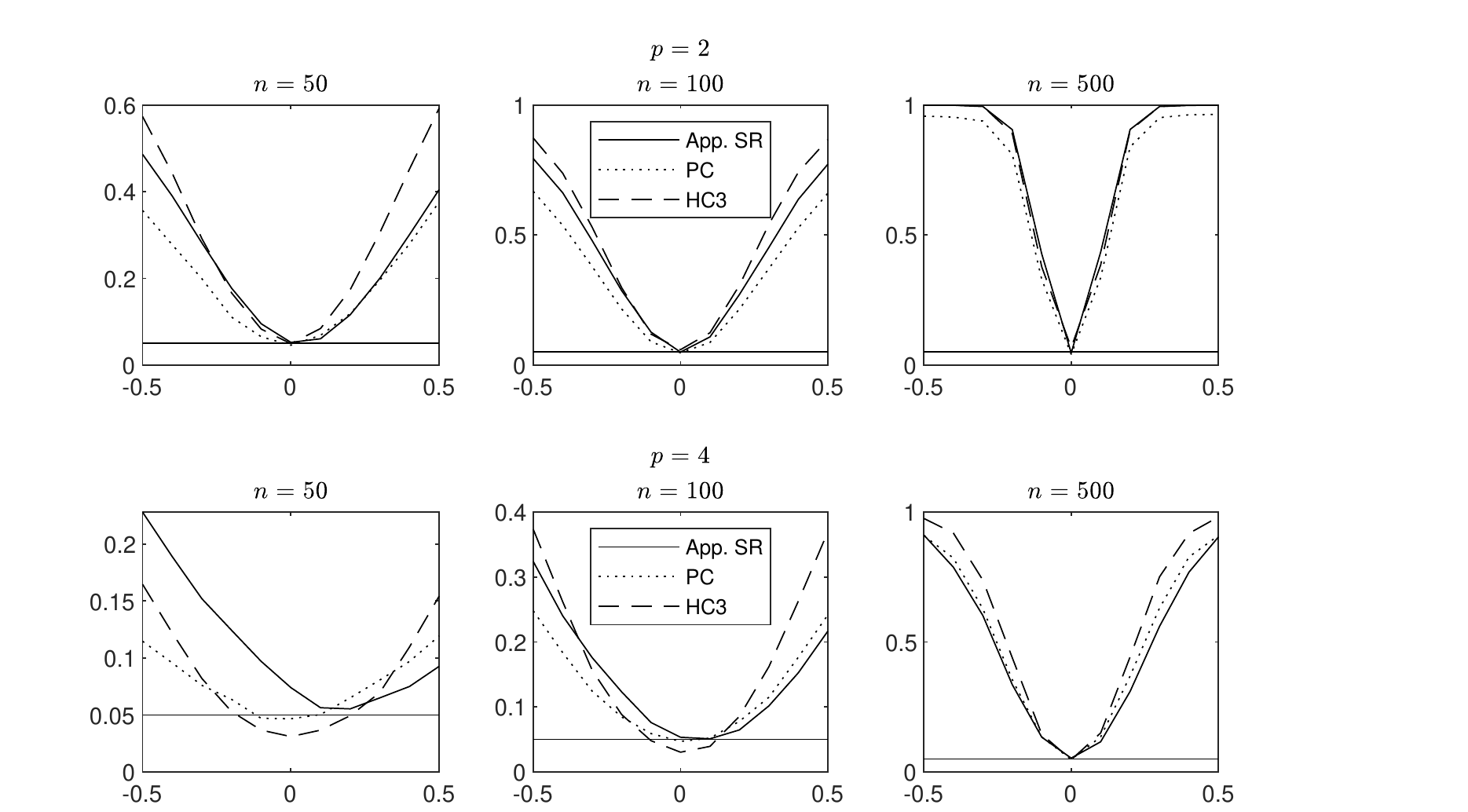}}

\textcolor{white}{.}\hspace{0.5cm} {\footnotesize Notes: same as in Figure \ref{fig:DGP1p}.}
	\label{fig:DGP2p}
\end{figure}

\begin{figure}[H]
\caption{Power curves: DGP3'}
{\centering
\includegraphics[scale=0.9, clip=true,trim=20mm 100mm 0mm 92mm]{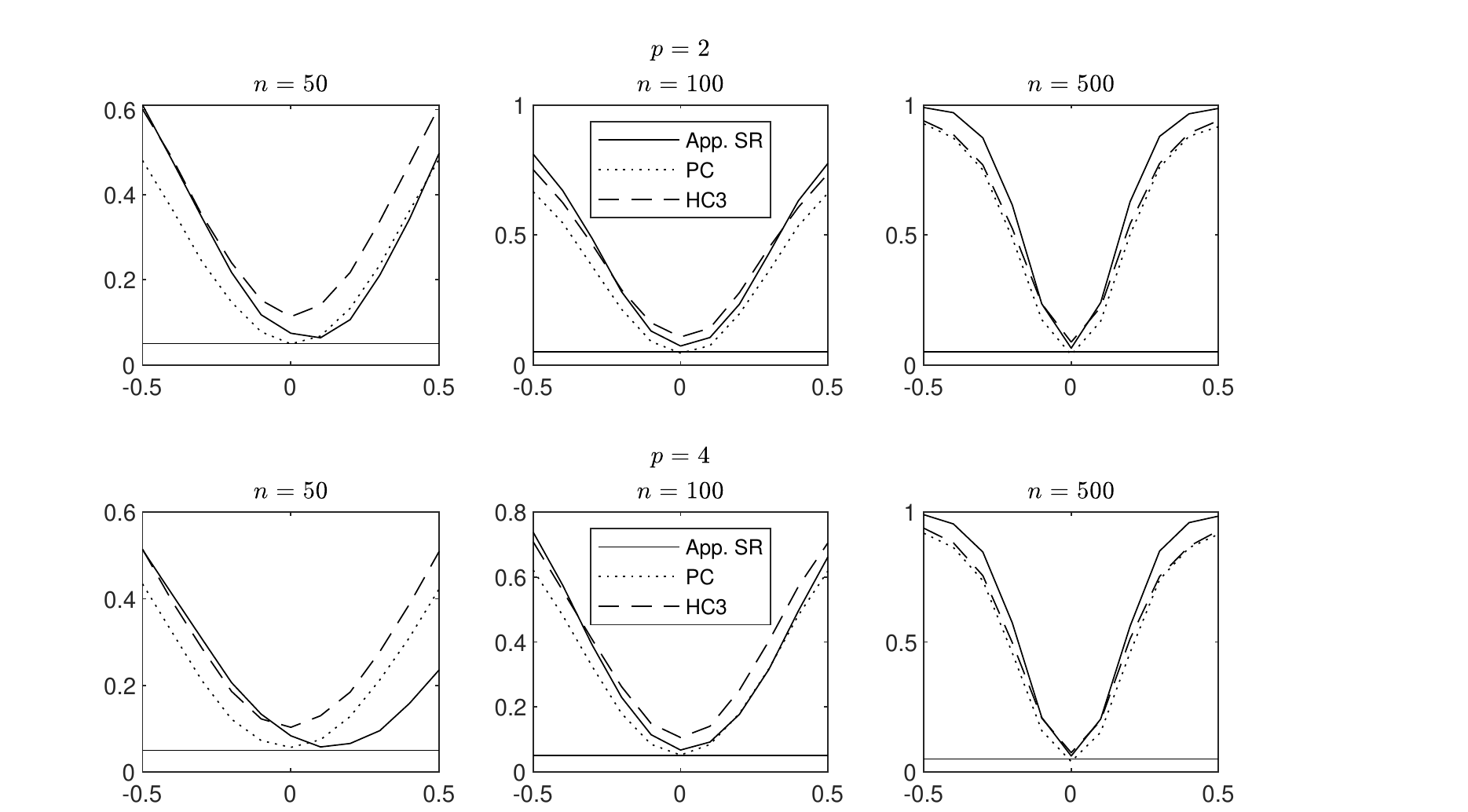}}

\textcolor{white}{.}\hspace{0.5cm} {\footnotesize Notes: same as in Figure \ref{fig:DGP1p}.}
	\label{fig:DGP3p}
\end{figure}

The power curves are displayed in Figures \ref{fig:DGP1p}-\ref{fig:DGP3p}; here we compare our test with the PC and HC3 tests. In DGP1', the test exhibits almost no distortion. It is slightly less powerful than the two other tests, but the difference gets attenuated as $n$ increases. In DGP2', the test is hardly distorted with $p=2$, with a level of at most 5.2\%, and has, in general, better power than the PC test. It exhibits some distortion when $p=4$ and $n=50$, with a level of 7.4\%, but this distortion quickly vanishes as $n$ increases. Its power is higher than that of the PC test for $\beta<0$ and $n=100$ but slightly lower otherwise. In DGP3', the test  appears to have good power compared to the PC and HC3 tests, even with $p=4$. It is also less distorted than the HC3 test but more than the PC test, with an average over the six cases of 7.1\% vs respectively 9.9\% and 4.7\%. Finally, additional simulations with higher $p$ (e.g., $p=10$), not presented here, suggest that the approximate SR test is not really sensitive to the dimension of $Z$. 

\par
Overall, the results suggest that with the data-driven choice of $S$ above, the test is asymptotically valid and consistent, though we leave this question for future research.

\vspace{-0.5cm}


\section{Applications} 
\label{sec:applications}

\subsection{Driving regulations and traffic fatalities} 
\label{sub:driving_regulations_and_traffic_fatalities}

We first apply the proposed randomization inference method to analyse the effect of driving regulations on traffic fatalities in the US, using  the same data and model as \citeauthor{Wooldridge(2015)} (\citeyear{Wooldridge(2015)}, Chapter 13). Specifically, we consider the linear model
\begin{equation}\label{eq:model_appli}
\Delta\, dthrte_i=const + \Delta\,open_i\,\beta+\Delta\,admn_i\,\gamma+u_i,
\end{equation}
where for any US state $i$, year $t$ and random variable $A_{i,t}$, we let $\Delta A_i=A_{i,1990}-A_{i,1985}$. In \eqref{eq:model_appli}, $dthrte$ denotes traffic fatality rate, $open$ is a dummy variable for having an open container law, which illegalizes for passengers to have open containers of alcoholic beverages, and $admn$ is a dummy variable for having administrative per se laws, allowing courts to suspend licenses after a driver is arrested for drunk driving but before the driver is convicted. The OLS estimate is $(\widehat{\beta},\widehat{\gamma})= (-0.42,-0.15)$, pointing towards a deterrent effect of the two types of laws on alcohol consumption by drivers.

\par
The coefficient $\gamma$ is never significant for any usual level, so we focus below on $\beta$. We compute the confidence intervals based on the SR test inversion (SR confidence interval hereafter) and those based on the CP and PC test inversion. We also consider the inversion of the test of the full vector of parameters, considered by \cite{DR2017} in their Section 3 (PR confidence interval hereafter). By projecting the corresponding confidence region over $\beta$, we obtain a confidence interval that is conservative under independence, or asymptotically conservative under weaker conditions. Finally, we consider the standard, non-robust and robust confidence intervals.\par

For all confidence intervals based on permutation tests, we invert the tests of $\beta=\beta_0$ for $\beta_0\in\{-1.7,-1.69,\dots,0.3\}$. As recommended above, we use the same set of permutations for all values of $\beta_0$ that we test. This way, we obtain proper intervals for the four permutation methods. We draw $\mathbb{S}'_n$ as explained in Footnote \ref{foot:Sprime}. For the other tests, we draw uniformly and with replacement $N'$ permutations from $\mathbb{G}_n$. To limit the effect of randomness, we use $N'=99,999$ instead of 499 as in the simulations.\footnote{In this application,  $n=51$, $S=3$ with $\max_{s} n_s=41$, so the set $\mathbb{S}_n$ is large ($|\mathbb{S}_n|\simeq 1.2\times 10^{55}$).} Even though we invert 201 tests and use a large number of permutations, the SR and PC confidence intervals take on our computer just a few seconds to compute (see the last column of Table \ref{Traffic1} for computational times). The CP and PR confidence intervals, on the other hand, are more computationally intensive. The reason for the CP method is that improving its power requires finding an optimal ordering, a difficult optimization problem \citep[see Section 2.4 in][]{Lei-Bickel(2019)}. The PR confidence interval is very costly to compute because it requires testing for values of $(\beta,\gamma)$, and thus considering a grid in $\R^2$ instead of $\R$.

\par
The results are reported in Table \ref{Traffic1}. We consider confidence intervals with nominal levels of 90\% and 95\%. The CP confidence intervals are by far the largest. Not surprisingly, the PR confidence interval is also large for the 95\% nominal level confidence interval, though it remains much shorter than the CP confidence interval. And actually, it is close to the SR, PC and robust confidence intervals when considering a nominal level of 90\%. For both nominal levels, the SR and PC confidence intervals are very close. They are also close to the robust confidence interval for the nominal level of 90\%, but around 11\% larger than this confidence interval for the nominal level of 95\%. Finally, the non-robust confidence intervals are the shortest of all intervals, being roughly 13\% smaller than the robust confidence intervals.

\par
The SR confidence interval may be more reliable than the robust confidence interval. To see why, note first that there is no evidence of heteroskedasticity: the White and Breusch-Pagan tests have p-values of respectively 0.56 and 0.34, respectively. If independence holds, the SR confidence interval has exact coverage, whereas the robust confidence interval may exhibit some distortion. To evaluate this, we ran simulations, assuming that $u_i$ is independent of $W_i$ and is distributed according to the empirical distribution of the residuals $\widehat{u}_i$. For nominal coverage of 95\% and 90\%, the robust confidence interval includes the true parameter in only 89.7\% and 85.7\% of the samples, respectively. Finally, there is strong evidence of non-normal errors: the Shapiro-Wilk and Jarque-Bera tests of normality on residuals have p-values of 0.0015 and 0.0012, respectively. As a result, the non-robust confidence intervals based on normality may also exhibit distortion. If one drops the normality assumption but maintains independence, the SR confidence interval is the only one with exact coverage. Then, one cannot exclude even at the 10\% level that the open container law has no effect on traffic fatalities.

\begin{table}[H]
\caption{{Confidence intervals (CIs) for $\beta$ in the traffic fatalities data.}}%
\begin{center}
	\label{Traffic1}
\begin{tabular}{l|cc|cc|c}
		\toprule
		& \multicolumn{2}{c|}{95\% coverage} & \multicolumn{2}{c|}{90\% coverage} & Computational  \\
		Test statistic & CI & Length & CI & Length & time (in sec.) \\ \midrule
SR & [-0.83,\,0.24] &1.07 & [-0.76,\,0.05] &0.81 & 5.0 \\
CP & [-1.61,\,0.27] &1.88 & [-1.37,\,0.03] &1.40 &262.6 \\
PC & [-0.86,\,0.21] &1.07 & [-0.77,\,0.03] &0.80 &5.6 \\
PR & [-1.09,\,0.19] &1.28 & [-0.85,\,0.00] &0.85 & $6.1\times 10^4$ \\ 
Non-robust & [-0.83,\,-0.01] &0.83 & [-0.76,\,-0.07] &0.69 & $<0.05$  \\
Robust HC3 & [-0.90,\,0.06] &0.96 & [-0.82,\,-0.02] &0.80 & $<0.05$  \\
	\bottomrule
	\end{tabular}
\end{center}
\footnotesize{Notes: the confidence intervals are obtained by inverting tests. ``SR'' corresponds to the stratified randomization test, ``CP'' is the cyclic permutation test of \cite{Lei-Bickel(2019)}, ``PC'' is based on the partial correlation statistic in Section 4 of \cite{DR2017} and ``PR'' is a projection of the confidence region on $(\beta,\gamma)$ using \cite{DR2017}'s first test in Section 3. The SR, PC and PR tests use 99,999 permutations (this number corresponds to $N'$ for the SR test). ``Non-robust'' is the usual $F$-test, which is not heteroskedasticity-robust, and ``Robust HC3'' is the heteroskedasticity-robust Wald test with the HC3 sandwich covariance matrix. Computational times (for one confidence interval) are obtained on Matlab (R2022a), with a MacBook Air (M1, 2020) with 8Go of RAM.}
\end{table}



\subsection{Project STAR} 
\label{sub:additional_empirical_examples}

Finally, we apply the SR test to the well-known dataset of the Project STAR experiment. \cite{Imbens-Rubin(2015)} (Chapter 9) provide a detailed analysis of the data using several stratified randomization-based inference methods. We follow their regression analysis of stratified randomized experiments (Chapter 9.6) and focus on schools with at least two regular classes and two small classes ignoring classes with teacher's aides. In the specifications considered below, there are at most 25 such schools that define the strata. The majority of schools have exactly 4 classes, and only a few have 5 or more classes.

\par
The regression model considered by \cite{Imbens-Rubin(2015)} is as follows:
\begin{equation}
	y_i=X_{i} \beta + \sum_{s=1}^{S} Z_{i}^s\gamma_s+u_i,\quad i=1,\dots, n,
	\label{eq:IR}
\end{equation}
where the treatment variable $X_{i}$ is the indicator for small classes, and the nuisance regressors $Z_{i}^s=1(i\in s), s=1,\dots, S,$ are the school (strata) indicators, and the outcome variable is the class-level (teacher-level) average math test scores for kindergarten children. We consider the class-level average reading test scores in addition to the math scores, and Grade 1 and Grade 2 as well.
As pointed out by \cite{Imbens-Rubin(2015)}, restricting the analysis to class-level data avoids a possible violation of the no-interference part of the Stable Unit Treatment Assumption (SUTVA).

\par
We deviate in a minor way from \cite{Imbens-Rubin(2015)}'s analysis by not standardizing the outcome variables to have mean 0 and standard deviation 1, because doing so introduces a slight dependence in the observations although the exchangeability of $u^s, s=1,\dots, S,$ would still be preserved. The results for standardized test scores are nevertheless similar, see Table \ref{STAR3} in Appendix \ref{app:STAR}.\footnote{\label{foot:replic}Also, we were unable to obtain exactly the same sample, and thus the same results, as \cite{Imbens-Rubin(2015)}. There are 66 classes (34 small and 32 regular) and 15 schools (strata) in our sample, as opposed to 68 classes (36 small and 32 regular) and 16 schools in theirs. However, our estimate of $\beta$ (0.22) and standard error (0.09), calculated following the variance formula in Theorem 9.1 of \cite{Imbens-Rubin(2015)}, are close to theirs \citep[0.24 and 0.10, respectively, see Chapter 9.6.2 of][]{Imbens-Rubin(2015)}.} 
The tests implemented are the same as those in Table \ref{Traffic1}, except that we include the HC0 confidence interval, denoted as Robust HC0 (IR),\footnote{The HC0 variance estimate is numerically identical to a sample analog of the variance formula in Theorem 9.1 of \cite{Imbens-Rubin(2015)}.} but do not include the projection-based test PR. The latter is computationally prohibitive in the current application, as there are at least 15 regression coefficients not under the test. We invert the tests of $\beta=\beta_0$ for $\beta_0\in\{-40,-39.99,\dots, 39.99, 40\}$ using $N'=99,999$ for each point.

\par
The  95\% confidence intervals are reported in Table \ref{STAR}. The number of possible permutations is at least $\vert \mathbb{S}_n\vert=9.47\times 10^{24}$ and $S/n$ is at most $25/109=0.23$ in the specifications. As a general pattern, we can notice that the CP confidence intervals include all the tested points in all of the specifications and the robust HC3 confidence intervals are the second widest, while the HC0 and PC confidence intervals are the shortest. The SR confidence intervals, though wider than the HC0 and PC confidence intervals, are comparable with the non-robust confidence intervals and always shorter than the HC3 confidence intervals.

\begin{table}[htbp]
	\small
	\caption{{95\% Confidence intervals for $\beta$ in the Project STAR data.}}%
	\begin{center}
		\label{STAR}
		\begin{tabular}{l|cc|cc|cc}
			\toprule
			\multicolumn{7}{c}{Non-standardized math test scores}\\ \midrule
			&\multicolumn{2}{c|}{Kindergarten}&\multicolumn{2}{c|}{Grade 1} &\multicolumn{2}{c}{Grade 2}\\ \midrule
			Test statistic & CI & Length & CI & Length & CI & Length\\ \midrule
			SR           &[0.15,\,21.27] &21.12 & [8.85,\,23.72] &14.87 &[2.32,\,20.77] &18.45 \\
			CP           &$\supseteq [-40,\,40]$ &$\ge 80$ & $\supseteq [-40,\,40]$ &$\ge 80$  & $\supseteq [-40,\,40]$ &$\ge 80$  \\
			PC           &[1.43,\,19.98] &18.55  & [9.73,\,22.85] &13.12 & [3.42,\,19.65] &16.23 \\
			Non-robust   &[-0.16,\,21.55] &21.71 & [9.07,\,23.52] &14.45 &[2.41,\,20.66] &18.25\\
			Robust HC3      &[-1.16,\,22.56] &23.72 & [7.99,\,24.59] &16.61 & [1.24,\,21.83] &20.59\\ 
			Robust HC0 (IR)     &[1.73,\,19.67] &17.94 & [9.88,\,22.70] & 12.83 & [3.62,\, 19.46] & 15.84\\
			\midrule

			BP, JB, SW pval &\multicolumn{2}{c|}{0.32,\;   0.89,\; 0.53}&\multicolumn{2}{c|}{0.04,\; 0.04,\; 0.15} &\multicolumn{2}{c}{0.86,\; 0.02,\; 0.18}\\
			$n, S, |\mathbb{S}_n|$ &\multicolumn{2}{c|}{66,\; 15,\; $9.47\times 10^{24}$}&\multicolumn{2}{c|}{109,\; 25,\; $1.68\times 10^{41}$} &\multicolumn{2}{c}{79,\; 18,\; $9.16\times 10^{29}$}\\ \midrule
			\multicolumn{7}{c}{Non-standardized reading test scores}\\ \midrule
			&\multicolumn{2}{c|}{Kindergarten}&\multicolumn{2}{c|}{Grade 1} &\multicolumn{2}{c}{Grade 2}\\ \midrule
			Test statistic & CI & Length & CI & Length  & CI & Length\\ \midrule
			SR          &[-1.25,\,14.56] &15.81  & [15.12,\,30.93] &15.81 &[3.76,\,19.68] &15.92 \\
			CP          & $\supseteq [-40,\,40]$ &$\ge 80$ & $\supseteq [-40,\,40]$ &$\ge 80$ & $\supseteq [-40,\,40]$ &$\ge 80$\\
			PC           & [-0.40,\,13.64] &14.04 & [16.07,\,29.96] &13.89 &[4.71,\,18.74] &14.03\\
			Non-robust   & [-1.06,\,14.27] &15.32 &[15.12,\,30.97] &15.86 &[3.83,\,19.59] &15.76\\
			Robust HC3     & [-2.27,\,15.48] &17.75 &[14.10,\,31.99] &17.89 & [2.83,\,20.59] &17.76\\ 
			Robust HC0 (IR) &[-0.16,\, 13.38] & 13.54 & [16.26,\, 29.83] & 13.57 & [4.83,\, 18.58] & 13.75\\
			\midrule
			BP, JB, SW pval &\multicolumn{2}{c|}{0.23,\; 0.01,\; 0.10 }&\multicolumn{2}{c|}{0.00,\; 0.76,\; 0.93} &\multicolumn{2}{c}{0.77,\; 0.00,\; 0.02 }\\
			$n, S, |\mathbb{S}_n|$ &\multicolumn{2}{c|}{66,\; 15,\; $9.47\times 10^{24}$}&\multicolumn{2}{c|}{109,\; 25,\; $1.68\times 10^{41}$} &\multicolumn{2}{c}{79,\; 18,\; $9.16\times 10^{29}$}\\
			\bottomrule
		\end{tabular}
	\end{center}
\footnotesize{Notes: the confidence intervals are obtained by inverting tests. The descriptions of the tests are the same as in Table \ref{Traffic1}. The SR and PC tests use 99,999 permutations. Robust HC0 (IR) is a confidence interval based on the HC0 variance estimate, calculated following the population variance in Theorem 9.1 of \cite{Imbens-Rubin(2015)}. BP, JB and SW pval denote the p-values of Breusch-Pagan homoskedasticity test, and Jarque-Bera and Shapiro-Wilk normality tests, respectively.}
\end{table}
\par
The baseline results for the math test scores of kindergarten children analyzed by \cite{Imbens-Rubin(2015)} are noteworthy. In this case, there is no evidence of either heteroskedasticity or non-normality. The Breusch-Pagan test for heteroskedasticity has p-value 0.32, so homoskedasticity assumption is supported at the conventional significance levels. The Jarque-Bera and Shapiro-Wilk tests for the residuals have p-values 0.89 and 0.53, respectively, pointing towards Gaussian errors. As such, the tests with finite sample validity i.e. the non-robust and SR tests, should be more reliable. And in fact, they turn out to be very similar: if anything, the non-robust confidence interval is slightly longer. It also includes 0, contrary to the SR confidence interval. The HC0 and PC confidence intervals also show a significant treatment effect.

\par
For reading scores in kindergarten, all test results suggest that the class-size reduction has no signifcant effect, in contrast with the results for math test scores. But class size reduction does seem to have an effect on both test scores in Grade 1 and Grade 2: all tests suggest significant treatment effects. As in the first application, the Breusch-Pagan and normality tests point to homoskedasticity and non-normality of the error terms in Grade 2, in which case the SR test could be the most reliable.

\par
Inasmuch as the permutation tests should be more reliable than the non-permutation tests in experimental datasets such as the current one, and the analysis using teacher-level samples guards effectively against a possible spillover in the students' performance, the results suggest that the class-size reduction has a small but significant effect on the math test scores for kindergarten children, and a bigger effect on the math and reading test scores for Grade 1 and Grade 2 students but no effect on the reading test scores for kindergarten children at teacher-level.

\section{Conclusion}
\label{sec:concl}

We develop a new permutation test for subvector inference in linear regressions. The test has exact size in finite samples if the error terms $u_i$ are independent of the regressors of interest $X_i$, conditional on other regressors $Z_i$. If independence fails but  $\C(X_i,u_i|Z_i)=0$, the test remains asymptotically valid with power against local alternatives under some conditions. The main one is that the number $S$ of distinct rows of $(Z_1,\dots,Z_n)'$ is negligible compared to the sample size $n$. Monte Carlo simulations suggest that the test has good power compared to other tests when, indeed, $n^{-1}S$ is small, and that it exhibits limited distortion without conditional independence. The two applications confirm that in some realistic designs, the test is informative and can thus be an appealing alternative to existing methods.

\medskip
A few questions are left for future research. First, some simulations we conducted (not reported above) suggest that the condition $n^{-1}S\to 0$ could be replaced by the weaker condition that the effective sample size $n-S$ tends to infinity. Second, while we show that the test is asymptotically valid if $\C(X_i,u_i|Z_i)=0$, we do not establish finite sample guarantees in this set-up.\footnote{\label{foot:Toulis} See Theorem 2 in \cite{Toulis(2022)} for an example of such guarantees. Note however that his result is obtained under conditions for which our test is actually exact.} Finally, constructing a permutation test for subvectors that is both exact under independence and asymptotically heteroskedasticity-robust for any design remains an important challenge.

	\newpage
	\bibliographystyle{chicago}
	\bibliography{Permutation}

	\newpage
	\appendix
		
	\section{Proofs}\label{Proofs}
	
	\subsection{Notation and abbreviations} 
	\label{sub:notation}
	
	Hereafter, we let $\bm{1}$ denote the $n\times 1$ vector of ones, and $\bm{1}_s$ denote the $n_s\times 1$ vector of ones. Otherwise, for any matrix $A$ with $n$ rows, the submatrix corresponding to stratum $s$ is denoted by $A^s$.  Let $\overset{d}{=}$ denotes equality in distribution. We recall that $P^\pi$ denotes the probability measure of $\pi \sim \mathcal{U}(\mathbb{S}_n)$, conditional on the data. $\E_\pi[\cdot]$ and $\V_\pi[\cdot]$ then denote the expectation and variance operators corresponding to $P^\pi$.
	$\Phi(\cdot)$ denotes the cumulative distribution function of a standard real  normal distribution.
		$O_{a.s.}(\cdot)$ and $o_{a.s.}(\cdot)$ mean $O(\cdot)$ and $o()$ almost surely.
	\par
	We write ``$T_n^{\pi}\cond T$ in probability" if a permutation statistic $T_n^\pi\in\mathbb{R}^m$ converges in distribution to a random variable $T$ on a set with
	probability approaching to 1 i.e. $P^\pi[T_n^\pi\leq x]\conp P[T\leq x]$ as $n\to\infty$ for every $x$ at which
	$x\mapsto P[T\leq x]$ is continuous.
For any matrix $V$ (possibly a vector), we let $\Vert V\Vert$ denote its Frobenius norm. We write ``$U_n-V_n\conp 0$ in probability", if two random sequences $U_n$ and $V_n$ satisfy $P^\pi[\Vert U_n-V_n\Vert >\epsilon]\conp 0$ as $n\to\infty$ for any $\epsilon>0$.

\par
Throughout the appendix, we index all quantities in the main text that
implicitly depend on $n$ by $n$ (and thus replace, e.g., $u_i$, $S$, $\mathcal{S}$ and $b_{i}^s$ by $u_{ni}, S_n, \mathcal{S}_n$ and $b_{ni}^s$, respectively). Also, in the proofs of Theorems \ref{thm:behav_Wpi} and \ref{thm:behav_W}, $\E[A_{ni}^s]$ will be used as a shortcut for $\E[A_{ni}\vert Z_{ni}=z_{ns}]$, for any random variable $A_{ni}$.

 We will use the following stratum level notation in accordance with \eqref{eq: partitioning}:
	\begin{align*}
		y^s
		&=[y_{n1}^s,\dots, y_{nn_s}^s]',\quad X^{s}=[X_{n1}^{s},\dots, X_{nn_s}^{s}]',\quad u^s=[u_{n1}^s,\dots, u_{nn_s}^s]',\\
		\tilde{X}
		&\equiv [\tilde{X}^{1},\dots, \tilde{X}^{S_n}]',\quad
		\tilde{X}^{s}\equiv M_{\bm{1}_s}X^s=[\tilde{X}_{n1},\dots, \tilde{X}_{nn_s}]',\\
		\tilde{X}_{ni}^{s}
		&\equiv {X}_{ni}^{s}-\bar{X}^s,\quad \bar{X}^s\equiv n_s^{-1}\sum_{i=1}^{n_s}X_{ni}^{s},\\
	    \tilde{u}
		&=[\tilde{u}^{1\prime},\dots, \tilde{u}^{S_n\prime}]^{\prime}=[\tilde{u}_{n1},\dots, \tilde{u}_{nn}]',\quad \tilde{u}^s\equiv M_{\bm{1}_s}u^s
		=[\tilde{u}_{n1}^s,\dots, \tilde{u}_{nn_s}^s]^{\prime},\\
		\tilde{u}_{ni}^s
		&=u_{ni}^s-\bar{u}^s,\quad \bar{u}^s\equiv n_s^{-1}\sum_{i=1}^{n_s}u_{ni}^s,\quad v^s=X^s(\beta-\beta_0)+u^s=[v_{n1}^s,\dots, v_{nn_s}^s]',\\
			\tilde{v}
		&=[\tilde{v}^{1\prime},\dots, \tilde{v}^{S_n\prime}]^{\prime}=[\tilde{v}_{n1},\dots, \tilde{v}_{nn}]',\quad \tilde{v}^s\equiv M_{\bm{1}_s}v^s
		=[\tilde{v}_{n1}^s,\dots, \tilde{v}_{nn_s}^s]^{\prime},\\
		\tilde{v}_{ni}^s
		&=v_{ni}^s-\bar{v}^s,\quad \bar{v}^s\equiv n_s^{-1}\sum_{i=1}^{n_s}v_{ni}^s.
	\end{align*}
	
Finally, we use the abbreviations SLLN for the strong law of large of numbers, WLLN for the weak law of large numbers, CLT for the central limit theorem, CMT for the continuous mapping theorem, and LHS and RHS for left-hand side and right-hand side respectively.
	

	\subsection{Theorem \ref{RPS}} 
	\label{sub:theorem_ref_rps}
	
	We reason conditional on $(W, N, \pi_1,\dots,\pi_N)$ hereafter and let, without loss of generality, $\pi_1=\Id$. For any $\bm{w}=(w^1,\dots,w^N)$, let $w^{(1)}<\dots<w^{(N)}$ be the corresponding ordered vector, $N^+(\bm{w})=|\{i\in\{1,\dots,N\}: w^{(i)}> w^{(q)}\}|$ and $N^0(\bm{w})=|\{i\in\{1,\dots,N\}: w^{(i)}= w^{(q)}\}|$. Let us also define $\widetilde{\phi}_\alpha(t,w^1,\dots,w^N)=1$ if $t>w^{(q)}$, $(N\alpha - N^+(\bm{w}))/N^0(\bm{w})$ if $t=w^{(q)}$ and $0$ otherwise. Then,
	$$\phi_\alpha=\widetilde{\phi}_\alpha\left(\mathcal{W}^{\pi_1}, \mathcal{W}^{\pi_1},\dots,\mathcal{W}^{\pi_N}\right).$$
	Now, we already showed in the text that $\mathcal{W}^\pi=g(W,u_\pi)$ for all $\pi\in \mathbb{S}_n$. By Assumption \ref{A1}\ref{1ex} and \ref{1indep}, the variables $(u_{\pi_1},\dots,u_{\pi_N})$ are exchangeable. Therefore, $(\mathcal{W}^{\pi_1},\dots,\mathcal{W}^{\pi_N})$ are also exchangeable. As a result, because $\widetilde{\phi}_\alpha$ is symmetric in its last $N$ arguments, we have, for all $k\geq 1$,
	$$\phi_\alpha \eqd \widetilde{\phi}_\alpha\left(\mathcal{W}^{\pi_k}, \mathcal{W}^{\pi_1},\dots,\mathcal{W}^{\pi_N}\right).$$
	Therefore,
	\begin{align*}
		\E[\phi_\alpha \mid W, N, \pi_1,...,\pi_N]
		&=\frac{1}{N}\sum_{k=1}^N\E\left[\widetilde{\phi}_\alpha\left(\mathcal{W}^{\pi_k}, \mathcal{W}^{\pi_1},\dots,\mathcal{W}^{\pi_N}\right)\mid W, N, \pi_1,...,\pi_N \right]\\
		&=\frac{1}{N}\E\left[\sum_{k=1}^N \widetilde{\phi}_\alpha\left(\mathcal{W}^{\pi_k}, \mathcal{W}^{\pi_1},\dots,\mathcal{W}^{\pi_N}\right) \mid W, N, \pi_1,...,\pi_N \right]\\
		&=\frac{1}{N}\E\left[N^{+}+ \frac{N\alpha - N^+}{N^0} N^{0}\mid W, N, \pi_1,...,\pi_N \right]\\
		&=\alpha.\;_\square
	\end{align*}

	\subsection{Lemma \ref{lem:suff_cond_stratasize}} 
	\label{sub:lemma_stratasize}
Because the support of $Z_{ni}=(Z_{n1i},\dots,Z_{npi})'$ is a subset of $\N^p$, we have
$$S_n \leq \prod_{j=1}^p \max_{i=1,\dots,n}Z_{nji}.$$
As a result,
\begin{equation}\label{eq:ineg_for_strata}
\frac{S_n}{n} \leq \prod_{j=1}^p \left(\frac{\max_{i=1,\dots,n}Z_{nji}}{n^{1/p}}\right).
\end{equation}
Now, by Lemma \ref{lem:max2} applied to $|Z_{nji}|^{p/2}$, we have
$n^{-1/2} \max_{i=1,\dots,n} |Z_{nji}|^{p/2} \conas 0$. As a result, $\max_{i=1,\dots,n} |Z_{nji}|/n^{1/p} \conas 0$. The result follows in view of \eqref{eq:ineg_for_strata}.

	\subsection{Theorem \ref{thm:behav_Wpi}} 
	\label{sub:theorem_ref_sps}

Remark that $\mathcal{W}^\pi =\left(n^{-1/2}\tilde{X}^{\prime}\tilde{v}_{\pi}\right)' (\widehat{V}^{\pi})^{-1}\left(n^{-1/2}\tilde{X}^{\prime}\tilde{v}_{\pi}\right)$, where
\begin{equation*}
\hat{V}^\pi\equiv n^{-1}\sum_{s=1}^{S_n}\sum_{i=1}^{n_s}\tilde{X}_{ni}^s\tilde{X}_{ni}^{s\prime}\tilde{v}_{n\pi(i)}^{s2}.
\end{equation*}
Moreover, $n^{-1/2}\tilde{X}^{\prime}\tilde{v}_\pi=n^{-1/2}\sum_{s\in\mathcal{I}_n}\tilde{X}^{s\prime}\tilde{v}^s_{\pi}+n^{-1/2}\sum_{s\in\mathcal{J}_n} \tilde{X}^{s\prime}\tilde{v}^s_{\pi}$. We prove the result in four steps.\par
First, we show the conditional asymptotic normality of $n^{-1/2}\sum_{s\in\mathcal{I}_n}\tilde{X}^{s\prime}\tilde{v}^s_{\pi}$, suitably normalized, assuming $\liminf_{n\to\infty}\vert\mathcal{I}_n\vert\geq 1$ a.s.. The case $\liminf_{n\to\infty}\vert\mathcal{I}_n\vert= 0$ is treated in the final step.\par
Second, we prove the same result on $n^{-1/2}\sum_{s\in\mathcal{J}_n} \tilde{X}^{s\prime}\tilde{v}^s_{\pi}$ when $|\mathcal{J}_n|\conas\infty$. The cases $\limsup_{n\to\infty}|\mathcal{J}_n|=\infty$ but $\liminf_{n\to\infty}|\mathcal{J}_n|<\infty$, and $\limsup_{n\to\infty}|\mathcal{J}_n|<\infty$ are treated in the final step.\par
Third, we prove the convergence of $\hat{V}^\pi$, in a sense that will be clarified below. Finally, we prove the conditional convergence in distribution of $\mathcal{W}^\pi$. Note that due to the demeaning within each stratum, the strata of size $1$ are discarded. So hereafter, we assume without loss of generality that for all $s\in\{1,\dots,S_n\}$, $n_s\geq 2$.

	\subsubsection*{Step 1: Conditional asymptotic normality of $n^{-1/2}\sum_{s\in\mathcal{I}_n} \tilde{X}^{s\prime}\tilde{v}^s_{\pi}$ when $\liminf_{n\to\infty}\vert\mathcal{I}_n\vert\geq 1$ a.s.} 
	\label{par: CANP_S1}

Assuming $\liminf_{n\to\infty}\vert\mathcal{I}_n\vert\geq 1$ a.s., we prove hereafter that
\begin{equation}
n^{-1/2}\tilde{V}_{n\mathcal{I}}^{-1/2} \sum_{s\in\mathcal{I}_n}\sum_{i=1}^{n_s}\tilde{X}^s_{ni}\tilde{v}^s_{n\pi(i)}
\cond \Norm{0,I_k}\quad\text{in probability},\label{eq: sum YI con 1}
\end{equation}
where
\begin{equation}\label{eq: tildeVI}
\tilde{V}_{n\mathcal{I}}\equiv n^{-1}\sum_{s\in\mathcal{I}_n}\frac{1}{n_s-1}\sum_{i=1}^{n_s}\tilde{X}_{ni}^s\tilde{X}_{ni}^{s\prime}\left(\sum_{i=1}^{n_s}\tilde{v}_{ni}^{s2}\right).
\end{equation}
We have
	\begin{align}
		\tilde{V}_{n\mathcal{I}} &=n^{-1}\sum_{s\in\mathcal{J}_n}\left(\sum_{i=1}^{n_s}\tilde{X}_{ni}^{s} \tilde{X}_{ni}^{s\prime}\right)\left(n_s^{-1} \sum_{i=1}^{n_s}\tilde{v}_{ni}^{s2}\right) \notag\\
		&\quad+n^{-1} \sum_{s\in\mathcal{J}_n}\frac{1}{n_s-1}\left(\sum_{i=1}^{n_s}\tilde{X}_{ni}^{s}\tilde{X}_{ni}^{s\prime}\right)\left(n_s^{-1}\sum_{i=1}^{n_s}\tilde{v}_{ni}^{s2}\right).
		\label{eq: tilde VI2}
	\end{align}
Since $2(n_s-1)\geq n_s$, by Cauchy-Schwarz inequality and the fact that by Assumption \ref{A2}\ref{stratasize}, $n^{-1}\vert\mathcal{I}_n\vert\conas 0$,
\begin{align}
& \E\left[\left\Vert n^{-1}\sum_{s\in\mathcal{I}_n}\frac{1}{n_s-1}\left(\sum_{i=1}^{n_s}\tilde{X}_{ni}^{s}\tilde{X}_{ni}^{s\prime}\right)\left(n_s^{-1}\sum_{i=1}^{n_s}\tilde{v}_{ni}^{s2}\right)\right\Vert\right] \notag \\
	&\leq 2n^{-1}\sum_{s\in\mathcal{I}_n}\E\left[\left(n_s^{-1}\sum_{i=1}^{n_s}\Vert \tilde{X}_{ni}^{s}\Vert^2\right)\left(n_s^{-1}\sum_{i=1}^{n_s}\tilde{v}_{ni}^{s2}\right)\right]\notag\\
	&\conas 0. \label{eq: tildeVI con2}
\end{align}
By Markov's inequality, the second term in \eqref{eq: tilde VI2} is $o_p(1)$. We establish in \eqref{eq: hVIJ con} in Step 3 below that
$$n^{-1}\sum_{s\in\mathcal{I}_n}\left(\sum_{i=1}^{n_s}\tilde{X}_{ni}^{s} \tilde{X}_{ni}^{s\prime}\right)\left(n_s^{-1} \sum_{i=1}^{n_s}\tilde{v}_{ni}^{s2}\right)-V_{n\mathcal{I}}^{*} \conp 0,$$
where
\begin{equation}\label{def: VI*}
	{V}_{n\mathcal{I}}^{*}
	\equiv n^{-1}\sum_{s\in\mathcal{I}_n}n_sQ_n^s\left(n_s^{-1}\sum_{i=1}^{n_s}\E[v_{ni}^{s2}]
	\right).
\end{equation}
Therefore,
\begin{equation}
\tilde{V}_{n\mathcal{I}}-{V}_{n\mathcal{I}}^{*}\conp 0.\label{eq: VtildeI con}
\end{equation}
Since
${V}_{n\mathcal{I}}^{*}
= n^{-1}\sum_{s\in\mathcal{I}_n}n_sQ_n^s\left(n_s^{-1}\sum_{i=1}^{n_s}\E[{u}_{ni}^{s2}]+n_s^{-1}\sum_{i=1}^{n_s}\E[(X_{ni}^{s\prime}(\beta-\beta_0))^2]\right)$ by Assumption \ref{A2}\ref{2cu}, for $n$ large, we have
\begin{align}\label{eq: eval VI*}
	\lambda_{\min}\left(V_{n\mathcal{I}}^{*}\right)
	&\geq
	\lambda_{\min}\left(V_{n\mathcal{I}}\right)+
	\lambda_{\min}\left(n^{-1}\sum_{s\in\mathcal{I}_n}n_sQ_n^s
	n_s^{-1}\sum_{i=1}^{n_s}\E\left[(X_{ni}^{s\prime}(\beta-\beta_0))^2\right]\right)>\lambda.
\end{align}
Hence, with probability approaching one, $\lambda_{\min}(\tilde{V}_{n\mathcal{I}})>\lambda>0$. As a result, $\tilde{V}_{n\mathcal{I}}^{-1/2}$ is well-defined.
By the Cram{\'e}r-Wold device it suffices to show that
for $\bm{\tau}\in\mathbb{R}^k\setminus\{0_{k\times 1}\}$ fixed
\begin{equation}
\frac{n^{-1/2}\bm{\tau}'\tilde{V}_{n\mathcal{I}}^{-1/2} \sum_{s\in\mathcal{I}_n}\sum_{i=1}^{n_s} \tilde{X}^s_{ni}\tilde{v}^s_{n\pi(i)}}
	{\left(1-n^{-1}\right)^{1/2}(\bm{\tau}'\bm{\tau})^{1/2}}\cond \Norm{0,1}\quad\text{in probability}.\label{eq:CLT_I}
\end{equation}
We verify the conditions of Lemma \ref{HoeffdingCLT} with  $b_{ni}^s=n^{-1/2}\bm{\tau}'\tilde{V}_{n\mathcal{I}}^{-1/2}\tilde{X}_{ni}^s$, $c_{ni}^s=\tilde{v}_{ni}^s$ and
$\mathcal{S}_n=|\mathcal{I}_n|$.
Condition \ref{cond: centered} holds because $\sum_{i=1}^{n_s}\tilde{X}_{ni}^s=0$
and $\sum_{i=1}^{n_s}\tilde{v}_{ni}^s=0$ for each $s$. Furthermore,
\begin{align*}
	\sigma_n^2
&=\sum_{s\in \mathcal{I}_n}\frac{1}{n_s-1}\left(\sum_{i=1}^{n_s}b_{ni}^{s2}\right)\left(\sum_{i=1}^{n_s}c_{ni}^{s2}\right)\notag\\
&=\sum_{s\in\mathcal{I}_n}\frac{1}{n_s-1}\sum_{i=1}^{n_s}\left(n^{-1/2}\bm{\tau}'\tilde{V}_{n\mathcal{I}}^{-1/2}\tilde{X}_{ni}^s\right)^2\left(\sum_{i=1}^{n_s}\tilde{v}_{ni}^{s2}\right)\notag\\
&=\bm{\tau}'\tilde{V}_{n\mathcal{I}}^{-1/2}n^{-1}\sum_{s\in\mathcal{I}_n}\frac{1}{n_s-1}\sum_{i=1}^{n_s}\tilde{X}_{ni}^s\tilde{X}_{ni}^{s\prime}\left(\sum_{i=1}^{n_s}\tilde{v}_{ni}^{s2}\right)\tilde{V}_{n\mathcal{I}}^{-1/2}\bm{\tau}\notag\\
&=\bm{\tau}'\bm{\tau}\notag\\
&>0,
\end{align*}
where the last equality is by the definition of $\tilde{V}_{n\mathcal{I}}$ in \eqref{eq: tildeVI}.
Thus, Condition \ref{cond: sigcon}  holds.
By H{\"o}lder's, Jensen's and $c_r$-inequalities,
\begin{align*}
	\left(n_s^{-1}\sum_{i=1}^{n_s}|\tilde{v}_{ni}^s|^3\right)^{4/3}
	&\leq n_s^{-1}\sum_{i=1}^{n_s}\tilde{v}_{ni}^{s4}\\
	&\leq 16n_s^{-1}\sum_{i=1}^{n_s}{v}_{ni}^{s4}\\
	&\leq 16n_s^{-1}\sum_{i=1}^{n_s}2^{3}\left(\Vert X_{ni}^s\Vert^{4}\Vert \beta-\beta_0\Vert^{4}+u_{ni}^{s4}\right).
\end{align*}
Hence $\E\left[\left(n_s^{-1}\sum_{i=1}^{n_s}|\tilde{v}_{ni}^s|^3\right)^{4/3}\right]<C_1$ for some constant $C_1$ and
$n_s^{-1}\sum_{i=1}^{n_s}|\tilde{v}_{ni}^s|^3, s=1,\dots,|\mathcal{I}_n|$, are uniformly integrable.
By Theorem 9.7 of \cite{Hansen(2021)},
\begin{equation}\label{eq: max con}
	|\mathcal{I}_n|^{-1}\max_{s\in\mathcal{I}_n}n_s^{-1}\sum_{i=1}^{n_s}|\tilde{v}_{ni}^s|^3\conp 0.
\end{equation}
Furthermore, by Jensen's inequality
\begin{align}
	n^{-1}\sum_{s\in\mathcal{I}_n}\sum_{i=1}^{n_s}\Vert\tilde{X}_{ni}^s\Vert^3
	&\leq 8n^{-1}\sum_{s\in\mathcal{I}_n}\sum_{i=1}^{n_s}\Vert{X}_{ni}^s\Vert^3\notag\\
	&= 8n^{-1}\sum_{i=1}^{n}\Vert{X}_{ni}\Vert^3\notag\\
	&=O_p(1),\label{eq: X3}
\end{align}
where the last equality is by the WLLN. Similarly,
\begin{align}
	|\mathcal{I}_n|^{-1}\max_{s\in\mathcal{I}_n}\left(n_s^{-1}\sum_{i=1}^{n_s}
	\tilde{v}_{ni}^{s4}\right)
	&\conp 0,\label{eq: max con2}\\
	n^{-1}\sum_{s\in\mathcal{I}_n}\sum_{i=1}^{n_s}\Vert \tilde{X}_{ni}^s\Vert^4
	&=O_p(1).\label{eq: X4}
\end{align}
Condition \ref{cond: 3mom} holds because
\begin{align*}
&\sum_{s\in\mathcal{I}_n} \left(n_s^{-1}\sum_{i=1}^{n_s} |b_{ni}^s|^3\right) \left(n_s^{-1}\sum_{i=1}^{n_s} |c_{ni}^s|^3\right)\notag\\
&=\sum_{s\in\mathcal{I}_n} n_s^{-1}\sum_{i=1}^{n_s} |n^{-1/2}\bm{\tau}'\tilde{V}_{n\mathcal{I}}^{-1/2}\tilde{X}_{ni}^s|^3\sum_{i=1}^{n_s}|\tilde{v}_{ni}^s|^3\notag\\
&\leq \Vert \bm{\tau}'\tilde{V}_{n\mathcal{I}}^{-1/2}\Vert^3n^{-1}\sum_{s\in\mathcal{I}_n} \sum_{i=1}^{n_s}\Vert\tilde{X}_{ni}^s\Vert^3 n^{-1/2}\max_{s\in\mathcal{I}_n}n_s^{-1}\sum_{i=1}^{n_s}|\tilde{v}_{ni}^s|^3\notag\\
&\leq (\lambda_{\min}(\tilde{V}_{n\mathcal{I}}))^{-3/2}\Vert \bm{\tau}\Vert^{3}n^{-1}\sum_{i=1}^{n}\Vert\tilde{X}_{ni}\Vert^3n^{-1/2}|\mathcal{I}_n|  |\mathcal{I}_n|^{-1}\max_{s\in\mathcal{I}_n}n_s^{-1}\sum_{i=1}^{n_s}|\tilde{v}_{ni}^s|^3\notag\\
&\conp 0,
\end{align*}
where the convergence holds by the CMT, the fact that $\lambda_{\min}(\tilde{V}_{n\mathcal{I}})-\lambda_{\min}({V}_{n\mathcal{I}})\conp 0$, $\lambda_{\min}({V}_{n\mathcal{I}})>\lambda>0$, \eqref{eq: max con}, \eqref{eq: X3} and
\begin{equation}\label{eq: card In bound}
n^{-1/2}|\mathcal{I}_n|\leq n^{-1/2}\frac{n}{\min_{s\in\mathcal{I}_n}n_s}\leq \frac{n^{1/2}}{c_n}\leq 1.
\end{equation}
Then, we have
\begin{align}
& n^{-2}\sum_{s\in\mathcal{I}_n} n_s^{-1}
\left(\sum_{i=1}b_{ni}^{s4}\right)\left(\sum_{i=1}c_{ni}^{s4}\right) \notag\\
&\leq  \Vert \bm{\tau}'\tilde{V}_{n\mathcal{I}}^{-1/2}\Vert^4
\sum_{s\in\mathcal{I}_n}
n^{-2}
\sum_{i=1}^{n_s}\Vert\tilde{X}_{ni}^s\Vert^4n_s^{-1}\sum_{i=1}^{n_s}\tilde{v}_{ni}^{s4}\notag\\
&\leq \Vert \bm{\tau}\Vert^4(\lambda_{\min}(\tilde{V}_{n\mathcal{I}}))^{-2} \left(n^{-1}\sum_{i=1}^{n}\Vert\tilde{X}_{ni}\Vert^4\right) \left(n^{-1}|\mathcal{I}_n|\right) \left(|\mathcal{I}_n|^{-1}\max_{s\in\mathcal{I}_n}n_s^{-1}\sum_{i=1}^{n_s}\tilde{v}_{ni}^{s4}\right)\notag\\
& \conp  0,\notag 
\end{align}
where the first inequality is by Cauchy-Schwarz inequality, the second inequality is by the inequality $\Vert \bm{\tau}'\tilde{V}_{n\mathcal{I}}^{-1/2}\Vert^2\leq \Vert \bm{\tau}\Vert^2/\lambda_{\min}(\tilde{V}_{n\mathcal{I}})$ followed by taking the maximum over $s\in\mathcal{I}_n$, and finally
the convergence follows from \eqref{eq: max con2} and \eqref{eq: X4}. Thus, Condition \ref{cond: Uvarcon} holds. Hence, Lemma \ref{HoeffdingCLT} applies and \eqref{eq:CLT_I} holds.

	\subsubsection*{Step 2: Conditional asymptotic normality of $n^{-1/2}\sum_{s\in\mathcal{J}_n} X^{s\prime}M_{\bm{1}_s}v^s_{\pi}$ when $|\mathcal{J}_n|\conas\infty$}
	\label{par: CANP_S3}

First, rewrite
\begin{equation}\label{eq: sum J1}
	n^{-1/2}\sum_{s\in\mathcal{J}_n} X^{s\prime}M_{\bm{1}_s}v^s_{\pi}
	=n^{-1/2}\sum_{s\in\mathcal{J}_n}\sum_{i=1}^{n_s}\tilde{X}_{ni}^{s}\tilde{v}_{n\pi(i)}^s.
\end{equation}
Let us define $\tilde{V}_{n\mathcal{J}}
	\equiv n^{-1}\sum_{s\in\mathcal{J}_n}\V_\pi\left[\sum_{i=1}^{n_s}\tilde{X}_{ni}^{s}\tilde{v}_{n\pi(i)}^s\right]$. We will show that when $|\mathcal{J}_n|\conas\infty$,
\begin{align}
	\tilde{V}_{n\mathcal{J}}^{-1/2}n^{-1/2}\sum_{s\in\mathcal{J}_n}X^{s\prime}M_{\bm{1}_s}v_\pi^s\cond \Norm{0, I_k}\ \text{in probability}.\label{eq: sum YI con 2}
\end{align}	
Observe that by Lemma S.3.4 of \cite{DR2017},
\begin{align*}
	\E_\pi\left[\sum_{i=1}^{n_s}\tilde{X}_{ni}^{s}\tilde{v}_{n\pi(i)}^s\right]
	&=
	\left(n_s^{-1}\sum_{i=1}^{n_s}\tilde{X}_{ni}^{s}\right)\left(\sum_{i=1}^{n_s}\tilde{v}_{ni}^s\right)=0. 
\end{align*}
Conditional on the observables, due to the stratified permutation,
the sum in \eqref{eq: sum J1} consists of mean-zero and independent but not necessarily identically distributed terms. Hence, to show \eqref{eq: sum YI con 2}, we verify the conditions of a multivariate Lindeberg CLT \citep[e.g.][Theorem 9.3]{Hansen(2021b)}. These conditions are: for any $\epsilon>0$, as $n\to\infty$,
\begin{align}
	&\frac{1}{n\lambda_{\min}(\tilde{V}_{n\mathcal{J}})}\sum_{s\in \mathcal{J}_n}\E_\pi\left[\big\Vert \sum_{i=1}^{n_s}\tilde{X}_{ni}^{s}\tilde{v}_{n\pi(i)}^s\big\Vert^2
	1\left(\big\Vert \sum_{i=1}^{n_s}\tilde{X}_{ni}^{s}\tilde{v}_{n\pi(i)}^s\big\Vert^2\geq n\epsilon\lambda_{\min}(\tilde{V}_{n\mathcal{J}}) \right)\right]\conas 0,\label{eq: Lindeberg cond}\\
	& \liminf_n \lambda_{\min}(\tilde{V}_{n\mathcal{J}}) > \lambda \; \text{a.s..} \label{eq: tilde VJ eval}
\end{align}
Actually, we only prove below in-probability versions of these conditions, and then invoke a subsequence argument to conclude.  First, let
\begin{equation}
V_{n\mathcal{J}}^{*}\equiv n^{-1}\sum_{s\in\mathcal{J}_n}n_sQ_n^s \left(n_s^{-1}\sum_{i=1}^{n_s}\E[v_{ni}^{s2}]\right).
\label{def: VJ*}
\end{equation}
By arguments similar to \eqref{eq: tilde VI2}, \eqref{eq: tildeVI con2} and \eqref{eq: VtildeI con} (see also \eqref{eq: hVIJ con} below),
\begin{equation}
\tilde{V}_{n\mathcal{J}}-V_{n\mathcal{J}}^{*}\conp 0.	
	\label{eq: Vtilde con}
\end{equation}
By the CMT, $\lambda_{\min}(\tilde{V}_{n\mathcal{J}})-\lambda_{\min}(V_{n\mathcal{J}}^{*})\conp 0$.
Since $0<\lambda<\lambda_{\min}(V_{n\mathcal{J}})$ by Assumption \ref{A2}\ref{2ns},
an argument analogous to \eqref{eq: eval VI*} yields $\lambda_{\min}\left(V_{n\mathcal{J}}^{*}\right)>\lambda$ for $n$ large. Hence, with probability approaching one, $\lambda_{\min}(\tilde{V}_{n\mathcal{J}})>\lambda>0$.

Next we verify an in-probability version of \eqref{eq: Lindeberg cond}. By convexity of $x\mapsto x^{4+\delta}, x>0,$ and Jensen's inequality,
	\begin{align}
		n_s^{-1}\sum_{i=1}^{n_s}\Vert \tilde{X}_{ni}^{s}\Vert^{4+\delta}
		&\leq 2^{3+\delta}\left(n_s^{-1}\sum_{i=1}^{n_s}\Vert {X}_{ni}^{s}\Vert^{4+\delta}
		+\left\Vert \bar{X}^s\right\Vert^{4+\delta}\right) \notag \\
		& \leq 2^{4+\delta}n_s^{-1}\sum_{i=1}^{n_s}\Vert {X}_{ni}^{s}\Vert^{4+\delta},\label{eq: Xtilde ineq}
	\end{align}
and similarly,		
\begin{equation}
n_s^{-1}\sum_{i=1}^{n_s}|\tilde{v}_{ni}^{s}|^{4+\delta} \leq 2^{4+\delta}n_s^{-1}\sum_{i=1}^{n_s} |v_{ni}^s|^{4+\delta}
\leq 2^{4+\delta}n_s^{-1}\sum_{i=1}^{n_s} 2^{3+\delta}(|u_{ni}^s|^{4+\delta}+|X_{ni}^{s\prime}(\beta-\beta_0)|^{4+\delta}).
	\label{eq: utilde ineq}
\end{equation}
Moreover, since $n_s \leq c_n$ for all $s\in\mathcal{J}_n$ and $\sum_{s\in\mathcal{J}_n} n_s \leq n$, we have
\begin{equation}
n^{-1-\delta/4}	\sum_{s\in\mathcal{J}_n} n_s^{1+\delta/4} \leq n^{-1-\delta/4}\left(\max_{s\in\mathcal{J}_n}n_s\right)^{\delta/4} \sum_{s\in\mathcal{J}_n} n_s \leq (c_n/n)^{\delta/4}\to 0. \label{eq: suffcon n_s}
\end{equation}
	Applying Lemma \ref{lem:MZ} with $a_i=\tilde{X}_{ni}^s$, $b_i=\tilde{v}_{ni}^s$, $n=n_s$ and $r=2+\delta/2$,  we obtain
	\begin{align}
		&n^{-1-\delta/4}\sum_{s\in\mathcal{J}_n}\E_\pi\left[\left\Vert \sum_{i=1}^{n_s}
		\tilde{X}_{ni}^s\tilde{v}_{n\pi(i)}^s\right\Vert^{2+\delta/2}\right]\notag\\
		&\leq M_{2+\delta/2} 	n^{-1-\delta/4}\sum_{s\in\mathcal{J}_n}n_s^{1+\delta/4}	\left(n_s^{-1}\sum_{i=1}^{n_s}\Vert \tilde{X}_{ni}^{s}\Vert^{2+\delta/2}\right)
		\left(n_s^{-1}\sum_{i=1}^{n_s}|\tilde{v}_{ni}^s|^{2+\delta/2}\right)\notag\\
		&\leq 0.5 M_{2+\delta/2} 	n^{-1-\delta/4}\sum_{s\in\mathcal{J}_n}n_s^{1+\delta/4}	\left(n_s^{-1}\sum_{i=1}^{n_s}\Vert \tilde{X}_{ni}^{s}\Vert^{4+\delta}+
		n_s^{-1}\sum_{i=1}^{n_s}|\tilde{v}_{ni}^{s}|^{4+\delta}\right)\notag\\
		&\leq 2^{3+\delta}M_{2+\delta/2} 	n^{-1-\delta/4}\sum_{s\in\mathcal{J}_n}n_s^{1+\delta/4}	\left(n_s^{-1}\sum_{i=1}^{n_s}\Vert {X}_{ni}^{s}\Vert^{4+\delta}+
		n_s^{-1}\sum_{i=1}^{n_s}|v_{ni}^s|^{4+\delta}\right)\notag\\
		&=o_p(1),\label{eq: Lindeberg sum J con}
	\end{align}
	where the second inequalty is by the convexity of $x\mapsto x^2$ and the inequality $a^2+b^2\geq 2ab$, the third inequality is by \eqref{eq: Xtilde ineq} and \eqref{eq: utilde ineq}, and the last equality is by Markov's inequality and \eqref{eq: suffcon n_s}. Note that
	\begin{align}
		&\frac{1}{n\lambda_{\min}(\tilde{V}_{n\mathcal{J}})}\sum_{s\in \mathcal{J}_n}\E_\pi\left[\frac{\big\Vert \sum_{i=1}^{n_s}\tilde{X}_{ni}^{s}\tilde{v}_{n\pi(i)}^s\big\Vert^{2+\delta/2}}
		{\big\Vert \sum_{i=1}^{n_s}\tilde{X}_{ni}^{s}\tilde{v}_{n\pi(i)}^s\big\Vert^{\delta/2}}
		1\left(\left\Vert \sum_{i=1}^{n_s}\tilde{X}_{ni}^{s}\tilde{v}_{n\pi(i)}^s\right\Vert^{\delta/2}\geq \left(n\epsilon\lambda_{\min}(\tilde{V}_{n\mathcal{J}})\right)^{\delta/4}\right)\right]\notag\\
		&\leq \frac{1}{\left(n\lambda_{\min}(\tilde{V}_{n\mathcal{J}})\right)^{1+\delta/4}\epsilon^{\delta/4}}\sum_{s\in \mathcal{J}_n}\E_\pi\left[\left\Vert \sum_{i=1}^{n_s}\tilde{X}_{ni}^{s}\tilde{v}_{n\pi(i)}^s\right\Vert^{2+\delta/2}\right]\notag\\
		&\conp 0,\label{eq: Lindeberg cond verif}
	\end{align}
	where the first inequality is the Lyapunov's inequality, and the convergence is by \eqref{eq: Lindeberg sum J con}.

	From \eqref{eq: Vtilde con} and \eqref{eq: Lindeberg cond verif}, the convergences in \eqref{eq: Lindeberg cond} and \eqref{eq: tilde VJ eval} hold almost surely along a subsequence $\{n_l\}$, see e.g. \cite{Durrett(2010)}, Theorem 2.3.2.
	Therefore,
	\begin{equation}\label{eq: YJ con0}
		\tilde{V}_{n\mathcal{J}}^{-1/2}n_l^{-1/2}\sum_{s\in\mathcal{J}_n} X^{s\prime}M_{\bm{1}_s}v^s_{\pi}
		=\tilde{V}_{n\mathcal{J}}^{-1/2}n_l^{-1/2}\sum_{s\in\mathcal{J}_n}\sum_{i=1}^{n_s}\tilde{X}_{ni}^{s}\tilde{v}_{n\pi(i)}^s
		\cond \Norm{0, I_k}\ \text{a.s..}
	\end{equation}
	Since \eqref{eq: YJ con0} holds for any subsequence of $\{n_l\}$, using \cite{Durrett(2010)}, Theorem 2.3.2 in the reverse direction, we obtain \eqref{eq: sum YI con 2}.
	\subsubsection*{Step 3: Consistency of $\hat{V}^\pi$} 
	\label{subsub:CovH0}

Let $V^{*}\equiv V_{n\mathcal{I}}^{*}+V_{n\mathcal{J}}^{*}
=n^{-1}\sum_{s=1}^{S_n}n_sQ_n^s\left(n_s^{-1}\sum_{i=1}^{n_s}\E[v_{ni}^{s2}]\right)$, where
$V_{n\mathcal{I}}^{*}$ and $V_{n\mathcal{J}}^{*}$ are defined in \eqref{def: VI*} and \eqref{def: VJ*}. We prove the convergence of $\hat{V}^\pi- (V_{n\mathcal{I}}^{*}+V_{n\mathcal{J}}^{*})$ to $0$ in two steps. First, we show $\E_\pi[\hat{V}^\pi]-{V}^{*}\conp0$.  Second, we show  $\hat{V}^\pi-\E_\pi[\hat{V}^\pi]\conp0$.
To that end, let $\mathcal{H}_n\in\{\mathcal{I}_n, \mathcal{J}_n\}$, $\hat{V}_{n\mathcal{H}}^\pi\equiv n^{-1}\sum_{s\in\mathcal{H}_n}\sum_{i=1}^{n_s}\tilde{X}_{ni}^{s}\tilde{X}_{ni}^{s\prime}\tilde{v}_{n\pi(i)}^{s2}$
and ${V}_{n\mathcal{H}}^{*}= n^{-1}\sum_{s\in\mathcal{H}_n}n_sQ_n^s\left(n_s^{-1}\sum_{i=1}^{n_s}\E[v_{ni}^{s2}]\right)$.
\paragraph*{Substep 1: $\E_\pi[\hat{V}^\pi]-{V}^{*}\conp0$} 
	\label{par:CovH0_S1}
It suffices show that  $\E_\pi[\hat{V}_{n\mathcal{H}}^\pi]-{V}_{n\mathcal{H}}^{*}\conp0$.
		By Lemma S.3.4 of \cite{DR2017}, we have
	\begin{align}
		\E_\pi\left[\hat{V}_{n\mathcal{H}}^\pi\right]&=\E_\pi\left[n^{-1}\sum_{s\in\mathcal{H}_n}\sum_{i=1}^{n_s}\tilde{X}_{ni}^{s}\tilde{X}_{ni}^{s\prime}\tilde{v}_{n\pi(i)}^{s2}\right]\notag\\
		&=\sum_{s\in\mathcal{H}_n}\frac{n_s}{n}\left(n_s^{-1}
		\sum_{i=1}^{n_s}\tilde{X}_{ni}^{s}\tilde{X}_{ni}^{s\prime}\right)
		\left(n_s^{-1}\sum_{i=1}^{n_s}\tilde{v}_{ni}^{s2}\right) \notag\\
		&=\sum_{s\in\mathcal{H}_n}\frac{n_s}{n}\left(n_s^{-1}\sum_{i=1}^{n_s}
		X_{ni}^{s}X_{ni}^{s\prime}-\bar{X}^{s}\bar{X}^{s\prime}\right)\left(n_s^{-1}\sum_{i=1}^{n_s}{v}_{ni}^{s2}-\bar{v}^{s2}\right)\notag\\
		&=n^{-1}\sum_{s\in\mathcal{H}_n}\left(\sum_{i=1}^{n_s}
		X_{ni}^sX_{ni}^{s\prime}\right)\left(n_s^{-1}\sum_{i=1}^{n_s}v_{ni}^{s2}\right) - n^{-1}\sum_{s\in\mathcal{H}_n}n_s\bar{X}^{s}\bar{X}^{s\prime}\left(n_s^{-1}\sum_{i=1}^{n_s}v_{ni}^{s2}\right) \notag \\
		&\quad-n^{-1}\sum_{s\in\mathcal{H}_n}\sum_{i=1}^{n_s}\tilde{X}_{ni}^s\tilde{X}_{ni}^{s\prime}\bar{v}^{s2}.\label{eq: EhVI3}
	\end{align}
	Consider the first summand in \eqref{eq: EhVI3}. We have
	\begin{align}
		&\E\left[\frac{1}{n}\left\Vert \sum_{s\in\mathcal{H}_n}\left(\sum_{i=1}^{n_s}
		X_{ni}^sX_{ni}^{s\prime}\right)\left(n_s^{-1}\sum_{i=1}^{n_s}v_{ni}^{s2}\right)
		-
		\sum_{s\in\mathcal{H}_n}\left(\sum_{i=1}^{n_s}
		\E[X_{ni}^sX_{ni}^{s\prime}]\right)\left(n_s^{-1}\sum_{i=1}^{n_s}\E[v_{ni}^{s2}]\right)\right\Vert\right]\notag\\
		&\leq
		\sum_{s\in\mathcal{H}_n}
		\E\left[\left\Vert n^{-1}\sum_{i=1}^{n_s}
		\left(X_{ni}^sX_{ni}^{s\prime}
		-
		\E[X_{ni}^sX_{ni}^{s\prime}]\right)
		\right\Vert
		\left(n_s^{-1}\sum_{i=1}^{n_s}v_{ni}^{s2}\right)
		\right]
		\notag\\
		&\quad+
		\sum_{s\in\mathcal{H}_n}\left(n_s^{-1}\sum_{i=1}^{n_s}
		\left\Vert\E[X_{ni}^sX_{ni}^{s\prime}]\right\Vert\right)
		\E\left[\left\vert n^{-1}\sum_{i=1}^{n_s}(v_{ni}^{s2}-\E[v_{ni}^{s2}])\right\vert\right]\notag\\
		&\leq
		\sum_{s\in\mathcal{H}_n}
		\left\{\frac{1}{n^2}\E\left[\left\Vert\sum_{i=1}^{n_s}\left(
		X_{ni}^sX_{ni}^{s\prime}
		-
		\E[X_{ni}^sX_{ni}^{s\prime}]\right)
		\right\Vert^2\right]\right\}^{1/2}
		\left\{\E\left[\left(n_s^{-1}\sum_{i=1}^{n_s}v_{ni}^{s2}\right)^2
		\right]\right\}^{1/2}\notag\\
		&\quad +\sum_{s\in\mathcal{H}_n}\left(n_s^{-1}\sum_{i=1}^{n_s}
		\E[\Vert X_{ni}^s\Vert^2]\right)
		n^{-1}\left\{\E\left[\left( n^{-1}\sum_{i=1}^{n_s}(v_{ni}^{s2}-\E[v_{ni}^{s2}])\right)^2\right]\right\}^{1/2}\notag\\
		&\leq
		\sum_{s\in\mathcal{H}_n}
		\left\{\frac{1}{n^2}\sum_{i=1}^{n_s}\E\left[\left\Vert
		X_{ni}^sX_{ni}^{s\prime}
		-
		\E[X_{ni}^sX_{ni}^{s\prime}]\right\Vert^2\right]\right\}^{1/2}
		\left\{\E\left[n_s^{-1}\sum_{i=1}^{n_s}v_{ni}^{s4}
		\right]\right\}^{1/2}\notag\\
		&\quad+\sum_{s\in\mathcal{H}_n}\left(n_s^{-1}\sum_{i=1}^{n_s}
		\E[\Vert X_{ni}^s\Vert^2]\right)
		n^{-1}\left\{\E\left[\sum_{i=1}^{n_s}(v_{ni}^{s2}-\E[v_{ni}^{s2}])^2\right]\right\}^{1/2}\notag\\
		&=O_{a.s.}\left(\sum_{s\in\mathcal{H}_n}
\frac{n_s^{1/2}}{n}\right)\notag \\
&=o_{a.s.}(1).\label{eq: EhVI1b}
	\end{align}
	The first and second inequalities hold by the triangle and Cauchy-Schwarz inequalities
	coupled with convexity of the Frobenius norm, respectively. The third inequality follows by independence and Cauchy-Schwarz. The first equality is by Assumption \ref{A2}\ref{2mom} and the last by
\begin{equation}\label{eq: sum n_s^1/2/n}
n^{-1}\sum_{s\in\mathcal{H}_n}
 n_s^{1/2}\leq n^{-1}\sum_{s=1}^{S_n}
 n_s^{1/2} \leq n^{-1}S_n^{1/2}\left(\sum_{s=1}^{S_n} n_s\right)^{1/2}=n^{-1/2}
S_n^{1/2}=o_{a.s.}(1).
\end{equation}
 By Markov's inequality and \eqref{eq: EhVI1b}, we obtain
	\begin{align}
		\frac{1}{n}\sum_{s\in\mathcal{H}_n}\left\{
\sum_{i=1}^{n_s}
		X_{ni}^sX_{ni}^{s\prime}\left(n_s^{-1}\sum_{i=1}^{n_s}v_{ni}^{s2}\right)-
		\sum_{i=1}^{n_s}\E[
		X_{ni}^sX_{ni}^{s\prime}]\left(n_s^{-1}\sum_{i=1}^{n_s}\E[v_{ni}^{s2}]\right)\right\}
		\conp 0.\label{eq: EhVI1 con}
	\end{align}
	To find the limit of the second summand in \eqref{eq: EhVI3}, first note that by the triangle inequality
		\begin{align}
		&\frac{1}{n}\left\Vert		\sum_{s\in\mathcal{H}_n}
		\left\{n_s\bar{X}^{s}\bar{X}^{s\prime}\left(n_s^{-1}\sum_{i=1}^{n_s}v_{ni}^{s2}\right)
		-n_s\E[\bar{X}^{s}]\E[\bar{X}^{s\prime}]
		\left(n_s^{-1}\sum_{i=1}^{n_s}\E[v_{ni}^{s2}]\right)\right\}
		\right\Vert\notag\\
		&\leq \left\Vert		\sum_{s\in\mathcal{H}_n}
		n^{-1}\left\{n_s\bar{X}^{s}\bar{X}^{s\prime}
		-n_s\E[\bar{X}^{s}]\E[\bar{X}^{s\prime}]\right\}
		\left(n_s^{-1}\sum_{i=1}^{n_s}v_{ni}^{s2}\right)\right\Vert \notag\\
		&\quad+\left\Vert\sum_{s\in\mathcal{H}_n}
		n^{-1}n_s\E[\bar{X}^{s}]\E[\bar{X}^{s\prime}]
		\left(n_s^{-1}\sum_{i=1}^{n_s}(v_{ni}^{s2}-\E[v_{ni}^{s2}])\right)
		\right\Vert\notag\\
		&\leq 		\sum_{s\in\mathcal{H}_n}
		n^{-1}\left\Vert
		\left(n_s^{-1}\sum_{i=1}^{n_s}(X_{ni}^{s}-\E[X_{ni}^{s}])\right)
		\sum_{i=1}^{n_s}X_{ni}^{s\prime}\left(n_s^{-1}\sum_{i=1}^{n_s}v_{ni}^{s2}\right)
		\right\Vert\notag\\
		&\quad+
		\sum_{s\in\mathcal{H}_n}
		n^{-1}\left\Vert
		\sum_{i=1}^{n_s}\E[X_{ni}^{s}]\left(n_s^{-1}\sum_{i=1}^{n_s}(X_{ni}^{s}
		-\E[X_{ni}^{s}])'\right)
		\left(n_s^{-1}\sum_{i=1}^{n_s}v_{ni}^{s2}\right)
		\right\Vert\notag\\
		&\quad+\sum_{s\in\mathcal{H}_n}
		n^{-1}\left\Vert\left(n_s^{-1}\sum_{i=1}^{n_s}\E[X_{ni}^{s}]\right)
		\left(\sum_{i=1}^{n_s}\E[X_{ni}^{s\prime}]\right)
		\left(n_s^{-1}\sum_{i=1}^{n_s}(v_{ni}^{s2}-\E[v_{ni}^{s2}])\right)
		\right\Vert.\label{eq: EhVI23}
	\end{align}
We bound each summand in \eqref{eq: EhVI23} in turn. First,
	\begin{align}
		&		\sum_{s\in\mathcal{H}_n}
n^{-1}\E\left[\left\Vert
		\sum_{i=1}^{n_s}(X_{ni}^{s}-\E[X_{ni}^{s}])
		\left(n_s^{-1}\sum_{i=1}^{n_s}X_{ni}^{s\prime}\right)\left(n_s^{-1}\sum_{i=1}^{n_s}v_{ni}^{s2}\right)
		\right\Vert\right]\notag\\
		&\leq
		n^{-1}		\sum_{s\in\mathcal{H}_n}
\left\{\E\left[\left\Vert \sum_{i=1}^{n_s}(X_{ni}^{s}-\E[X_{ni}^{s}])
		\left(n_s^{-1}\sum_{i=1}^{n_s}X_{ni}^{s\prime}\right)\right\Vert^2\right]\right\}^{1/2}\left\{\E\left[\left(n_s^{-1}\sum_{i=1}^{n_s}v_{ni}^{s2}\right)^2\right]\right\}^{1/2}\notag\\
		&\leq
		n^{-1}		\sum_{s\in\mathcal{H}_n}
\left\{\E\left[\left\Vert \sum_{i=1}^{n_s}(X_{ni}^{s}-\E[X_{ni}^{s}])\right\Vert^4\right]\right\}^{1/4}
		\left\{n_s^{-1}\sum_{i=1}^{n_s}\E[\Vert X_{ni}^{s}\Vert^4]\right\}^{1/4}\left\{n_s^{-1}\sum_{i=1}^{n_s}\E[v_{ni}^{s4}]\right\}^{1/2}, \label{eq:term1_EhVI23}
	\end{align}
	where the inequalities follow respectively by the Cauchy-Schwarz and Jensen's inequalities. Let $X_{ni}^s=[X_{ni1}^s, \dots, X_{nik}^s]'$. By the Cauchy-Schwarz inequality again,
\begin{equation}\label{eq: E4 ineq}
	\E[\Vert \sum_{i=1}^{n_s}(X_{ni}^s-\E[X_{ni}^s])\Vert^4]
	\leq k\sum_{l=1}^k\E\left[\left\{\sum_{i=1}^{n_s}(X_{nil}^s-\E[X_{nil}^s])\right\}^4\right].
\end{equation}
Using independence, the Cauchy-Schwarz inequality and the fact $\sup_{n,s,i}\E[(X_{nil}^s-\E[X_{nil}^s])^4]<C_0$ for some constant $C_0<\infty$, for $l=1,\dots, k$,
\begin{align}
	&\E\left[\left\{\sum_{i=1}^{n_s}(X_{nil}^s-\E[X_{nil}^s])\right\}^4\right]\notag\\
	&=\sum_{i=1}^{n_s}\E[(X_{nil}^s-\E[X_{nil}^s])^4]+3\sum_{i\neq j}
	\E[(X_{nil}^s-\E[X_{nil}^s])^2(X_{njl}^s-\E[X_{njl}^s])^2]\notag\\
	&\leq \sum_{i=1}^{n_s}\E[(X_{nil}^s-\E[X_{nil}^s])^4]+3\sum_{i\neq j}
	\left(\E[(X_{nil}^s-\E[X_{nil}^s])^4]\right)^{1/2}\left(\E[(X_{njl}^s-\E[X_{njl}^s])^4]\right)^{1/2}\notag\\
	&=O_{a.s.}(n_s^2).\label{eq: E4 order}
\end{align}
From \eqref{eq: E4 ineq} and \eqref{eq: E4 order}, we obtain
\begin{equation}
\E\left[\left\Vert\sum_{i=1}^{n_s}({X}_{ni}^{s}-\E[{X}_{ni}^s])\right\Vert^4\right]=O_{a.s.}(n_s^2).	
	\label{eq: sum E4 order}
\end{equation}
Combining this with \eqref{eq:term1_EhVI23} and using Assumption \ref{A2}\ref{2mom}, we obtain
	\begin{align}
		\sum_{s\in\mathcal{H}_n}
n^{-1}\E\left[\left\Vert
		\sum_{i=1}^{n_s}(X_{ni}^{s}-\E[X_{ni}^{s}])
		\left(n_s^{-1}\sum_{i=1}^{n_s}X_{ni}^{s\prime}\right)\left(n_s^{-1}\sum_{i=1}^{n_s}v_{ni}^{s2}\right)
		\right\Vert\right]&=O_{a.s.}\left(n^{-1}		\sum_{s\in\mathcal{H}_n}
n_s^{1/2}\right)\notag\\
		&=o_{a.s.}(1).\label{eq: EhVI21 con}
	\end{align}
	Similarly, the second summand in \eqref{eq: EhVI23} satisfies
	\begin{equation}
				\sum_{s\in\mathcal{H}_n}
n^{-1}\left\Vert
		\sum_{i=1}^{n_s}\E[X_{ni}^{s}]\left(n_s^{-1}\sum_{i=1}^{n_s}(X_{ni}^{s}
		-\E[X_{ni}^{s}])'\right)
		\left(n_s^{-1}\sum_{i=1}^{n_s}v_{ni}^{s2}\right)
		\right\Vert=o_{a.s.}(1).\label{eq: EhVI22 con}
	\end{equation}
	Consider the last summand in \eqref{eq: EhVI23}. By the triangle and Cauchy-Schwarz inequalities,
	\begin{align}
		&		\sum_{s\in\mathcal{H}_n}
		n^{-1}\E\left[
		\left\Vert
		\left(n_s^{-1}\sum_{i=1}^{n_s}\E[X_{ni}^{s}]\right)
		\left(\sum_{i=1}^{n_s}\E[X_{ni}^{s\prime}]\right)
		\left(n_s^{-1}\sum_{i=1}^{n_s}(v_{ni}^{s2}-\E[v_{ni}^{s2}])\right)
		\right\Vert
		\right]\notag\\
		&\leq 		\sum_{s\in\mathcal{H}_n}
n^{-1}
		\left(
		n_s^{-1}\sum_{i=1}^{n_s}\E[\left\Vert X_{ni}^{s}\right\Vert]
		\right)
		\left(\sum_{i=1}^{n_s}\E[\left\Vert X_{ni}^{s}\right\Vert]\right)
		\E\left[\left\vert n_s^{-1}\sum_{i=1}^{n_s}(v_{ni}^{s2}-\E[v_{ni}^{s2}])\right\vert\right]
		\notag\\
		&\leq 		\sum_{s\in\mathcal{H}_n}
n^{-1}
		\left(
		n_s^{-1}\sum_{i=1}^{n_s}\E[\left\Vert X_{ni}^{s}\right\Vert]
		\right)
		\left(n_s^{-1}\sum_{i=1}^{n_s}\E[\left\Vert X_{ni}^{s}\right\Vert]\right)
		\left\{\E\left[\sum_{i=1}^{n_s}(v_{ni}^{s2}-\E[v_{ni}^{s2}])^2\right]\right\}^{1/2}\notag\\
		&=O_{a.s.}\left(n^{-1}		\sum_{s\in\mathcal{H}_n}
n_s^{1/2}\right)\notag\\
		&=o_{a.s.}(1).\label{eq: EhVI23 con}
	\end{align}
	Combining \eqref{eq: EhVI23}, \eqref{eq: EhVI21 con}, \eqref{eq: EhVI22 con} and \eqref{eq: EhVI23 con},
	\begin{align}
	\frac{1}{n}\left\Vert	\sum_{s\in\mathcal{H}_n}
	\left\{n_s\bar{X}^{s}\bar{X}^{s\prime}\left(n_s^{-1}\sum_{i=1}^{n_s}v_{ni}^{s2}\right)
	-n_s\E[\bar{X}^{s}]\E[\bar{X}^{s\prime}]
	\left(n_s^{-1}\sum_{i=1}^{n_s}\E[v_{ni}^{s2}]\right)\right\}
	\right\Vert	
		&\conp 0.\label{eq: EhVI2 con}
	\end{align}
	Consider the last summand in \eqref{eq: EhVI3}. Since $n_s \geq 1$, $n^{-1} \sum_{s=1}^{S_n} n_s^{-1} \leq n^{-1}S_n$. On the other hand, by convexity of $x\mapsto x^{-1}$, $S_n^{-1}\sum_{s=1}^{S_n} n_s^{-1} \geq \frac{1}{S_n^{-1}\sum_{s=1}^{S_n} n_s} = n^{-1}S_n$, so $n^{-1}S_n\geq n^{-1}\sum_{s=1}^{S_n} n_s^{-1} \geq n^{-2}S_n^2$. 	
Thus, $n^{-1}S_n\conas 0$ is equivalent to
\begin{equation}\label{stratasize2}
n^{-1}\sum_{s=1}^{S_n} n_s^{-1}\conas 0.
\end{equation}
	
	From Assumption \ref{A2}\ref{2mom} and the WLLN for triangular array of random variables
	\citep[see, e.g.,][Theorem 1]{Hansen-Lee(2019)}, we have
	\begin{align}
		n^{-1}\sum_{i=1}^n\Vert {X}_{ni}\Vert^4-n^{-1}\sum_{i=1}^n\E[\Vert {X}_{ni}\Vert^4]\conp 0.\label{X4Z4}
	\end{align}
	By the triangle and Jensen's inequalities
	\begin{align}
		n^{-1}\sum_{s=1}^{S_n}\sum_{i=1}^{n_s}\Vert \tilde{X}_{ni}^{s}\Vert^4
		&\leq 8n^{-1}\sum_{s=1}^{S_n}\sum_{i=1}^{n_s}\left(\Vert X_{ni}^s\Vert^4+\left\Vert n_s^{-1}\sum_{i=1}^{n_s}X_{ni}^s\right\Vert^4\right)\notag\\
		&\leq 8n^{-1}\sum_{s=1}^{S_n}\sum_{i=1}^{n_s}\left(\Vert X_{ni}^s\Vert^4+n_s^{-1}\sum_{i=1}^{n_s}\Vert X_{ni}\Vert^4\right)\notag\\
		&\leq 16n^{-1}\sum_{s=1}^{S_n}\sum_{i=1}^{n_s}\Vert X_{ni}^s\Vert^4\notag\\
		&=16n^{-1}\sum_{i=1}^{n}\Vert X_{ni}\Vert^4\notag\\
		&=O_p(1).\label{eq: X*4 expansion}
	\end{align}
	where the last equality is due to \eqref{X4Z4}. Moreover,
	\begin{align}
		\E\left[n^{-1}		\sum_{s=1}^{S_n}
\sum_{i=1}^{n_s}\bar{v}^{s4}\right]
		&=\E\left[n^{-1}\sum_{s=1}^{S_n}{n_s}\bar{v}^{s4}\right]\notag\\
		&=n^{-1}		\sum_{s=1}^{S_n}
n_s^{-1}\E\left[\left(n_s^{-1}\sum_{i=1}^{n_s}v_{ni}^{s2}\right)^2\right]\notag\\
		&\leq n^{-1}		\sum_{s=1}^{S_n}
n_s^{-1}\E\left[n_s^{-1}\sum_{i=1}^{n_s}v_{ni}^{s4}\right]\notag\\
		&=O_{a.s.}\left(n^{-1}		\sum_{s=1}^{S_n}
n_s^{-1}\right)\notag\\
		&\conas 0, \label{eq: sum_sI_bar_u_s^4 op1}
	\end{align}
	where the second equality is by independence, the first inequality follows by Jensen's inequality, the second equality holds by Assumption \ref{A2}\ref{2mom} and the convergence is by \eqref{stratasize2}.
	
	Then, by the triangle and Cauchy-Schwarz inequalities
	\begin{align}
		\left\Vert n^{-1}		\sum_{s\in\mathcal{H}_n}
\sum_{i=1}^{n_s}\tilde{X}_{ni}^{s}\tilde{X}_{ni}^{s\prime}\bar{v}^{s2}\right\Vert
		&\leq n^{-1}		\sum_{s=1}^{S_n}
\sum_{i=1}^{n_s}\left\Vert \tilde{X}_{ni}^{s}\right\Vert^2\bar{v}^{s2}\notag\\
		&\leq \left(n^{-1}		\sum_{s=1}^{S_n}
\sum_{i=1}^{n_s}\left\Vert \tilde{X}_{ni}^{s}\right\Vert^4\right)^{1/2}\left(n^{-1}		\sum_{s=1}^{S_n}
\sum_{i=1}^{n_s}\bar{v}^{s4}\right)^{1/2}\notag\\
		&=o_p(1),\label{eq: EhVI3 con}
	\end{align}
	where we used \eqref{eq: X*4 expansion},  \eqref{eq: sum_sI_bar_u_s^4 op1} and Markov's inequality to obtain \eqref{eq: EhVI3 con}. Therefore, by \eqref{eq: EhVI1 con}, \eqref{eq: EhVI2 con}, \eqref{eq: EhVI3 con}, and Markov's inequality, we obtain
	\begin{equation}\label{eq: EhVH con}
	\Vert \E_\pi[\hat{V}^\pi_{n\mathcal{H}}]-{V}_{n\mathcal{H}}^{*}\Vert\conp 0,
	\end{equation}
	 hence $\Vert \E_\pi[\hat{V}^\pi]-{V}^{*}\Vert\conp 0$.
		\paragraph*{Substep 2: $\hat{V}^\pi-\E_\pi[\hat{V}^\pi]\conp0$}
	\label{par:CovH0_S2}

We will show that the variance of each element of $\hat{V}_{n\mathcal{H}}^\pi\in\{\hat{V}_{n\mathcal{I}}^\pi, \hat{V}_{n\mathcal{J}}^\pi\}$ converges in probability to $0$.
	Using the fact that the permutations are independent across different strata, that $n_s>1$ for all $s$ and	Lemma S.3.4 of \cite{DR2017}, we have for $(j, l) \in \{1,\dots, k\}^2$,
\begin{align}
	&\V_\pi\left[{n}^{-1}\sum_{s\in\mathcal{H}_n}\sum_{i=1}^{n_s}\tilde{X}_{nij}^{s}\tilde{X}_{nil}^{s}\tilde{v}_{n\pi(i)}^{s2}\right]\notag\\
	&=\sum_{s\in\mathcal{H}_n}\frac{n_s^2}{n^2}\V_\pi\left[n_s^{-1}\sum_{i=1}^n\tilde{X}_{nij}^{s}\tilde{X}_{nil}^{s}\tilde{v}_{n\pi(i)}^{s2}\right]\notag\\
	&=\sum_{s\in\mathcal{H}_n}\frac{n_s^2}{n^2}\frac{1}{n_s-1}\left(n_s^{-1}\sum_{i=1}^{n_s}\tilde{X}_{nij}^{s2}\tilde{X}_{nil}^{s2}-\left(n_s^{-1}\sum_{i=1}^{n_s}\tilde{X}_{nij}^{s}\tilde{X}_{nil}^{s}\right)^2\right) \left(n_s^{-1}\sum_{i=1}^{n_s}\tilde{v}_{ni}^{s4} -\left(n_s^{-1}\sum_{i=1}^{n_s} \tilde{v}_{ni}^{s2}\right)^2\right) \notag\\
	&\leq \sum_{s\in\mathcal{H}_n}\frac{n_s^2}{n^2}\frac{1}{n_s-1}\left(n_s^{-1}\sum_{i=1}^{n_s} \tilde{X}_{nij}^{s2} \tilde{X}_{nil}^{s2}\right)\left(n_s^{-1}\sum_{i=1}^{n_s}\tilde{v}_{ni}^{s4}\right) \notag \\
	&\leq 2\sum_{s\in\mathcal{H}_n}\frac{n_s}{n^2}\left(n_s^{-1}\sum_{i=1}^{n_s}\tilde{X}_{nij}^{s2} \tilde{X}_{nil}^{s2}\right)16\left(n_s^{-1}\sum_{i=1}^{n_s}{v}_{ni}^{s4}\right) \notag\\
	&\leq 32\sum_{s\in\mathcal{H}_n}\frac{n_s}{n^2}\left(n_s^{-1}\sum_{i=1}^{n_s}\tilde{X}_{nij}^{s2} \tilde{X}_{nil}^{s2}\right)\max_{1\leq i\leq n_s}{v}_{ni}^{s4}\notag\\
	&\leq 32\left(n^{-1}\sum_{s\in\mathcal{H}_n}\sum_{i=1}^{n_s}\tilde{X}_{nij}^{s2}\tilde{X}_{nil}^{s2} \right)\left(n^{-1}\max_{1\leq i\leq n}{v}_{ni}^{4}\right),\label{eq:for_Vcon1}
\end{align}
	where the second inequality is due to $n_s^{-1}\sum_{i=1}^{n_s}\tilde{v}_{ni}^{s4}\leq 16n_s^{-1}\sum_{i=1}^{n_s}v_{ni}^{s4}
	$ which follows by the triangle and Jensen's inequalities, and
	$v_{ni}\equiv X_{ni}'(\beta-\beta_0)+u_{ni}$. Now, since  $$\sup_{n, i}\E[\vert{v}_{ni}\vert^{4+\delta}]<2^{3+\delta}\left(\E\left[\vert{u}_{ni}\vert^{4+\delta}\right]+\E\left[\Vert {X}_{ni}\Vert^{4+\delta}\right]\Vert \beta-\beta_0\Vert^{4+\delta}\right)<\infty$$ for some $\delta>0$, we have
\begin{equation}
n^{-1}\max_{1\leq i\leq n}{v}_{ni}^{4}=o_p(1),	
	\label{eq:for_Vcon2}
\end{equation}	
see, e.g., \cite{Hansen(2021b)}, Theorem 9.7. Also,
	\begin{align}
		n^{-1}\sum_{s\in\mathcal{H}_n}\sum_{i=1}^{n_s}\tilde{X}_{nij}^{s2}\tilde{X}_{nil}^{s2}
		&\leq \left(n^{-1}\sum_{s\in\mathcal{H}_n}\sum_{i=1}^{n_s}\tilde{X}_{nij}^{s4}\right)^{1/2}\left(n^{-1}\sum_{s\in\mathcal{H}_n}\sum_{i=1}^{n_s}\tilde{X}_{nil}^{s4}\right)^{1/2}\notag\\
		&\leq \left(n^{-1}\sum_{i=1}^{n}\tilde{X}_{nij}^{4}\right)^{1/2}\left(n^{-1}\sum_{i=1}^{n}\tilde{X}_{nil}^{4}\right)^{1/2}\notag\\
		&=O_p(1),\label{eq:for_Vcon3}
	\end{align}
	where the equality holds by \eqref{eq: X*4 expansion}. Combining \eqref{eq:for_Vcon1}-\eqref{eq:for_Vcon3} with the Chebyshev's inequality, we obtain
	\begin{equation}\label{eq: hVH con}
	\hat{V}_{n\mathcal{H}}^\pi-\E_\pi[\hat{V}_{n\mathcal{H}}^\pi]\conp 0\quad\text{in probability,}
	\end{equation}
hence $\hat{V}^\pi-\E_\pi[\hat{V}^\pi]\conp 0$ in probability.	
	\par
Finally, \eqref{eq: EhVH con} and \eqref{eq: hVH con} together give
	\begin{align}\label{eq: hVIJ con}
	n^{-1}\sum_{s\in\mathcal{I}_n}\sum_{i=1}^{n_s}\tilde{X}_{ni}^{s}\tilde{X}_{ni}^{s\prime}\tilde{v}_{n\pi(i)}^{s2}-V_{n\mathcal{I}}^{*}\conp 0,\quad
	n^{-1}\sum_{s\in\mathcal{J}_n}\sum_{i=1}^{n_s}\tilde{X}_{ni}^{s} \tilde{X}_{ni}^{s\prime}\tilde{v}_{n\pi(i)}^{s2}
	-V_{n\mathcal{J}}^{*}\conp 0
	\end{align}
in probability.

	\subsubsection*{Step 4: Asymptotic distribution of $\mathcal{W}^\pi, \pi\sim \mathcal{U}(\mathbb{S}_n)$} 
	\label{subsub:asymptotic_validity_of_phi__alpha_mathcal_w_pi___pi_in_mathbb_s__n}
From \eqref{eq: VtildeI con} and \eqref{eq: Vtilde con} $\tilde{V}_{n\mathcal{I}}+\tilde{V}_{n\mathcal{J}}-({V}_{n\mathcal{I}}^{*}+V_{n\mathcal{J}}^{*})\conp 0$ in probability, so by the CMT $\lambda_{\min}(\tilde{V}_{n\mathcal{I}}+\tilde{V}_{n\mathcal{J}})-\lambda_{\min}({V}_{n\mathcal{I}}^{*}+V_{n\mathcal{J}}^{*})
	\conp 0$ in probability. Combining the latter with $\lambda_{\min}(V_{n\mathcal{I}}^{*}+V_{n\mathcal{J}}^{*})\geq \lambda_{\min}(V_{n\mathcal{I}}^{*})+\lambda_{\min}(V_{n\mathcal{J}}^{*})>\lambda>0$, with probability tending to $1$
\begin{equation}\label{eq: Vpi eigen}
\lambda_{\min}(\tilde{V}_{n\mathcal{I}}+\tilde{V}_{n\mathcal{J}})>\lambda.
\end{equation}	
To determine the asymptotic distribution of $\mathcal{W}^{\pi}$, we will show that
\begin{equation}\label{AN}
A_n\equiv (\tilde{V}_{n\mathcal{I}}+\tilde{V}_{n\mathcal{J}})^{-1/2}n^{-1/2}\sum_{s=1}^{S_n}X^{s\prime}M_{\bm{1}_s} u_\pi^s
	\cond \Norm{0, I_k}\quad\text{in probability.}
\end{equation}	
We will complete the proof by considering the following four cases:
\begin{align*}
\text{\bf Case 1:}&\ \liminf_{n\to\infty}|\mathcal{I}_n|\geq 1\ \text{a.s. and}\ |\mathcal{J}_n|\conas\infty,\\
\text{\bf Case 2:}&\ \liminf_{n\to\infty}|\mathcal{I}_n|\geq 1\ \text{a.s. and}\ \limsup_{n\to\infty} |\mathcal{J}_n|<\infty\ \text{a.s.},\\
\text{\bf Case 3:}&\ \liminf_{n\to\infty}|\mathcal{I}_n|\geq 1\ \text{a.s. and}\ \limsup_{n\to\infty} |\mathcal{J}_n|=\infty\ \text{a.s., but}\ \liminf_{n\to\infty} |\mathcal{J}_n|<\infty\ \text{a.s.},\\
\text{\bf Case 4:}&\ \liminf_{n\to\infty}|\mathcal{I}_n|=0\ \text{a.s..}
\end{align*}
\par
\paragraph*{Case 1: $\liminf_{n\to\infty}|\mathcal{I}_n|\geq 1\ \text{a.s. and}\ |\mathcal{J}_n|\conas\infty$}
Set in Lemma \ref{lem:CLT_sum}
\begin{align*}
	t_n
	&=
	\begin{bmatrix}
		\tilde{V}_{n\mathcal{I}}^{1/2}(\tilde{V}_{n\mathcal{I}}+\tilde{V}_{n\mathcal{J}})^{-1/2}\\ \tilde{V}_{n\mathcal{J}}^{1/2}(\tilde{V}_{n\mathcal{I}}+\tilde{V}_{n\mathcal{J}})^{-1/2}
	\end{bmatrix},\quad
	X_n
	=
	\begin{bmatrix}
	\tilde{V}_{n\mathcal{I}}^{-1/2}n^{-1/2}\sum_{s\in\mathcal{I}_n}X^{s\prime}M_{\bm{1}_s}u_\pi^s\\
	\tilde{V}_{n\mathcal{J}}^{-1/2}n^{-1/2}\sum_{s\in\mathcal{J}_n}X^{s\prime}M_{\bm{1}_s}u_\pi^s
	\end{bmatrix}.
\end{align*}
The two components of $X_n$ are independent as they belong to different strata. \eqref{AN} then follows from \eqref{eq: sum YI con 1}, \eqref{eq: sum YI con 2} and Lemma \ref{lem:CLT_sum}.\par
\paragraph*{Case 2: $\liminf_{n\to\infty}|\mathcal{I}_n|\geq 1\ \text{a.s. and}\ \limsup_{n\to\infty} |\mathcal{J}_n|<\infty\ \text{a.s.}$}

Since $n^{-1}\sum_{s\in\mathcal{J}_n}n_s\leq n^{-1/2}|\mathcal{J}_n|\conas 0$,
\begin{align}
\Vert {V}_{n\mathcal{J}}^{*}\Vert
&\leq n^{-1}\sum_{s\in\mathcal{J}_n}n_s\Vert Q_n^s\Vert \left(n_s^{-1}\sum_{i=1}^{n_s}\E[{v}_{ni}^{s2}]\right)\notag\\
&=O_{a.s.}\left(n^{-1}\sum_{s\in\mathcal{J}_n}n_s\right)\notag\\
&=o_{a.s.}(1).\label{eq: VJstar op1}
\end{align}
The latter combined with \eqref{eq: hVIJ con} gives $\Vert \tilde{V}_{n\mathcal{J}}\Vert\conp 0$ in probability.
 By Chebyshev's inequality,
for any $\epsilon>0$
\begin{align}
P^\pi\left[\Vert n^{-1/2}\sum_{s\in\mathcal{J}_n} X^{s\prime}M_{\bm{1}_s}v^s_{\pi}\Vert>\epsilon\right]
&\leq \epsilon^{-2}\Vert \tilde{V}_{n\mathcal{J}}\Vert\notag\\
&\conp 0.\label{eq: sum Jn op1}
\end{align}
Moreover, from the fact that $\Vert \tilde{V}_{n\mathcal{J}}\Vert\conp 0$ in probability, and \eqref{eq: Vpi eigen}
\begin{align}
\Vert (\tilde{V}_{n\mathcal{I}}+\tilde{V}_{n\mathcal{J}})^{-1/2}\tilde{V}_{n\mathcal{I}}^{1/2}-I_k\Vert
&\leq \Vert (\tilde{V}_{n\mathcal{I}}+\tilde{V}_{n\mathcal{J}})^{-1/2}\Vert \Vert \tilde{V}_{n\mathcal{I}}^{1/2}-(\tilde{V}_{n\mathcal{I}}+\tilde{V}_{n\mathcal{J}})^{1/2}\Vert\notag\\
&\leq k^{1/2}\{\lambda_{\min}(\tilde{V}_{n\mathcal{I}}+\tilde{V}_{n\mathcal{J}})\}^{-1/2}
\{\mathrm{tr}(\tilde{V}_{n\mathcal{J}})\}^{1/2}\notag\\
&\conp 0\quad\text{in probability,}\label{eq: cov equiv}
\end{align}
where the second inequality uses the inequality $\Vert B\Vert\leq {k}^{1/2}
\{\lambda_{\min}((B'B)^{-1})\}^{-1/2}$ for $k\times k$ invertible matrix $B$, and the Powers-St\o{}rmer inequality $\Vert B^{1/2}-C^{1/2}\Vert^2\leq \mathrm{tr}\{((B-C)'(B-C))^{1/2}\}$ for $k\times k$ positive definite matrices $B$ and $C$. By Slutsky's lemma, \eqref{eq: sum YI con 1}, \eqref{eq: sum Jn op1} and \eqref{eq: cov equiv},
\begin{align}
	A_n
	&=(\tilde{V}_{n\mathcal{I}}+\tilde{V}_{n\mathcal{J}})^{-1/2}\tilde{V}_{n\mathcal{I}}^{-1/2}\tilde{V}_{n\mathcal{I}}^{-1/2}n^{-1/2}\sum_{s\in\mathcal{I}}X^{s\prime}M_{\bm{1}_s} u_\pi^s+(\tilde{V}_{n\mathcal{I}}+\tilde{V}_{n\mathcal{J}})^{-1/2}\sum_{s\in\mathcal{J}}X^{s\prime}M_{\bm{1}_s} u_\pi^s\notag\\
	&\cond \Norm{0, I_k}\quad\text{in probability.}\label{AN J op1}
\end{align}	
\eqref{AN} follows.\par
\paragraph*{Case 3: $\liminf_{n\to\infty}|\mathcal{I}_n|\geq 1\ \text{a.s. and}\ \limsup_{n\to\infty} |\mathcal{J}_n|=\infty\ \text{a.s., but}\ \liminf_{n\to\infty} |\mathcal{J}_n|<\infty\ \text{a.s.}$}

Take a subsequence $\{n_l\}$. If $|\mathcal{J}_{n_l}|$ is bounded, then \eqref{eq: sum Jn op1} holds for $\{n_l\}$. As shown above, this entails \eqref{AN}. If the subsequence $\{n_l\}$ is not bounded, there exists a further subsequence $n_l(m)$ for which $|\mathcal{J}_{n_l(m)}|\conas \infty$. Then, as shown above \eqref{AN} holds along $\{n_l(m)\}$. Finally, fix $\bm{t}\in \mathbb{R}^k$ and $\epsilon>0$, and consider
\begin{equation*}
b_n\equiv P[|U_n(t)|>\epsilon],\quad U_n(t)\equiv P^\pi[A_n \leq \bm{t}] - \Phi_k(\bm{t}),
\end{equation*}
where the inequality is understood element-wise.
We proved that every subsequence $\{b_{n_l}\}$ of $\{b_n\}$ admits a further subsequence $\{b_{n_l(m)}\}$ tending to 0. Hence, $b_n$ tends to 0 by Urysohn's subsequence principle. \eqref{AN} follows.\par
\paragraph*{Case 4: $\liminf_{n\to\infty}|\mathcal{I}_n|=0\ \text{a.s.}$}
Take any subsequence $\{n_l\}$. If $|\mathcal{I}_{n_l}|=0$ for all $l$, then $\tilde{V}_{n_l\mathcal{I}}=0$
and $\tilde{V}_{n_l\mathcal{J}}=\tilde{V}_{n_l\mathcal{I}}+\tilde{V}_{n_l\mathcal{J}}$. Since
$|\mathcal{J}_n|\conas \infty$ in this case, \eqref{AN} follows from \eqref{eq: sum YI con 2}.
If $|\mathcal{I}_{n_l}|\neq 0$ for some $l$, there exists a further subsequence $n_l(m)$ such that
$|\mathcal{I}_{n_l(m)}|\conas 0$.
Then, since $n_{l}(m)^{-1}\sum_{s\in\mathcal{I}_{n_l(m)}}n_s\leq n_l(m)^{-1}n_l(m)|\mathcal{I}_{n_l(m)}|\conas 0$, similarly to \eqref{eq: VJstar op1}
\begin{align*}
	\Vert {V}_{n_l(m)\mathcal{I}}^{*}\Vert
	&=o_{a.s.}(1).
\end{align*}
Proceeding similarly to the second case above, the analogs of \eqref{eq: sum Jn op1} and \eqref{eq: cov equiv} hold:
\begin{align}
\Vert(\tilde{V}_{n_l(m)\mathcal{I}}+\tilde{V}_{n_l(m)\mathcal{J}})^{-1/2}\tilde{V}_{n_l(m)\mathcal{J}}^{1/2}-I_k\Vert
&\conp 0\quad\text{in probability,}\label{eq: cov equiv2}\\
P^\pi\left[\Vert n_l(m)^{-1/2}\sum_{s\in\mathcal{I}_{n_l(m)}} X^{s\prime}M_{\bm{1}_s}v^s_{\pi}\Vert>\epsilon\right]
&\leq \epsilon^{-2}\Vert \tilde{V}_{n_l(m)\mathcal{I}}\Vert\notag\\
&\conp 0.\label{eq: sum In op1}
\end{align}
As in \eqref{AN J op1}, from Slutsky's lemma, \eqref{eq: sum YI con 2},
\eqref{eq: cov equiv2} and \eqref{eq: sum In op1}, $A_{n_l(m)}\cond \Norm{0, I_k}$ in probability.
\eqref{AN} again follows from Urysohn's subsequence principle.\par

From \eqref{eq: Vpi eigen} and the consistency of $\hat{V}^\pi$, we obtain
	\begin{align*}
		& \left\Vert
		(\tilde{V}_{n\mathcal{I}}+\tilde{V}_{n\mathcal{J}})^{-1/2}\hat{V}^\pi
		(\tilde{V}_{n\mathcal{I}}+\tilde{V}_{n\mathcal{J}})^{-1/2}-I_k\right\Vert \\
&= \left\Vert
		(\tilde{V}_{n\mathcal{I}}+\tilde{V}_{n\mathcal{J}})^{-1/2}(\hat{V}^\pi-\tilde{V}_{n\mathcal{I}}-\tilde{V}_{n\mathcal{J}})
		(\tilde{V}_{n\mathcal{I}}+\tilde{V}_{n\mathcal{J}})^{-1/2}\right\Vert\\
		&\leq \left\Vert
		(\tilde{V}_{n\mathcal{I}}+\tilde{V}_{n\mathcal{J}})^{-1/2}\right\Vert^2\left\Vert\hat{V}^\pi-\tilde{V}_{n\mathcal{I}}-\tilde{V}_{n\mathcal{J}}\right\Vert\\
		&= \mathrm{tr}\left\{
		(\tilde{V}_{n\mathcal{I}}+\tilde{V}_{n\mathcal{J}})^{-1}\right\}\left\Vert\hat{V}^\pi-\tilde{V}_{n\mathcal{I}}-\tilde{V}_{n\mathcal{J}}\right\Vert\\
		&\leq k\left(\lambda_{\min}(\tilde{V}_{n\mathcal{I}}+\tilde{V}_{n\mathcal{J}})\right)^{-1}\left\Vert \hat{V}^\pi
		-\tilde{V}_{n\mathcal{I}}-\tilde{V}_{n\mathcal{J}}\right\Vert\\
		&\conp 0\quad\text{in probability.}
	\end{align*}
	Then, by the CMT
	\begin{equation}\label{eq: covm consistency}
		\left\Vert
		(\tilde{V}_{n\mathcal{I}}+\tilde{V}_{n\mathcal{J}})^{1/2}(\hat{V}^\pi)^{-1}
		(\tilde{V}_{n\mathcal{I}}+\tilde{V}_{n\mathcal{J}})^{1/2}-I_k\right\Vert\conp 0\quad\text{in probability.}
	\end{equation}
	By the CMT (e.g. \cite{Hansen(2021b)}, Theorem 10.5), \eqref{AN} and \eqref{eq: covm consistency} for $\pi\sim\mathcal{U}(\mathbb{S}_n)$
$$\mathcal{W}^{\pi}\cond  \mathcal{W}^{\pi}_{\infty}\sim \chi^2_k\quad\text{in probability.}$$


\subsection{Lemma \ref{HoeffdingCLT}} 
\label{sub:lemma_ref_hoeffding_clt}

	The proof is divided into two steps.
	In Step 1, an exchangeable pair is constructed. In Step 2, the asympotic normality is derived by
	showing that the moment bounds on Wasserstein distance converges in probability to 0.
	\paragraph*{Step 1: Exchangeable pair}
	
	Write $\pi$ as $(\pi_1,\dots,\pi_{\mathcal{S}_n})$. Let $S^e$ be a r.v. in $\{1,\dots, \mathcal{S}_n\}$,  with $P(S^e=s)=p_s\equiv(n_s-1)/(n-\mathcal{S}_n)$ and $(I, J)$ be a uniformly chosen transposition from $n_{S^e}(n_{S^e}-1)$ distinct pairs of indices $(i,j), i, j=1,\dots,n_s$ in stratum $S^e$. Both $S^e$ and $(I,J)$ are assumed independent of $\pi$. Then,  let $\pi' = (\pi'_1,\dots,\pi'_{\mathcal{S}_n})$ with $\pi'_s=\pi_s$ if $s\ne S^e$ and $ \pi'_{S_e}=\pi_{S^e}\circ (I, J)$. We first show that $(\pi, \pi')$ is an exchangeable pair:
	\begin{equation}\label{eq: expair}
		(\pi, \pi')\overset{d}{=}(\pi',\pi).
	\end{equation}
	Let $\pi^{*}\equiv \pi\circ (i, j)$.
	If $\pi\sim\mathcal{U}(\mathbb{S}_n)$ and
	$(i, j)$ is a uniformly chosen transposition from the indices $i=1,\dots,n_s$ in stratum $s$, then
	$\pi^{*}\sim\mathcal{U}(\mathbb{S}_n)$ and
	$\pi\overset{d}{=} \pi^{*}$. Also, $\pi=\pi^{*}\circ (i,j)$ because of the transposition.
	Then, for all $(A,B)\in\mathbb{S}_n^2$,
	\begin{align*}
		P(\pi\in A, \pi'\in B)
		&=\sum_{s=1}^{\mathcal{S}_n}P(\pi\in A, \pi'\in B| S^e=s)p_s\\
		&=\sum_{s=1}^{\mathcal{S}_n}P(\pi\in A, \pi\circ (I, J)\in B| s)p_s\\
		&=\sum_{s=1}^{\mathcal{S}_n}\frac{1}{n_s(n_s-1)}\sum_{i,j=1; i\neq j}^{n_s}P(\pi\in A, \pi\circ (i, j)\in B| s)p_s\\
		&=\sum_{s=1}^{\mathcal{S}_n}\frac{1}{n_s(n_s-1)}\sum_{i,j=1; i\neq j}^{n_s}P(\pi^{*}\circ(i,j)\in A, \pi^{*}\in B| s)p_s\\
		&=\sum_{s=1}^{\mathcal{S}_n}\frac{1}{n_s(n_s-1)}\sum_{i,j=1; i\neq j}^{n_s}P(\pi\circ(i,j)\in A, \pi\in B| s)p_s\\
		&=\sum_{s=1}^{\mathcal{S}_n}P(\pi\circ (I, J)\in A, \pi\in B| s)p_s\\
		&=P(\pi\circ (I, J)\in A, \pi\in B),
	\end{align*}
	where the first equality is by the iterated expectations, the second is by the definition of $\pi'$, the third is by the fact that $(I, J)$ is a uniformly chosen transposition, the fourth is by the definition of $\pi^{*}$, the fifth equality is by $\pi\overset{d}{=} \pi^{*}$ and $(i, j)$ is independent of $\pi$, the sixth is again by the fact that $(I, J)$ is a uniformly chosen transposition, and the seventh is by the iterated expectations. Therefore, \eqref{eq: expair} holds.

	\paragraph*{Step 2: Asymptotic normality}

	Note first that $\E_\pi[T^\pi]=0$ and $\V_\pi[T^\pi]=1$. Let $\pi'$ be as above, $a_{nij}^s\equiv b_{ni}^s c_{nj}^s$ and $\bm{a}_n=(a^s_{nij})_{s=1,\dots,\mathcal{S}_n, (i,j)\in\{1,\dots,n_s\}^2}$. Then $T^{\pi'}-T^\pi=\sigma_n^{-1}(a^{S^e}_{nI\pi(J)} +a^{S^e}_{nJ\pi(I)} - a^{S^e}_{nI\pi(I)} - a^{S^e}_{nJ\pi(J)})$. Thus, letting $\sum_{i\neq j}$ denote $\sum_{i,j=1; i\neq j}^{n_s}$
	\begin{align*}
		\E\left[T^{\pi'}-T^\pi|\pi,\bm{a}_n\right]
		&=\E\left[\E[\sigma_n^{-1}(a^{S^e}_{nI\pi(J)}
		+a^{S^e}_{nJ\pi(I)}-a^{S^e}_{nI\pi(I)}-a^{S^e}_{nJ\pi(J)})|S^e,\pi,\bm{a}_n]|\pi,\bm{a}_n\right]\\
		&=\sigma_n^{-1}\sum_{s=1}^{\mathcal{S}_n}\frac{n_s-1}{n-\mathcal{S}_n}
		\frac{1}{n_s(n_s-1)}\sum_{i\neq j}(a^s_{ni\pi(j)}
		+a^s_{nj\pi(i)}-a^s_{ni\pi(i)}-a^s_{nj\pi(j)})\\
		&=\sigma_n^{-1}\sum_{s=1}^{\mathcal{S}_n}\frac{1}{n_s(n-\mathcal{S}_n)}
		\left(-\sum_{i=1}^{n_s}a^s_{ni\pi(i)}
		-\sum_{i=1}^{n_s}a^s_{ni\pi(i)}-2(n_s-1)\sum_{i=1}^{n_s}a^s_{ni\pi(i)}\right)\\
		&=-\frac{2}{n-\mathcal{S}_n}\sigma_n^{-1}\sum_{s=1}^{\mathcal{S}_n}\sum_{i=1}^{n_s}a^s_{ni\pi(i)}\\
		&=-\lambda T^\pi,
	\end{align*}
	where $\lambda=2/(n-\mathcal{S}_n)$. As a result, we also get $\E\left[T^{\pi'}-T^\pi|T^\pi,\bm{a}_n\right]=-\lambda T^\pi$.\par
	The (conditional) Wasserstein distance between $T^\pi$ and $Z\sim \Norm{0,1}$ is defined as follows \citep[see, e.g.][Chapter 4]{CGS(2011)}:
$$d_{\mathrm{W}}(T^\pi, Z|\bm{a}_n)\equiv  \sup_{h\in\mathcal{L}}\left\vert\E_\pi[h(T^\pi)-h(Z)]\right\vert,$$
where $\mathcal{L}\equiv \{h:\mathbb{R}\to\mathbb{R}: \forall x, y\in\mathbb{R},\ |h(y)-h(x)|\leq |y-x|\}$ denotes the collection of Lipschitz functions with Lipschitz constant $1$.\footnote{\label{foot:Wasserstein} We use the Wasserstein distance because of its common usage in the literature on Stein's method \citep[e.g.][]{Chatterjee(2008), CGS(2011), Chen(2021)}, but we refer to Chapter 6 of \cite{CGS(2011)} for combinatorial CLTs with $\pi\sim \mathcal{U}(\mathbb{G}_n)$ that uses the Kolmogorov distance.} From Corollary 4.3 of \cite{CGS(2011)}, 
	\begin{equation}\label{eq: Wassd bound}
		d_{\text{W}}(T^\pi, Z|\bm{a}_n)\leq
		\sqrt{\frac{2}{\mathrm{\uppi}}\V_\pi
			\left(\E\left[\frac{1}{2\lambda}(T^{\pi'}-T^\pi)^2|\pi,\bm{a}_n\right]\right)}
		+\frac{1}{3\lambda}\E_\pi[|T^{\pi'}-T^\pi|^3].
	\end{equation}
	For the second term on the RHS of \eqref{eq: Wassd bound},
	\begin{align*}
		\lambda^{-1}\E[|T^{\pi'}-T^\pi|^3|\pi,\bm{a}_n]
		&=\sigma_n^{-3}\lambda^{-1}\sum_{s=1}^{\mathcal{S}_n}
		\frac{n_s-1}{n-\mathcal{S}_n}
		\frac{1}{n_s(n_s-1)}
		\sum_{i\neq j}| a_{ni\pi(j)}^s+a_{nj\pi(i)}^s-a_{ni\pi(i)}^s-a_{nj\pi(j)}^s|^3\\
		&\leq\sigma_n^{-3}\lambda^{-1}\sum_{s=1}^{\mathcal{S}_n}
		\frac{16}{n_s(n-\mathcal{S}_n)}
		\sum_{i\neq j}\left(| a_{ni\pi(j)}^s|^3+|a_{nj\pi(i)}^s|^3+|a_{ni\pi(i)}^s|^3+|a_{nj\pi(j)}^s|^3\right)\\
		&\leq 16\sigma_n^{-3}\sum_{s=1}^{\mathcal{S}_n} n_s^{-1} \sum_{i,j}|a_{nij}^s|^3,
	\end{align*}
	where the first equality is by iterated expectations, the first inequality is by convexity of $x\mapsto |x|^3$ and the second inequality uses $\sum_{i\neq j}| a_{ni\pi(j)}^s|^3=\sum_{i, j\ne \pi(i)} |a_{nij}^s|^3$. Hence, by Conditions \ref{cond: sigcon}, \ref{cond: 3mom} and the CMT,
	\begin{align}
		\lambda^{-1}\E_\pi[|T^{\pi'}-T^\pi|^3]
		&\leq 16\sigma_n^{-3}\sum_{s=1}^{\mathcal{S}_n}
		n_s^{-1} \sum_{i,j}|a_{nij}^s|^3\notag\\
		&=16\sigma_n^{-3}\sum_{s=1}^{\mathcal{S}_n}n_s^{-1} \left(\sum_{i=1}^{n_s} |b_{ni}^s|^3\right)
		\left(\sum_{i=1}^{n_s} |c_{ni}^s|^3\right)\notag\\
		&\conp 0.\label{eq:conv_RHS2}
	\end{align}
	Next we will show that $\V_\pi\left[\frac{1}{2\lambda}\E[(T^{\pi\prime}-T^\pi)^2\vert \pi, \bm{a}_n]\right]\conp 0$. Note first that
	\begin{align}
		\E[(T^{\pi'}-T^\pi)^2|\pi,\bm{a}_n]
		&=\frac{1}{\sigma_n^2}\sum_{s=1}^{\mathcal{S}_n}
		\frac{n_s-1}{n-\mathcal{S}_n}\frac{1}{n_s(n_s-1)}
		\sum_{i\neq j}( a_{ni\pi(j)}^s+a_{nj\pi(i)}^s-a_{ni\pi(i)}^s-a_{nj\pi(j)}^s)^2\label{eq: 2mom re}.
	\end{align}
    Furthermore, 
	\begin{align}
		&\sum_{i\neq j}( a_{ni\pi(j)}^s+a_{nj\pi(i)}^s-a_{ni\pi(i)}^s-a_{nj\pi(j)}^s)^2\notag\\
		&=\sum_{i\neq j}
		\bigg(
		a_{ni\pi(i)}^{s2}+a_{nj\pi(j)}^{s2}+a_{ni\pi(j)}^{s2}+a_{nj\pi(i)}^{s2}
		+2a_{ni\pi(i)}^{s}a_{nj\pi(j)}^s-2a_{ni\pi(i)}^{s}a_{ni\pi(j)}^s\notag\\
		&\quad-2a_{ni\pi(i)}^{s}a_{nj\pi(i)}^s
		-2a_{nj\pi(j)}^{s}a_{ni\pi(j)}^s-2a_{nj\pi(j)}^{s}a_{nj\pi(i)}^s+2a_{ni\pi(j)}^{s}a_{nj\pi(i)}^s
		\bigg)\notag\\
		&=
		2n_s\sum_{i=1}^{n_s}a_{ni\pi(i)}^s+2\left(\sum_{i=1}^{n_s}a_{ni\pi(i)}^s\right)^2
		+2\sum_{i,j}a_{ni\pi(j)}^{s2}+2\sum_{i\neq j}a_{ni\pi(j)}^sa_{nj\pi(i)}^s, \label{eq: 2mom re b}
	\end{align}
	where the second equality uses the following identities
	\begin{align*}
		\sum_{i\neq j}
		(a_{ni\pi(i)}^{s2}+a_{nj\pi(j)}^{s2})
		&=2(n_s-1)\sum_{i=1}^{n_s}a_{ni\pi(i)}^{s2},\\
		\sum_{i\neq j}
		a_{ni\pi(j)}^{s2}+\sum_{i\neq j}a_{nj\pi(i)}^{s2}
		&=2\sum_{i\neq j}a_{ni\pi(j)}^{s2},\\
		\sum_{i\neq j}
		a_{ni\pi(i)}^sa_{nj\pi(j)}^s
		&=\sum_{j=1}^na_{nj\pi(j)}^s\left(\sum_{i=1}^na_{ni\pi(i)}^s-a_{nj\pi(j)}^s\right)
		=\left(\sum_{i=1}^na_{ni\pi(i)}^s\right)^2-\sum_{i=1}^{n_s}a_{ni\pi(i)}^{s2},\\
		\sum_{i\neq j} a_{ni\pi(i)}^sa_{ni\pi(j)}^s & = \sum_{i\neq j} a_{nj\pi(j)}^sa_{nj\pi(i)}^s
		=\sum_{i=1}^na_{ni\pi(i)}^s\left(\sum_{j=1}^na_{ni\pi(j)}^s-a_{ni\pi(i)}^s\right)=-\sum_{i=1}^n
		a_{ni\pi(i)}^{s2},\\
		\sum_{i\neq j}a_{ni\pi(i)}^sa_{nj\pi(i)}^s &= \sum_{i\neq j}
		a_{nj\pi(j)}^sa_{ni\pi(j)}^s=-\sum_{i=1}^{n_s}a_{ni\pi(i)}^{s2}.
	\end{align*}
	From \eqref{eq: 2mom re} and \eqref{eq: 2mom re b}, we obtain
	\begin{align}
		&	\V_\pi\left[\frac{1}{2\lambda}\E[(T^{\pi\prime}-T^\pi)^2\vert \pi, \bm{a}_n]\right]\notag\\
		&=\V_\pi\left[
		\frac{1}{2\lambda}\sum_{s=1}^{\mathcal{S}_n}
		\frac{n_s-1}{n-|\mathcal{S}_n|}
		\frac{1}{n_s(n_s-1)}
		\left(2n_s\sum_{i=1}^{n_s}a_{ni\pi(i)}^s+2\left(\sum_{i=1}^{n_s}a_{ni\pi(i)}^s\right)^2+2\sum_{i,j}a_{ni\pi(j)}^{s2}\right.\right.\notag\\
		& \hspace{1.5cm}\left.\left. +2\sum_{i\neq j}a_{ni\pi(j)}^sa_{nj\pi(i)}^s\right)
		\right]\notag\\
		&=\V_\pi\left[
		\frac{1}{2}\sum_{s=1}^{\mathcal{S}_n}
		\frac{1}{n_s}
		\left(n_s\sum_{i=1}^{n_s}a_{ni\pi(i)}^{s2}+\left(\sum_{i=1}^{n_s}a_{ni\pi(i)}^s\right)^2
		+\sum_{i,j}a_{nij}^{s2}+\sum_{i\neq j}a_{ni\pi(j)}^sa_{nj\pi(i)}^s\right)
		\right]\notag\\
		&\leq \frac{3}{4}\left(
		\sum_{s=1}^{\mathcal{S}_n}
		\V_\pi\left[\sum_{i=1}^{n_s}a_{ni\pi(i)}^{s2}\right]
		+\sum_{s=1}^{\mathcal{S}_n}
		\frac{1}{n_s^2}
		\V_\pi\left[\left(\sum_{i=1}^{n_s}a_{ni\pi(i)}^{s}\right)^2\right]\right.\notag\\
		& \hspace{1cm} \left. +\sum_{s=1}^{\mathcal{S}_n}\frac{1}{n_s^2}
		\V_\pi\left[\sum_{i\neq j}a_{ni\pi(j)}^{s}a_{nj\pi(i)}^s\right]
		\right),\label{eq: 3 var bound}
	\end{align}
	where the inequality follows by $\V[X+Y+Z]\leq 3(\V[X]+\V[Y]+\V[Z])$.
	\par
	Consider the first summand in \eqref{eq: 3 var bound}. Let $\overline{b^{s2}}\equiv n_s^{-1}\sum_{j=1}^{n_s}b_{ni}^{s2}$ and $\overline{c^{s2}}\equiv n_s^{-1}\sum_{j=1}^{n_s}c_{ni}^{s2}$. Then, we have
	\begin{align}
		\sum_{s=1}^{\mathcal{S}_n}
		\V_\pi\left[\sum_{i=1}^{n_s}a_{ni\pi(i)}^{s2}\right]
		&=
		\sum_{s=1}^{\mathcal{S}_n}
		\frac{1}{n_s-1} \sum_{i,j}\left[a_{nij}^{s2}- \frac{1}{n_s}\sum_{g=1}^{n_s}a_{ngj}^{s2}- \frac{1}{n_s}\sum_{h=1}^{n_s}a_{nih}^{s2}+ \frac{1}{n^2_s}\sum_{g,h}^{n_s}a_{ngh}^{s2}\right]^2\notag\\
		&= \sum_{s=1}^{\mathcal{S}_n} \frac{1}{n_s-1} \sum_{i,j}[(b^{s2}_i - \overline{b^{s2}})(c^{s2}_j - \overline{c^{s2}})]^2\notag\\
		&\leq	\sum_{s=1}^{\mathcal{S}_n}
		\frac{1}{n_s-1}
		\sum_{i=1}^{n_s}b_{ni}^{s4}\sum_{i=1}^{n_s}c_{ni}^{s4}\notag\\
		&\conp 0,\label{eq: VB1}
	\end{align}
	where the first equality is by Theorem 2 of \cite{Hoeffding(1951)},
	the second uses $a^s_{nij}=b^s_{ni}c^s_{nj}$ and some algebra, the first inequality follows using $\V(X)\le \E(X^2)$ and the convergence follows from Condition \ref{cond: Uvarcon}.\par
	Next, for the second summand in \eqref{eq: 3 var bound},
	\begin{align}
		\sum_{s=1}^{\mathcal{S}_n}
		n_s^{-2}\V_\pi\left[\left(\sum_{i=1}^{n_s}a_{ni\pi(i)}^{s}\right)^2\right]
			&\leq \sum_{s=1}^{\mathcal{S}_n}
		n_s^{-2}\E_\pi\left[
		\left(\sum_{i=1}^{n_s}b_{ni}^sc_{n\pi(i)}^{s}\right)^4\right]\notag\\
		&\leq \sum_{s=1}^{\mathcal{S}_n} \frac{M_4}{n_s^2}\left(\sum_{i=1}^{n_s} b_{ni}^{s4}\right) \left(\sum_{i=1}^{n_s}c_{ni}^{s4}\right)\notag\\
		&\conp 0,\label{eq: VB2}
	\end{align}
	where the first inequality follows by $\V[X]\leq \E[X^2]$ and the definition of $a_{ni\pi(i)}^s$, the second inequality is
	by Lemma \ref{lem:MZ} above and the convergence holds by Condition \ref{cond: Uvarcon}.\par

	Finally, consider the third summand \eqref{eq: 3 var bound}.
Remark that
$$\sum_{i\neq  j}a_{ni\pi(j)}^{s}a_{nj\pi(i)}^s = \left(\sum_{i=1}^{n_s} b_{ni}^s c^s_{n\pi(i)}\right)^2 - \sum_{i=1}^{n_s} b^{s2}_{ni} c^{s2}_{\pi(i)}=
\left(\sum_{i=1}^{n_s} a^s_{ni\pi(i)}\right)^2 - \sum_{i=1}^{n_s} a^{s2}_{ni\pi(i)}.$$
As a result,
	\begin{align}
		\sum_{s=1}^{\mathcal{S}_n} \frac{1}{n_s^2} \V_\pi\left[\sum_{i\neq  j}a_{ni\pi(j)}^{s}a_{nj\pi(i)}^s\right]
		&\leq 2\sum_{s=1}^{\mathcal{S}_n}
		\frac{1}{n_s^2} \left\{\V_\pi\left[\left(\sum_{i=1}^{n_s} a^s_{ni\pi(i)}\right)^2\right]+\V_\pi\left[\sum_{i=1}^{n_s} a^{s2}_{ni\pi(i)}\right]\right\} \notag\\
		&\conp 0,\label{eq: VB3}
	\end{align}
	where we used $\V(X+Y) \leq 2(\V(X)+\V(Y))$ and \eqref{eq: VB1} and \eqref{eq: VB2} to obtain the convergence.  Combining \eqref{eq: 3 var bound},
	\eqref{eq: VB1}, \eqref{eq: VB2}, \eqref{eq: VB3} and CMT, we obtain
	\begin{align}
		\V_\pi
		\left(\E\left[\frac{1}{2\lambda}(T^{\pi'}-T^\pi)^2|\pi,\bm{a}_n\right]\right)
		&\conp 0. \label{eq:conv_RHS1}
	\end{align}
	Combining \eqref{eq: Wassd bound}, \eqref{eq:conv_RHS2} and \eqref{eq:conv_RHS1}, we obtain $d_{\text{W}}(T^\pi, Z|\bm{a}_n)\conp 0$. By Theorem 2.3.2 of \cite{Durrett(2010)}, for any subsequence $\{n_l\}$, there exists a further subsequence $\{n_l(m)\}$ such that $d_{\text{W}}(T^\pi, Z|\bm{a}_n)\conas 0$ along $\{n_l(m)\}$.
	Since the Wasserstein distance bounds the bounded Lipschitz distance from above:
	\begin{equation}
		d_{\mathrm{BL}}(T^\pi, Z|\bm{a}_n)\leq d_{\mathrm{W}}(T^\pi, Z|\bm{a}_n),
	\end{equation}
	where
	\begin{align*}
		d_{\mathrm{BL}}(T^\pi, Z|\bm{a}_n)&\equiv  \sup_{h\in\mathcal{BL}}\left\vert\E_\pi[h(T^\pi)-h(Z)]\right\vert,\\
		\mathcal{BL}&\equiv \left\{h:\mathbb{R}\to\mathbb{R}: \max\left\{\sup_{x, y \in\mathbb{R}, x\neq y}\frac{|h(x)-h(y)|}{|x-y|}, \sup_{x\in\mathbb{R}}|f(x)|\right\}\leq 1\right\},
	\end{align*}
 $\E_\pi[h(T^\pi)-h(Z)]\conas 0$ along $\{n_l(m)\}$	for any Lipschitz function $h:\mathbb{R}\to\mathbb{R}$ with Lipschitz constant $1$. By the Portmanteau theorem,
	$P^\pi(T^\pi\leq t)\conas \Phi(t)$ along $\{n_l(m)\}$.
	Now using Theorem 2.3.2 of \cite{Durrett(2010)} in the reverse direction gives
	$P^\pi(T^\pi\leq t)\conp \Phi(t)$ along the full sequence $\{n\}$.
	$\square$\par


\subsection{Lemma \ref{lem:MZ}} 
\label{sub:lemma_lem_MZ}

Let $\eps_1,\dots, \eps_n$ be independent Rademacher variables. Using Lemma \ref{lem:Sym} with $a_i=a_i$ and $\xi_i=b_{\pi(i)}$, and the $c_r$ inequality, we obtain
\begin{align}
\E_\pi\left[\left\Vert\sum_{i=1}^na_ib_{\pi(i)}\right\Vert^r\right] &\leq
\E_\pi\left[\max_{1\leq k\leq n}\left\Vert\sum_{i=1}^ka_ib_{\pi(i)}\right\Vert^r\right]\notag\\
&\leq 2^{r-1}\left\{\left(\frac{6r^2}{(r-1)^2}\right)^r\E\left[\left\Vert \sum_{i=1}^na_ib_{\pi(i)}\eps_i\right\Vert^r\right] \right. \notag \\
& \left. \qquad +\left(\frac{8(4r-1)}{(9r-1)(n-1)}\right)^r
\E\left[\left|\sum_{i=1}^nb_{\pi(i)}\eps_i\right|^r\right]\left(\sum_{i=1}^n\Vert a_i\Vert\right)^r\right\}.\label{MZW0}
\end{align}
Consider the term $\E\left[\Vert \sum_{i=1}^na_ib_{\pi(i)}\eps_i\Vert^r\right]$.
By Lemma \ref{lem:KK},
\begin{align}
\E\left[\left\Vert \sum_{i=1}^n a_ib_{\pi(i)}\eps_i\right\Vert^r\right]
&=
\E_\pi\left\{\E_\eps\left[\left\Vert\sum_{i=1}^na_ib_{\pi(i)}\eps_i\right\Vert^r\right]\right\}\notag\\
&\leq
K_r\E_\pi\left\{
\left(\sum_{i=1}^n\Vert a_i\Vert^2b_{\pi(i)}^2\right)^{r/2}
\right\}\notag\\
&\leq K_rn^{(r/2)\vee 1}\E_\pi\left[n^{-1}\sum_{i=1}^n\Vert a_i\Vert^r|b_{\pi(i)}|^r\right]\notag\\
&=K_rn^{(r/2)\vee 1}\left(n^{-1}\sum_{i=1}^n\Vert a_i\Vert^r\right)\left(n^{-1}\sum_{i=1}^n|b_{i}|^r\right).\label{MZW1}
\end{align}
where $\E_\eps[\cdot]$ denotes the expectation with respect to $\eps_i, i=1,\dots, n$ conditional on $\pi$, the second inequality holds by convexity for $r\geq 2$ and using $(x+y)^r \leq x^r +y^r$ if $r<2$ and the last line holds since $\pi\sim\mathcal{U}(\mathbb{G}_n)$.  Similarly, by Lemma \ref{lem:KK},
\begin{align}
\E\left[\left|\sum_{i=1}^nb_{\pi(i)}\eps_i\right|^r\right]
&=
\E_\pi\left\{\E_\eps\left[\left|\sum_{i=1}^nb_{\pi(i)}\eps_i\right|^r\right]\right\}\notag\\
&\leq
K_r\E_\pi\left\{
\left(\sum_{i=1}^nb_{\pi(i)}^2\right)^{r/2}
\right\}\notag\\
&=
K_r
\left(\sum_{i=1}^nb_{i}^2\right)^{r/2}\notag\\
&\leq
K_rn^{(r/2) \vee 1}\left(n^{-1}\sum_{i=1}^n|b_{i}|^r\right).\label{MZW2}
\end{align}
Combining \eqref{MZW0}, \eqref{MZW1} and \eqref{MZW2} and noting that $\frac{n}{n-1}\leq 2$, we obtain \eqref{eq: PMZW ineq} with
$$M_r\equiv
2^{r-1}\left\{\left(\frac{6r^2}{(r-1)^2}\right)^r
+\left(\frac{8(4r-1)}{9r-1}\right)^r2^r\right\}K_r \; _\square$$


	\subsection{Theorem \ref{thm:behav_W}} 
	\label{sub:theorem_ref_power}

	We prove the result in three steps. First, we derive the asymptotic distribution of $n^{-1/2}\tilde{X}'u$ in $n^{-1/2}\tilde{X}'(y-X\beta_0)=n^{-1/2}\tilde{X}'\tilde{X}(\beta-\beta_0)+n^{-1/2}\tilde{X}'u$.
	Then, we show the consistency of the covariance matrix estimate. The third step concludes.
	
	\paragraph*{Step 1: Asymptotic distribution of $n^{-1/2}\tilde{X}'u$} 
	\label{paragraph:power_AWald_S1}

Rewrite
\begin{align}
	\Omega_n^{-1/2}
	n^{-1/2}\sum_{s=1}^{S_n}
	X^{s}M_{\bm{1}_s}u^s
	&=		\Omega_n^{-1/2}
	n^{-1/2}\sum_{s=1}^{S_n}
	\sum_{i=1}^{n_s}\left(X_{ni}^s-\E\left[n_s^{-1}\sum_{i=1}^{n_s}X_{ni}^s\right]\right)u_{ni}^s\notag\\
	&-		\Omega_n^{-1/2}
	n^{-1/2}\sum_{s=1}^{S_n}
	n_s^{-1}\sum_{i=1}^{n_s}\left(X_{ni}^s-\E\left[X_{ni}^s\right]\right)\left(\sum_{i=1}^{n_s}u_{ni}^s\right)
	\label{eq: XMu decomp}
\end{align}
Remark that $\{X^{s\prime}M_{\bm{1}_s}u^s\}_{s=1,\dots, S_n}$ are independent conditional on the $\sigma$-field generated by $Z_{n1},\dots, Z_{nn}$. We first derive the asymptotic normality of the first summand of
\eqref{eq: XMu decomp}.

By Assumption \ref{A2}\ref{2ns}, \eqref{cond: Ceval} holds, with $\bar{\Sigma}_n=\Omega_n$. Let $\lambda_n\equiv\lambda_{\min}(\Omega_n)$. By Lyapunov and Cauchy-Schwarz inequalities for any $\eps>0$, as $n\to \infty$,
\begin{align*}
	&\frac{1}{n\lambda_n}\sum_{s=1}^{S_n}\sum_{i=1}^{n_s}\E\left[\left\Vert X_{ni}^{s}-n_s^{-1}\sum_{i=1}^{n_s}\E[X_{ni}^s]\right\Vert^2u_{ni}^{s2}1\left(\left\Vert X_{ni}^{s}-n_s^{-1}\sum_{i=1}^{n_s}\E[X_{ni}^s]\right\Vert^2u_{ni}^{s2}>n\eps\right)\right]\\
	&\leq
	\frac{1}{\eps^{\delta/4}\lambda_n^{1+\delta/4}n^{1+\delta/4}}\sum_{s=1}^{S_n}\sum_{i=1}^{n_s}\E\left[\left\Vert X_{ni}^{s}-n_s^{-1}\sum_{i=1}^{n_s}\E[X_{ni}^s]\right\Vert^{2+\delta/2}|u_{ni}^s|^{2+\delta/2}\right]\\
	&\leq \frac{1}{\eps^{\delta/4}\lambda_n^{1+\delta/4}n^{1+\delta/4}}\sum_{s=1}^{S_n}\sum_{i=1}^{n_s}\left(\E\left[\left\Vert X_{ni}^{s}-n_s^{-1}\sum_{i=1}^{n_s}\E[X_{ni}^s]\right\Vert^{4+\delta}\right]\right)^{1/2}\left(\E\left[|u_{ni}^{s}|^{4+\delta}\right]\right)^{1/2}\\
	&\conas 0.
\end{align*}	
Hence, \eqref{cond: CLindeberg} holds, with $u_{ni}^s\left(X_{ni}^s - n_s^{-1}\sum_{i=1}^{n_s}\E[X_{ni}^s]\right)$ in place of $X_{ni}$. Then, Lemma \ref{lem: cond Lindeberg} yields
\begin{equation}\label{eq: XMu1 con}
	\Omega_n^{-1/2}
	n^{-1/2}\sum_{s=1}^{S_n}
	\sum_{i=1}^{n_s}\left(X_{ni}^s-\E\left[n_s^{-1}\sum_{i=1}^{n_s}X_{ni}^s\right]\right)u_{ni}^s
	\cond \Norm{0, I_{k}}.
\end{equation}
Consider the second summand of \eqref{eq: XMu decomp}. We have
\begin{align*}
	&\left\Vert \V\left[n^{-1/2}\sum_{s=1}^{S_n}
	n_s^{-1}\sum_{i=1}^{n_s}\left(X_{ni}^s-\E\left[X_{ni}^s\right]\right)\left(\sum_{i=1}^{n_s}u_{ni}^s\right)\right]\right\Vert\notag\\
	&=\left\Vert n^{-1}\sum_{s=1}^{S_n}n_s^{-2}
	\sum_{i,j=1}^{n_s}
	\sum_{k,l=1}^{n_s}
	\E\left[(X_{ni}^s-\E[X_{ni}^s])(X_{nj}^s-\E[X_{nj}^s])'u_{nk}^su_{nl}^s\right]\right\Vert\notag\\
	&=\left\Vert n^{-1}\sum_{s=1}^{S_n}n_s^{-2}
	\sum_{i=1}^{n_s}
	\sum_{j=1}^{n_s}
	\E\left[(X_{ni}^s-\E[X_{ni}^s])(X_{ni}^s-\E[X_{ni}^s])'u_{nj}^{s2}\right]\right\Vert\notag\\
	&\leq
	n^{-1}\sum_{s=1}^{S_n}n_s^{-2}
	\sum_{i=1}^{n_s}
	\sum_{j=1}^{n_s}
	\E\left[\Vert X_{ni}^s-\E[X_{ni}^s]\Vert^2u_{nj}^{s2}\right]\notag\\
	&\leq n^{-1}\sum_{s=1}^{S_n}n_s^{-2}
	\sum_{i=1}^{n_s}\sum_{j=1}^{n_s}
	(\E[\Vert X_{ni}^s-\E[X_{ni}^s]\Vert^4])^{1/2}(\E[u_{nj}^{s4}])^{1/2}\notag\\
	&= O_{a.s.}(n^{-1}S_n)\notag\\
	&\conas 0,
\end{align*}
where the first and second equalities hold by independence (Assumption \ref{A2}\ref{2ex}) and
Assumption \ref{A2}\ref{2cu}, the first inequality is by Jensen's inequality, the second inequality is by Cauchy-Schwarz, the third inequality is by Assumption \ref{A2}\ref{2mom} and
the last is by Assumption \ref{A2}\ref{2ns}. Then, by Chebyshev's inequality
\begin{align}
	&\left\Vert\Omega_n^{-1/2}
	n^{-1/2}\sum_{s=1}^{S_n}
	n_s^{-1}\sum_{i=1}^{n_s}\left(X_{ni}^s-\E\left[X_{ni}^s\right]\right)\left(\sum_{i=1}^{n_s}u_{ni}^s\right)\right\Vert\notag\\
	&\leq k^{1/2}\lambda_n^{-1/2}
	\left\Vert
	n^{-1/2}\sum_{s=1}^{S_n}
	n_s^{-1}\sum_{i=1}^{n_s}\left(X_{ni}^s-\E\left[X_{ni}^s\right]\right)\left(\sum_{i=1}^{n_s}u_{ni}^s\right)\right\Vert\notag\\
	&=o_p(1).\label{eq: XMu2 con}
\end{align}
From \eqref{eq: XMu1 con}, \eqref{eq: XMu2 con} and Slutsky's lemma, we obtain
\begin{equation}\label{eq: selfAN}
	\Omega_n^{-1/2}
	n^{-1/2}\sum_{s=1}^{S_n}X^{s\prime}M_{\bm{1}_s}u^s
	\cond \Norm{0, I_{k}}.
\end{equation}	

	\paragraph*{Step 2: Consistency of the covariance matrix estimator} 

Let $\hat{\Omega}_n
		\equiv n^{-1}\sum_{s=1}^{S_n}\sum_{i=1}^{n_s}\tilde{X}_{ni}^{s}\tilde{X}_{ni}^{s\prime}\tilde{v}_{ni}^{s2}$. We prove that $\hat{\Omega}_n - \Omega_n \conp 0$.  Let us define
$$\tilde{\Omega}_n\equiv n^{-1}\sum_{s=1}^{S_n}\sum_{i=1}^{n_s}
(X_{ni}^s-\E[n_s^{-1}\sum_{i=1}^{n_s}X_{ni}^s])(X_{ni}^s-\E[n_s^{-1}\sum_{i=1}^{n_s}X_{ni}^s])'v_{ni}^{s2}.$$ By the WLLN,  $\tilde{\Omega}_n-\Omega_n\conp 0$. Therefore, it suffices to show that
$$\hat{\Omega}_n-\tilde{\Omega}_n\conp 0.$$
Note that
\begin{equation}
		\hat{\Omega}_n=n^{-1}\sum_{s=1}^{S_n}
		\sum_{i=1}^{n_s}\tilde{X}_{ni}^{s}\tilde{X}_{ni}^{s\prime}v_{ni}^{s2}-2n^{-1}\sum_{s=1}^{S_n}\sum_{i=1}^{n_s}\tilde{X}_{ni}^{s}\tilde{X}_{ni}^{s\prime}v_{ni}^{s}\bar{v}^s+n^{-1}\sum_{s=1}^{S_n}
		\sum_{i=1}^{n_s}\tilde{X}_{ni}^{s}\tilde{X}_{ni}^{s\prime}\bar{v}^{s2}.\label{AC1}
\end{equation}
Moreover,  by the triangle inequality,
\begin{align}
\Vert\tilde{\Omega}_n
-n^{-1}\sum_{s=1}^{S_n}\sum_{i=1}^{n_s}\tilde{X}_{ni}^{s}\tilde{X}_{ni}^{s\prime}{v}_{ni}^{s2}
\Vert
&\leq 2
n^{-1}\sum_{s=1}^{S_n}\sum_{i=1}^{n_s}
\Vert\bar{X}^s-\E[\bar{X}^s]\Vert \Vert X_{ni}^{s}\Vert v_{ni}^{s2}\notag\\
&\quad+
n^{-1}\sum_{s=1}^{S_n}\sum_{i=1}^{n_s}
\Vert \bar{X}^s-\E[\bar{X}^s]\Vert\Vert\bar{X}^s\Vert v_{ni}^{s2}\notag\\
&\quad+n^{-1}\sum_{s=1}^{S_n}\sum_{i=1}^{n_s}
\Vert \bar{X}^s-\E[\bar{X}^s]\Vert\Vert\E[\bar{X}^s]\Vert v_{ni}^{s2}.\label{eq: tOm triangle}
\end{align}
From \eqref{eq: sum E4 order},
\begin{align}
n^{-1}\sum_{s=1}^{S_n}\sum_{i=1}^{n_s}\E[\Vert\bar{X}^{s}-\E[\bar{X}^s]\Vert^4]
&=n^{-1}\sum_{s=1}^{S_n}n_s^{-3}
\E\left[\left\Vert\sum_{i=1}^{n_s}({X}_{ni}^{s}-\E[{X}_{ni}^s])\right\Vert^4\right]\notag\\
&=O_{a.s.}(n^{-1}S_n)\notag\\
&=o_{a.s.}(1).\label{eq: EXbar4}
\end{align}
By the Cauchy-Schwarz inequality applied twice,
\begin{align*}
&n^{-1}\sum_{s=1}^{S_n}\sum_{i=1}^{n_s}
\Vert\bar{X}^s-\E[\bar{X}^s]\Vert \Vert X_{ni}^{s}\Vert v_{ni}^{s2}\\
&\leq \left(n^{-1}\sum_{i=1}^{n}\Vert {X}_{ni}\Vert^4\right)^{1/4}\left(n^{-1}\sum_{s=1}^{S_n}\sum_{i=1}^{n_s}\Vert\bar{X}^{s}-\E[\bar{X}^s]\Vert^4\right)^{1/4}
\left(n^{-1}\sum_{s=1}^{S_n}\sum_{i=1}^{n_s}\E[v_{ni}^{s4}]\right)^{1/2}
\end{align*}
Then, again by the Cauchy-Schwarz inequality and \eqref{eq: EXbar4},
\begin{align}
&n^{-1}\sum_{s=1}^{S_n}\sum_{i=1}^{n_s}
\E\left[\Vert\bar{X}^s-\E[\bar{X}^s]\Vert \Vert X_{ni}^{s}\Vert v_{ni}^{s2}\right]\notag\\
&\leq \left(n^{-1}\sum_{i=1}^{n}\E[\left\Vert {X}_{ni}\right\Vert^4]\right)^{1/4}\left(n^{-1}\sum_{s=1}^{S_n}\sum_{i=1}^{n_s}\E[\Vert\bar{X}^{s}-\E[\bar{X}^s]\Vert^4]\right)^{1/4}
\left(n^{-1}\sum_{s=1}^{S_n}\sum_{i=1}^{n_s}\E[v_{ni}^{s4}]\right)^{1/2} \notag\\
&=o_{a.s.}(1)\label{eq: EOm1}.
\end{align}		
Similarly,
\begin{align}
&n^{-1}\sum_{s=1}^{S_n}\sum_{i=1}^{n_s}
\E\left[\Vert\bar{X}^s-\E[\bar{X}^s]\Vert \Vert \bar{X}^{s}\Vert v_{ni}^{s2}\right]\notag\\
&\leq \left(n^{-1}\sum_{i=1}^{S_n}\sum_{i=1}^{n_s}\E[\Vert\bar{X}^s\Vert^4]\right)^{1/4}\left(n^{-1}\sum_{s=1}^{S_n}\sum_{i=1}^{n_s}\E[\Vert\bar{X}^{s}-\E[\bar{X}^s]\Vert^4]\right)^{1/4} \left(n^{-1}\sum_{s=1}^{S_n}\sum_{i=1}^{n_s}\E[v_{ni}^{s4}]\right)^{1/2}\notag\\
&=o_{a.s.}(1).\label{eq: EOm2}
\end{align}		
By Markov's inequality, \eqref{eq: EOm1} and \eqref{eq: EOm2}, the first and second terms on the RHS of \eqref{eq: tOm triangle} are $o_p(1)$. By analogous arguments, the third term of \eqref{eq: tOm triangle} is also an $o_p(1)$. Hence,
$$n^{-1}\sum_{s=1}^{S_n}\sum_{i=1}^{n_s}\tilde{X}_{ni}^s\tilde{X}_{ni}^{s\prime}v_{ni}^{s2}-\Omega_n\conp 0.$$
Now, we showed in \eqref{eq: EhVI3 con} that the third term in \eqref{AC1} is an $o_p(1)$.  Finally,  consider the second term in \eqref{AC1}.  By the triangle and Cauchy-Schwarz inequalities,
	\begin{align}
		\left\Vert n^{-1}\sum_{s=1}^{S_n}\sum_{i=1}^{n_s}\tilde{X}_{ni}^{s}\tilde{X}_{ni}^{s\prime}v_{ni}^s\bar{v}^{s}\right\Vert
		&\leq n^{-1}\sum_{s=1}^{S_n}\sum_{i=1}^{n_s}\left\Vert \tilde{X}_{ni}^{s}\right\Vert^2|v_{ni}^{s}\bar{v}^{s}|\notag\\
		&\leq \left(n^{-1}\sum_{s=1}^{S_n}\sum_{i=1}^{n_s}\left\Vert \tilde{X}_{ni}^{s}\right\Vert^4\right)^{1/2}\left(n^{-1}\sum_{s=1}^{S_n}\sum_{i=1}^{n_s}v_{ni}^{s2}\bar{v}^{s2}\right)^{1/2}\notag\\
		&\leq \left(n^{-1}\sum_{i=1}^{n}\left\Vert \tilde{X}_{ni}\right\Vert^4\right)^{1/2}\left(n^{-1}\sum_{s=1}^{S_n}\sum_{i=1}^{n_s}v_{ni}^{s4}\right)^{1/4}\left(n^{-1}\sum_{s=1}^{S_n}\sum_{i=1}^{n_s}\bar{v}^{s4}\right)^{1/4}\notag\\
		&\conp 0,\label{eq: sumI term3 op1}
	\end{align}
	where the convergence holds due to \eqref{eq: X*4 expansion}
	, the fact that
	$n^{-1}\sum_{s=1}^{S_n}\sum_{i=1}^{n_s}v_{ni}^{s4}=O_p(1)$ which holds by the WLLN, and \eqref{eq: sum_sI_bar_u_s^4 op1}.


	\paragraph*{Step 3: Asymptotic distribution of $\mathcal{W}$} 
	\label{paragraph:power_AWald_S3}
We first determine the limit of $n^{-1}\tilde{X}'\tilde{X}$. Rewrite
	\begin{align}
n^{-1}\tilde{X}'\tilde{X}=n^{-1}\sum_{s=1}^{S_n}\sum_{i=1}^{n_s}\tilde{X}_{ni}^{s}\tilde{X}_{ni}^{s\prime}
		&=n^{-1}\sum_{s=1}^{S_n}\sum_{i=1}^{n_s}
		X_{ni}^{s}X_{ni}^{s\prime}-
		n^{-1}\sum_{s=1}^{S_n}n_s\bar{X}^{s}\bar{X}^{s\prime}.\label{eq: I EXX}
	\end{align}
For the first summand of \eqref{eq: I EXX}, by the WLLN,
	\begin{align}
n^{-1}\sum_{s=1}^{S_n}\sum_{i=1}^{n_s}
		X_{ni}^sX_{ni}^{s\prime}-
		n^{-1}\sum_{s=1}^{S_n}\sum_{i=1}^{n_s}
		\E\left[X_{ni}^sX_{ni}^{s\prime}\right]
		\conp 0.\label{eq: EXX1 con}
	\end{align}	
Consider the second summand of \eqref{eq: I EXX}. By the triangle inequality,
	\begin{align}
		&n^{-1}\left\Vert\sum_{s=1}^{S_n}\left\{n_s\bar{X}^{s}\bar{X}^{s\prime}
		-n_s\E[\bar{X}^{s}]\E[\bar{X}^{s}]'\right\}
		\right\Vert\notag\\
		&\leq \sum_{s=1}^{S_n}n^{-1}\left\Vert
		\sum_{i=1}^{n_s}(X_{ni}^{s}-\E[X_{ni}^{s}])
		\bar{X}^{s\prime}
		\right\Vert+
		\sum_{s=1}^{S_n}n^{-1}\left\Vert
		\sum_{i=1}^{n_s}\E[X_{ni}^{s}]\left(\bar{X}^{s}
		-\E[\bar{X}^{s}]\right)'
		\right\Vert.\label{eq: EXX}
	\end{align}
By the Cauchy-Schwarz inequality and convexity of $x\mapsto \|x\|^2$, the first summand of \eqref{eq: EXX} satisfies
	\begin{align}
		&\sum_{s=1}^{S_n}n^{-1}\E\left[\left\Vert
		\sum_{i=1}^{n_s}(X_{ni}^{s}-\E[X_{ni}^{s}])
		\left(n_s^{-1}\sum_{i=1}^{n_s}X_{ni}^{s\prime}\right)
		\right\Vert\right]\notag\\
		&\leq
		n^{-1}\sum_{s=1}^{S_n}\left\{\E\left[\left\Vert \sum_{i=1}^{n_s}(X_{ni}^{s}-\E[X_{ni}^{s}])\right\Vert^2\right]\right\}^{1/2}
		\left\{n_s^{-1}\sum_{i=1}^{n_s}\E[\Vert X_{ni}^{s}\Vert^2]\right\}^{1/2}\notag\\
		&=O_{a.s.}\left(n^{-1}\sum_{s=1}^{S_n}n_s^{1/2}\right)\notag\\
		&=o_{a.s.}(1),\label{eq: EXX21}
	\end{align}
	where the first equality uses $\E\left[n_s^{-1}\left\Vert \sum_{i=1}^{n_s}(X_{ni}^{s}-\E[X_{ni}^{s}])\right\Vert^2\right]=O(1)$ which holds by the independence assumption and
	the last by \eqref{eq: sum n_s^1/2/n}. Similarly, for the second summand of \eqref{eq: EXX},
	\begin{equation}
		\sum_{s=1}^{S_n}n^{-1}\left\Vert
		\sum_{i=1}^{n_s}\E[X_{ni}^{s}]\left(n_s^{-1}\sum_{i=1}^{n_s}(X_{ni}^{s}
		-\E[X_{ni}^{s}])'\right)
		\right\Vert=o_{a.s.}(1).\label{eq: EXX22}
	\end{equation}

Combining \eqref{eq: EXX21} and \eqref{eq: EXX22}, we obtain
	\begin{align}
		n^{-1}\left\Vert\sum_{s=1}^{S_n}\left\{n_s\bar{X}^{s}\bar{X}^{s\prime}
		-n_s\E[\bar{X}^{s}]\E[\bar{X}^{s}]'\right\}
		\right\Vert
		\conp 0.\label{eq: I EXX con}
	\end{align}
	Moreover, by the triangle inequality and convexity of $x\mapsto \|x\|^2$,
\begin{align}
\left\Vert n^{-1}\sum_{s=1}^{S_n}n_s\left\{\E[\bar{X}^{s}]\E[\bar{X}^{s\prime}]
-\E[\bar{X}^{s}\bar{X}^{s\prime}]\right\}\right\Vert
&=	
\left\Vert n^{-1}\sum_{s=1}^{S_n}
n_s\E\left[(\bar{X}^{s}-\E[\bar{X}^s])(\bar{X}^s-\E[\bar{X}^s])'\right]\right\Vert\notag\\	
&\leq n^{-1}\sum_{s=1}^{S_n}
\E\left[n_s^{-1}\left\Vert \sum_{i=1}^{n_s}(X_{ni}^{s}-\E[X_{ni}^s])\right\Vert^2\right]\notag\\	
&= n^{-1}\sum_{s=1}^{S_n}
\E\left[n_s^{-1}\sum_{i=1}^{n_s}\left\Vert X_{ni}^{s}-\E[X_{ni}^s]\right\Vert^2\right]\notag\\
&=O_{a.s.}(n^{-1}S_n)\notag\\
&=o_{a.s.}(1).\label{eq: equiv}	
\end{align}	
\eqref{eq: I EXX con} and \eqref{eq: equiv}	together yield
$n^{-1}\left\Vert\sum_{s=1}^{S_n}\left\{n_s\bar{X}^{s}\bar{X}^{s\prime}
-n_s\E[\bar{X}^{s}\bar{X}^{s\prime}]\right\}
\right\Vert\conp 0$. The latter combined with \eqref{eq: EXX1 con} gives
\begin{equation}\label{eq: XXcon}
 n^{-1}\tilde{X}'\tilde{X}-\E[n^{-1}\tilde{X}'\tilde{X}]\conp 0.
\end{equation}
To determine the asymptotic distribution of $\mathcal{W}$, rewrite
$$\Omega_n^{-1/2}n^{-1/2}\tilde{X}'(y-X\beta_0)=\Omega_n^{-1/2}n^{-1/2}\tilde{X}'\tilde{X}(\beta_n-\beta_0)+\Omega_n^{-1/2}n^{-1/2}\sum_{s=1}^{S_n}X^{s\prime}M_{\bm{1}_s} u^s.$$
Using \eqref{eq: selfAN}, \eqref{eq: XXcon} and $G=\lim_{n\to\infty} \Omega_n^{-1/2} \E[n^{-1}\tilde{X}'\tilde{X}]$, we obtain, if $\beta_n=\beta_0+hn^{-1/2}$,
$$\Omega_n^{-1/2}n^{-1/2}\tilde{X}'(y-X\beta_0) \cond \Norm{Gh, I_k}.$$
By the same argument as in \eqref{eq: covm consistency}, $\Omega_n^{1/2}\hat{\Omega}_n^{-1} \Omega_n^{1/2}\conp I_k$. Then, by Slutsky's lemma and the CMT,
$$\mathcal{W}\cond  \mathcal{W}_{\infty}\sim \chi^2_k(\|Gh\|^2).$$
	Under the fixed alternative $H_1:\beta\neq\beta_0$, since $\Vert\Omega_n^{-1/2}\E[n^{-1}\tilde{X}'\tilde{X}]n^{1/2}(\beta-\beta_0)\Vert\to \infty$, we obtain
$\mathcal{W} \conp \infty$.

	
\subsection{Corollary \ref{cor:SR}} 

First, recall that $\mathbb{S}'_n$ includes $\pi_1=\Id$. Let $(\pi_2,...,\pi_{N_n})$ denote the other permutations in $\mathbb{S}'_n$, and $\pi_{N_n+1},...,\pi_{\vert\mathbb{S}_n\vert}$ be the remaining permutations in $\mathbb{S}_n$. By Theorem \ref{thm:behav_Wpi}, conditional  on the data and with probability tending to one, $\mathcal{W}^{\pi}\cond \chi^2_k$. Hence, for any $t\in\mathbb{R}$,
\begin{equation}\label{eq: ecdf con}
{F}_n(t)\equiv (\vert\mathbb{S}_n\vert-1)^{-1}\sum_{i=2}^{\vert\mathbb{S}_n\vert}
1(\mathcal{W}^{\pi_i}\leq t)\conp P[\chi^2_k\leq t].
\end{equation}
Take any subsequence of $\{n\}$. Since $N_n\conp\infty$, there exists a further subsequence $\{m\}$ such that $N_m\conas \infty$. From Corollary 4.1 of \cite{Romano(1989)}, $\sup_{t\in\mathbb{R}}\vert\frac{1}{N_m-1}\sum_{i=2}^{N_m}1(\mathcal{W}^{\pi_i}\leq t)-F_n(t)\vert\conp 0$, hence
by the triangle inequality and \eqref{eq: ecdf con}, for any $t\in\mathbb{R}$
\begin{align}
\left\vert\frac{1}{N_m-1}\sum_{i=2}^{N_m}1(\mathcal{W}^{\pi_i}\leq t)-P[\chi^2_k\leq t]\right\vert	
&\leq \sup_{t\in\mathbb{R}}\left\vert\frac{1}{N_m-1}\sum_{i=2}^{N_m}1(\mathcal{W}^{\pi_i}\leq t)-F_n(t)\right\vert\notag\\
&\quad+\vert F_n(t)-P[\chi^2_k\leq t]\vert\notag\\
&\conp 0.\label{eq: ecdf2 con}
\end{align}
As a result, the empirical cdf $\hat{F}_{N_m}(t)$ of $\mathcal{W}^\pi$ on $\mathbb{S}'_n$ satisfies
\begin{align}
\hat{F}_{N_m}(t)&=N_m^{-1}1(\mathcal{W}^{\pi_1}\leq t)
	+\frac{N_m-1}{N_m}\frac{1}{N_m-1}\sum_{i=2}^{N_m}1(\mathcal{W}^{\pi_i}\leq t)\conp P[\chi^2_k\leq t].\label{eq: Fhat con}
\end{align}

Since the cdf of $\chi^2_k$ distribution is continuous and strictly increasing at its $1-\alpha$ quantile $q_{1-\alpha}(\chi^2_k)$, by Lemma 11.2.1 of \cite{Lehmann-Romano(2005)}, along the subsequence $\{m\}$,
\begin{equation}
\mathcal{W}^{(q)}\conp q_{1-\alpha}(\chi^2_k).
	\label{eq:conv_qWpi}
\end{equation}

By definition, $\E[\phi_\alpha((\mathcal{W}^\pi)_{\pi\in\mathbb{S}_n})]=P[\mathcal{W}> \mathcal{W}^{(q)}]+\frac{N\alpha-N^{+}}{N^{0}}P[\mathcal{W}= \mathcal{W}^{(q)}]$, hence
\begin{align}\label{sandwich}
	P[\mathcal{W}>\mathcal{W}^{(q)}]\leq \E[\phi_\alpha((\mathcal{W}^\pi)_{\pi\in\mathbb{S}_n})]\leq P[\mathcal{W}\geq \mathcal{W}^{(q)}].
\end{align}
Now, suppose first that $\beta_n=\beta_0+n^{-1/2}h$, with  either $h=0$ or $h\ne 0$, fixed. By Point 1 of Theorem \ref{thm:behav_W}, Equation \eqref{eq:conv_qWpi} and Slutsky's lemma, $\mathcal{W}-\mathcal{W}^{(q)} \cond\mathcal{W}_\infty - q_{1-\alpha}(\chi^2_k)$. Then, by continuity of the cdf of the $\chi^2_k(\|Gh\|^2) $ distribution at all positive points,
$$	\lim_{n\rightarrow\infty}P[\mathcal{W}>\mathcal{W}^{(q)}]=\lim_{n\rightarrow\infty}P[\mathcal{W}\geq \mathcal{W}^{(q)}]=P[\mathcal{W}_{\infty}>q_{1-\alpha}(\chi^2_k)].$$
Therefore, by the sandwich theorem, along the subsequence $\{m\}$,
$$	\lim_{n\to\infty} \E[\phi_\alpha((\mathcal{W}^\pi)_{\pi\in\mathbb{S}_n})]=P[\mathcal{W}_{\infty}>q_{1-\alpha}(\chi^2_k)].$$
By Urysohn's subsequence principle, the convergence above holds along $\{n\}$. Points 1 and 2 follow. Now, suppose that $n^{1/2}\Vert\beta_n - \beta_0\Vert \to\infty$. By Point 2 of Theorem \ref{thm:behav_W} and \eqref{eq:conv_qWpi}, $\mathcal{W}-\mathcal{W}^{(q)} \conp\infty$ along $\{m\}$. Then, \eqref{sandwich} implies that
		\begin{align*}
		\lim_{n\to\infty} \E[\phi_\alpha((\mathcal{W}^\pi)_{\pi\in\mathbb{S}_{n}})]
		&=P[\infty>q_{1-\alpha}(\chi^2_k)]=1.
	\end{align*}	
Again, by Urysohn's subsequence principle, the above result holds along $\{n\}\; _\square$	

	\section{Technical lemmas} 
	\label{sub:key_lemmas}
		
The following lemmas are used in the proof of the permutation version of the Marcinkiewicz-Zygmund inequality.

		\begin{lemma}[Kahane-Khintchine inequality]\label{lem:KK}
For all $r\in[1,\infty)$, there exists a constant $K_r$ depending only on $r$ such that
for any $d\times 1$ vectors $x_1,\dots, x_n\in\mathbb{R}^d$ and independent Rademacher variables
$\eps_1,\dots, \eps_n$
\begin{equation*}
\E\left[\left\Vert \sum_{i=1}^nx_i\eps_i\right\Vert^r\right]\leq
	K_r\left(\sum_{i=1}^n\Vert x_i\Vert^2\right)^{r/2}.
\end{equation*}			
\end{lemma}	
This is a special case of the general Kahane-Khintchine inequality, see Theorem 1.3.1 of \cite{Pena&Gine(2012)}. We also use the following lemma, which is a particular case of Theorem 4.1 in \cite{Chobanyan-Salehi(2001)}.

\begin{lemma}\label{lem:Sym}
	Let $\xi_1,\dots, \xi_n$ be exchangeable real random variables  satisfying $\sum_{i=1}^n\xi_i=0$ and let $a_1,\dots, a_n$ be vectors in $\R^d$. Then, for any $1<r<\infty$, $n>1$, and independent Rademacher variables $\eps_1,\dots, \eps_n$ that are independent of $\xi_1,\dots, \xi_n$,
	\begin{align}
\left(\E\left[\max_{1\leq k\leq n}\left\Vert\sum_{i=1}^ka_i\xi_i\right\Vert^r\right]\right)^{1/r}
\leq &
\frac{6r^2}{(r-1)^2}
\left(\E\left[\left\Vert \sum_{i=1}^na_i\xi_i\eps_i\right\Vert^r\right]\right)^{1/r} \notag \\
& +\frac{8(4r-1)}{(9r-1)(n-1)}
\left(\E\left[\left| \sum_{i=1}^n\xi_i\eps_i\right|^r\right]\right)^{1/r}
\sum_{i=1}^n\Vert a_i\Vert.\label{eq: sym}
	\end{align}			
\end{lemma}

The first lemma used below is a SLLN for triangular array of row-wise independent random variables, which follows from Theorem 1 and Corollary 1 of \cite{Hu-Moricz-Taylor(1989)}.
\begin{lemma}[Triangular array SLLN]\label{lem:HMT}
	Let $\{Y_{ni}: i=1,\dots, n; n=1, 2,\dots\}$ be an array of row-wise independent random vectors that satisfies either
	\begin{enumerate}[label=(\alph*)]
		\item\label{cond_nid} for some $\delta>0$
		$\sup_{n, i}\E[\Vert Y_{ni}\Vert ^{2+\delta}]<\infty$; or
		\item\label{cond_iid}$\{Y_{ni}\}_{i=1}^n$ have identical marginal distributions with
		$\E[\Vert Y_{ni}\Vert ^{2}]<\infty$.
	\end{enumerate}
	Then,
	$$n^{-1}\sum_{i=1}^n(Y_{ni}-\E[Y_{ni}])\conas 0.$$
\end{lemma}
\noindent \textit{Proof:} Let us first suppose that Condition \ref{cond_nid} holds. Let $Y_{nij}$ be the $j$th element of $Y_{ni}$. We have
\begin{equation*}
	\E[|Y_{nij}-\E[Y_{nij}]|^{2+\delta}]\leq 2^{1+\delta}\left(\E[|Y_{nij}|^{2+\delta}]+\E[Y_{nij}]^{2+\delta}\right)<\infty.
\end{equation*}
The result follows by applying Corollary 1 of \cite{Hu-Moricz-Taylor(1989)} with $p=1$ and $X_{ni}=Y_{nij}-\E[Y_{nij}]$. Now suppose that Condition \ref{cond_iid} holds. Then, as above, $\E[|Y_{nij}-\E[Y_{nij}]|^2]<\infty$ and the result follows from Theorem 1 of \cite{Hu-Moricz-Taylor(1989)}.\;$_\square$
\par
The following lemma is useful to establish Lindeberg's condition for the CLT and to control the growth of the maximum of independent random variables.

\begin{lemma}\label{lem:max2}
	Let $\{X_{ni}: i=1,\dots,n; n=1, 2,\dots\}$ be an array of row-wise independent random variables that satisfies either
	\begin{enumerate}[label=(\alph*)]
		\item for some $\delta>0$
		$\sup_{n, i}\E[\vert X_{ni}\vert ^{4+\delta}]<\infty$; or
		\item $\{X_{ni}\}_{i=1}^n$ have identical marginal distributions with
		$\sup_{n, i} \E[X_{ni}^{4}]<\infty$.
	\end{enumerate}
	Let $\tilde{X}_{ni}\equiv X_{ni}-\bar{X}_n$ with $\bar{X}_n\equiv n^{-1}\sum_{i=1}^nX_{ni}$. Then, for any $\epsilon>0$, there exists $C>0$ and $n_1$ such that almost surely, for all $n\geq n_1$,
	\begin{align}
		n^{-1}\sum_{i=1}^nX_{ni}^21(\vert X_{ni}\vert>C)<\epsilon,\label{DUI 1}\\ 
		n^{-1/2} \max_{1\leq i\leq n}{\vert X_{ni}\vert}\conas 0.\label{max as}
	\end{align}
\end{lemma}
\noindent \textit{Proof:}
For any $C>0$,
\begin{align}
	n^{-1}X_{ni}^2 	&=n^{-1}X_{ni}^2 1\left(X_{ni}^2\leq C^2\right)+n^{-1}X_{ni}^21\left(X_{ni}^2>C^2\right) \notag \\
	&\leq n^{-1}C^2+n^{-1}X_{ni}^21(X_{ni}^2>C^2) \notag \\
	&\leq n^{-1}{C^2}+n^{-1}\sum_{j=1}^nX_{nj}^21(X_{nj}^2>C^2).\label{eq:decomp_X2}
\end{align}
Since the RHS of the inequality in the last line does not depend on $i$,
$$n^{-1}\max_{1\leq i\leq n}X_{ni}^2\leq n^{-1}C^2+n^{-1}\sum_{i=1}^nX_{ni}^21(X_{ni}^2>C^2).$$
By Lemma \ref{lem:HMT},
\begin{equation}
	n^{-1}\sum_{i=1}^nX_{ni}^21(X_{ni}^2>C^2)-n^{-1}\sum_{i=1}^n\E[X_{ni}^21(X_{ni}^2>C^2)]\conas 0.
	\label{eq:conv_truncated_mean}
\end{equation}
Fix $\epsilon>0$. By the Cauchy-Schwarz and Markov's inequalities,
\begin{align*}
	\E[X_{ni}^21(X_{ni}^2>C^2)] & \leq (\E[X_{ni}^4])^{1/2}(\E[1(X_{ni}^2>C^2)])^{1/2} \\
	& \leq (\E[X_{ni}^4])^{1/2}(\E[X_{ni}^2]/C^2)^{1/2}\\
	& <\epsilon/4,	
\end{align*}
where the last inequality follows for $C$ sufficiently large. Then, in view of \eqref{eq:conv_truncated_mean}, there exists $n_0\in\mathbb{N}$ and $C_0>0$ such that almost surely (a.s.) and for all $n\geq n_0$,
$$n^{-1}\sum_{i=1}^nX_{ni}^21(X_{ni}^2>C_0^2)<\epsilon/2.$$
Next, choose $n_1$ such that $n_1^{-1}C_0^2\leq \epsilon/2$. Then, by \eqref{eq:decomp_X2}, we have a.s., for any $n\geq \max(n_0,n_1)$, $n^{-1}\max_{1\leq i\leq n}X_{ni}^2<\epsilon$.  Thus,  \eqref{DUI 1} and \eqref{max as} hold.
$_\square$

\medskip
We will also use the following simple lemma.
	
	\begin{lemma}\label{lem:CLT_sum}
		Let $X_n=(X_{n1},\dots,X_{nK})'$ be a random vector satisfying $X_n \cond \Norm{0,I_K}$. Then, for any $t_n\in \mathbb{R}^{K\times L}$ such that $t_n't_n = I_L$ and $L\leq K$, $t_n' X_n \cond \Norm{0,I_L}$.
	\end{lemma}
	
	\noindent\textit{Proof:} By Skorokhod representation theorem, there exists $\tilde{X}_n=(\tilde{X}_{n1},\dots,\tilde{X}_{nK})$ and $\tilde{X}=(\tilde{X}_{1},\dots,\tilde{X}_{K})$ with $\tilde{X}_n\stackrel{d}{=}X_n$ for all $n$, $\tilde{X}\sim \Norm{0,I_K}$ and such that $\tilde{X}_n\conas \tilde{X}$. Let $t_n$ be as in the lemma. Because
	$$\left\Vert t_n'(\tilde{X}_n - \tilde{X})\right\Vert\leq  \Vert t_n\Vert \Vert \tilde{X}_{n} - \tilde{X}\Vert= L^{1/2}\Vert \tilde{X}_{n} - \tilde{X}\Vert\conas 0,$$
	we have $t_n'\tilde{X}_n= t_n'\tilde{X} + o_p(1)$. Moreover, $t_n'\tilde{X}\sim \Norm{0,I_L}$. Thus, by Slutsky's lemma,
	$t_n'\tilde{X}_n\cond \Norm{0,I_L}$. The result follows since $t_n'\tilde{X}_n$ has the same distribution as $t_n' X_n. \;_\square$

\bigskip
The conditional version of multivariate Lindeberg CLT stated in the next lemma is obtained
from Theorem 1 and Corollary 3 of \cite{Bulinski(2017)}.
\begin{lemma}\label{lem: cond Lindeberg}
	Let $\{X_{ns}: s=1,\dots, q_n; n\in\mathbb{N}\}$ be a triangular array of $d\times 1$ random vectors which are conditionally independent in each row given a $\sigma$-field $\mathcal{A}_n$ for all $n\in\mathbb{N}$
	with $\E[X_{ns}\vert \mathcal{A}_n]=0$ and $\Sigma_{ns}=\E[X_{ns}X_{ns}'\vert \mathcal{A}_n]$.
	Let $\bar{\Sigma}_n=\sum_{s=1}^{q_n}\Sigma_{ns}$ and $\lambda_n=\lambda_{\min}(\bar{\Sigma}_n)$.
	Suppose
	\begin{align}
	&\liminf_{n\to\infty}\lambda_n>0\ \text{a.s.},\label{cond: Ceval}\\
	&\forall\epsilon>0\quad\frac{1}{\lambda_n}\sum_{s=1}^{q_n}\E\left[\Vert X_{ns}\Vert^2 1(\Vert X_{ns}\Vert^2>\epsilon \lambda_n)\vert \mathcal{A}_n\right]\conp 0. \label{cond: CLindeberg}
	\end{align}
	Then,
	\begin{equation}
		\bar{\Sigma}_n^{-1/2}\sum_{s=1}^{q_n}X_{ns}\cond \Norm{0, I_d}.
	\end{equation}
\end{lemma}
\noindent\textit{Proof:} Let $Y_{ns}=t'\bar{\Sigma}_n^{-1/2}X_{ns}$ where $t\in\mathbb{R}^d, t\neq 0$. We verify the conditions of Theorem 1 of \cite{Bulinski(2017)} for $Y_{ns}$. First, $(Y_{ns})_{1\le s\le q_n}$ are independent conditional on $\mathcal{A}_n$. Second,
$\V[Y_{ns}\vert\mathcal{A}_n]=t'\bar{\Sigma}_n^{-1/2}\Sigma_{ns}\bar{\Sigma}_n^{-1/2}t\leq  t'\bar{\Sigma}_n^{-1/2}\bar{\Sigma}_n\bar{\Sigma}_n^{-1/2}t=t't<\infty$ a.s.. Third,
$\sigma_n^2\equiv \V[\sum_{s=1}^{q_n}Y_{ns}\vert\mathcal{A}_n]=t't>0$ a.s.. Fourth,
on noting that $\E[Y_{ns}\vert\mathcal{A}_n]=0$ and letting
$\epsilon\equiv \eps^2$ for any $\eps>0$,
we have
\begin{align*}
	& \frac{1}{\sigma_n^2}\sum_{s=1}^{q_n}
	\E\left[Y_{ns}^21(\vert Y_{ns}\vert\geq \eps\sigma_n)\vert\mathcal{A}_n\right] \\
	&\leq \frac{1}{\sigma_n^2}
	\sum_{s=1}^{q_n}\Vert t'\bar{\Sigma}_n^{-1/2}\Vert^2 \E\left[\Vert X_{ns}\Vert^21\left(\Vert t'\bar{\Sigma}_n^{-1/2}\Vert^2
	\Vert X_{ns}\Vert^2 >\eps^2\sigma_n^2\right)\vert\mathcal{A}_n\right]\\
	&\leq \frac{\Vert t\Vert^2}{\sigma_n^2\lambda_n}
		\sum_{s=1}^{q_n}\E\left[\Vert X_{ns}\Vert^21\left(
	\Vert X_{ns}\Vert^2 >\frac{\eps^2\sigma_n^2\lambda_n}{\Vert t\Vert^2 }\right)\vert\mathcal{A}_n\right]\\
	&=\frac{1}{\lambda_n}
	\sum_{s=1}^{q_n}\E\left[\Vert X_{ns}\Vert^21\left(
	\Vert X_{ns}\Vert^2 >\epsilon\lambda_n\right)\vert\mathcal{A}_n\right]\\
	&\conp 0,
\end{align*}
where the first inequality is by Cauchy-Schwarz, the second is by $t'\bar{\Sigma}_n^{-1}t\leq \Vert t\Vert^2/ \lambda_n$ , the equality uses $\Vert t\Vert^2=\sigma_n^2$, and the convergence uses \eqref{cond: CLindeberg}. The result then follows from the Cram{\'e}r-Wold device and Theorem 1 and Corollary 3 of \cite{Bulinski(2017)}.
$\;_\square$

\newpage
\thispagestyle{empty}

\section{Additional results on the Project STAR}
\label{app:STAR}
\begin{table}[H]
	\small
	\caption{{95\% Confidence intervals for $\beta$ with standardized individual student scores.}}%
	\begin{center}
		\label{STAR3}
		\begin{tabular}{l|cc|cc|cc}
			\toprule
			\multicolumn{7}{c}{Standardized math test scores}\\ \midrule
			&\multicolumn{2}{c|}{Kindergarten}&\multicolumn{2}{c|}{Grade 1} &\multicolumn{2}{c}{Grade 2}\\ \midrule
			Test statistic & CI & Length & CI & Length  & CI & Length\\ \midrule
			SR           & [0.01,\,0.43] &0.42 & [0.25,\,0.60] &0.35 & [0.07,\,0.53] &0.46\\
			CP           & $\supseteq [-4,\,4]$ & $\ge 8$ & $\supseteq [-4,\,4]$ & $\ge 8$ & $\supseteq [-4,\,4]$ & $\ge 8$\\
			PC           & [0.03,\,0.40] &0.37 & [0.27,\,0.58] &0.31  & [0.10,\,0.50] &0.40\\
			Non-robust   &[0.00,\,0.44] &0.44 & [0.25,\,0.60] &0.35 &[0.06,\,0.53] &0.47 \\ 
			Robust HC3      & [-0.02,\,0.46] &0.48 &[0.22,\,0.62] &0.40 &  [0.03,\,0.56] &0.53\\ 
			Robust HC0 (IR) &[0.03,\, 0.40] & 0.36 & [0.27,\, 0.58] & 0.31 & [0.10,\,0.50] & 0.40\\	\midrule
			BP, JB, SW pval &\multicolumn{2}{c|}{0.58,\; 0.63,\; 0.80}&\multicolumn{2}{c|}{0.23,\; 0.63,\; 0.54} &\multicolumn{2}{c}{0.84,\; 0.70,\; 0.84}\\
			$n, S, |\mathbb{S}_n|$ &\multicolumn{2}{c|}{66,\; 15,\; $9.47\times 10^{24}$}&\multicolumn{2}{c|}{109,\; 25,\; $1.68\times 10^{41}$} &\multicolumn{2}{c}{79,\; 18,\; $9.16\times 10^{29}$}\\ \midrule
			\multicolumn{7}{c}{Standardized reading test scores}\\ \midrule
			&\multicolumn{2}{c|}{Kindergarten}&\multicolumn{2}{c|}{Grade 1} &\multicolumn{2}{c}{Grade 2}\\ \midrule
			Test statistic & CI & Length & CI & Length & CI & Length\\ \midrule
			SR & [-0.07,\,0.44] &0.51   & [0.32,\,0.62] &0.30 & [0.13,\,0.52] &0.39 \\
			CP & $\supseteq [-4,\,4]$ & $\ge 8$ & $\supseteq [-4,\,4]$ & $\ge 8$ & $\supseteq [-4,\,4]$ & $\ge 8$\\
			PC &[-0.04,\,0.41] &0.45  & [0.34,\,0.60] &0.26 & [0.16,\,0.50] &0.34\\
			Non-robust & [-0.07,\,0.44] &0.50 & [0.32,\,0.62] &0.30 & [0.13,\,0.53] &0.40 \\
			Robust HC3 & [-0.10,\,0.47] &0.58 & [0.30,\,0.64] &0.34 & [0.10,\,0.55] &0.45 \\
			Robust HC0 (IR) &[-0.04,\,0.40] & 0.44 & [0.34,\,0.60] & 0.26 & [0.16,\,0.50] & 0.34\\
			\midrule
			BP, JB, SW pval &\multicolumn{2}{c|}{0.28,\; 0.16,\; 0.34}&\multicolumn{2}{c|}{0.02,\; 0.90,\; 0.91} &\multicolumn{2}{c}{0.77,\; 0.24,\; 0.57}\\
			$n, S, |\mathbb{S}_n|$ &\multicolumn{2}{c|}{66,\; 15,\; $9.47\times 10^{24}$}&\multicolumn{2}{c|}{109,\; 25,\; $1.68\times 10^{41}$} &\multicolumn{2}{c}{79,\; 18,\; $9.16\times 10^{29}$}\\
			\bottomrule
		\end{tabular}
	\end{center}
	\footnotesize{Notes: this table is similar to Table \ref{STAR}, except that the dependent variable is the class average test scores when individual student scores are normalized to have mean zero and standard deviation one across all the students in the schools with at least 2 regular and 2 small classes.} 
	\end{table}
		
\end{document}